\documentclass[11pt]{article}
\usepackage{setspace, graphicx, caption, subcaption, float, relsize, rotating, color, verbatim, multirow}
\usepackage{graphicx}
\usepackage[breaklinks=true,colorlinks=true, urlcolor=blue,linkcolor=blue, citecolor=magenta]{hyperref}
\usepackage{empheq}
\usepackage{enumerate}
\usepackage{dsfont}
\usepackage{mathtools}
\usepackage{bm}
\usepackage{amssymb}
\usepackage{amsmath}
\usepackage{amsfonts}
\usepackage{amsthm}
\usepackage{mathrsfs}
\usepackage{bigints}
\usepackage{enumerate}
\usepackage{authblk}
\usepackage{color}
\usepackage{sectsty}
\usepackage{amsfonts}
\usepackage{graphicx}
\usepackage{times}
\usepackage[plain,noend]{algorithm2e}

\usepackage{amsfonts}
\usepackage{amsmath}
\usepackage{graphicx}
\usepackage{xr-hyper}
\usepackage{mathrsfs}
\usepackage{amssymb}
\usepackage{amsthm}
\usepackage{color}
\usepackage{epsfig,latexsym,verbatim}
\usepackage{multirow}
\usepackage{multicol}
\usepackage{float}
\usepackage[authoryear]{natbib}
\usepackage{bm}
\usepackage{algorithmic}
\usepackage{tabularx}
\usepackage{booktabs}
\usepackage{multirow}
\usepackage{dsfont}
\usepackage[normalem]{ulem}
\usepackage[textwidth=25mm,textsize=footnotesize]{todonotes}

\renewcommand{\thefootnote}{\fnsymbol{footnote}}
\renewcommand{\arraystretch}{1.3}

\def\bs{\boldsymbol}

\newtheorem{thm}{Theorem}

\newtheorem{rk}{Remark}
\newtheorem{theorem}[thm]{Theorem}
\newtheorem{assumption}[]{Assumption}
\newtheorem{lemma}[thm]{Lemma}

\newcommand{\nc}{\newcommand}
\nc{\dps}{\displaystyle}
\nc{\tr}{\text{tr}}

\numberwithin{equation}{section}
\allowdisplaybreaks

\date{}
\begin{document}

\title{{\Large Causal Inference of General Treatment Effects using Neural Networks with A Diverging Number of Confounders}\thanks{This paper is a revised version of ``Efficient Estimation of General
			Treatment Effects using Neural
			Networks with A Diverging Number of
			Confounders", arxiv:2009.07055 (15 Sep 2020). This work has been presented since 2020 at various workshops and seminars including: UCSB, USC, UC-Irvine,  Guanghua School of Management at Peking University, University of Chicago, Tsinghua University, and Toulouse School of Economics. The authors are equal contributors and are alphabetically ordered.}}
\author[$^\dagger$]{Xiaohong Chen \thanks{E-mail:
		\texttt{xiaohong.chen@yale.edu} }}
\author[$^\star$]{Ying Liu \thanks{E-mail:
		\texttt{yliu364@ucr.edu} }}
\author[$^\star$]{Shujie Ma \thanks{\texttt{Corresponding Author} E-mail:
\texttt{shujie.ma@ucr.edu}}}
\author[$\ddagger$] {Zheng Zhang \thanks{\texttt{Corresponding Author}
E-mail: \texttt{zhengzhang@ruc.edu.cn}}}

\affil[$\dagger$]{Department of Economics, Yale University}

\affil[$\star$]{ Department of Statistics,
	University of California at Riverside}

\affil[$\ddagger$]{Center for Applied Statistics, Institute of Statistics \& Big Data, Renmin University of China}
\maketitle

\begin{abstract}
Semiparametric efficient estimation of various multi-valued causal effects, including quantile treatment effects, is important in economic, biomedical, and other social sciences. Under the unconfoundedness condition, adjustment for confounders requires estimating the nuisance functions relating outcome or treatment to confounders nonparametrically. This paper considers a generalized optimization framework for efficient estimation of general treatment effects using artificial neural networks (ANNs) to approximate the unknown nuisance function of growing-dimensional confounders. We establish a new approximation error bound for the ANNs to the nuisance function belonging to a mixed smoothness class without a known sparsity structure. We show that the ANNs can alleviate the \textquotedblleft curse of dimensionality\textquotedblright\ under this circumstance. We establish
the root-$n$ consistency and asymptotic normality of the proposed general treatment effects estimators, and apply a weighted bootstrap procedure for conducting inference. The proposed
methods are illustrated via simulation studies and a real data application.
\end{abstract}

\setcounter{section}{0}

\baselineskip0.6cm



\renewcommand\thefootnote{\arabic{footnote}}

\section{Introduction}
Estimation of and inference on various causal effects using observational data have been popular in economics and other sciences. Under the unconfoundedness assumption, semiparametric efficient estimation of multi-valued Treatment Effects (TEs), including quantile TEs and asymmetric TEs, requires nonparametric estimation of a nuisance function relating outcome and/or treatment to confounders. Various nonparametric linear smoothers such as kernels and splines have been used in Outcome Regression (OR) or Inverse Probability Weighting (IPW) based TE studies. In many applied works, researchers believe that the unconfoundedness assumption is likely to hold conditioning on many confounders/covariates. However, popular nonparametric linear smoothers estimated nuisance function(s) of many covariates suffer from the so-called ``curse of dimensionality". Artificial Neural Networks (ANNs) are nonlinear sieves that can approximate an unknown function of high dimensional covariates better than linear sieves such as splines, cosines, and polynomials. Moreover, recent computation advances have made the implementation of ANNs easier. This motivates our investigation of ANN-based efficient estimation and inference on general TEs with increasing dimensional confounders, without known sparsity.

In this paper, we propose an ANN-based, root-$n$ consistent, asymptotic normal, and efficient estimator of general multi-valued TE. Our TE estimator is obtained by directly optimizing a generalized objective function that involves an ANN-approximated nonparametric Propensity Score (PS) function, which is the only nuisance function to be estimated. Theoretically, we derive a new convergence rate of the ANN estimator for the nuisance function under mild conditions when the number of confounders is allowed to grow with the sample size ($n$). Our method can be naturally used to estimate general TEs, including the
average, quantile, and asymmetric least squares TEs. In addition, our optimization
procedure enables us to easily construct  convenient weighted bootstrapped confidence sets,
without the need of estimating the asymptotic variances that are of complicated forms for quantile TEs and asymmetric TEs.\footnote{See our online supplement for consistent estimation of the asymptotic variances. Nevertheless, our simulation studies indicate that bootstrapped CSs are more accurate than the CSs based on estimated asymptotic variance.}

Feedforward ANNs are effective tools for solving the classification and prediction problems with high dimensional covariates and big data sets.  The basic idea is to extract linear combinations of the inputs as features, and then model the target as a nonlinear function of these features. It has been shown in the literature %
\citep{barron1993universal,hornik1994degree,chen1999improved,klusowski2018approximation}
that when the unknown target function admits a Fourier representation with a bounded
moment, its ANNs approximator enjoys a fast approximation rate, making ANNs
a promising tool to potentially break the notorious \textquotedblleft curse
of dimensionality\textquotedblright\ in nonparametric multivariate
regression. This Fourier function class is recently named ``Barron class'' by  \cite{e2022barron}, who claim it is one right function space to address the curse of dimensionality problem. Nevertheless, it is of interest to investigate how the Barron class is related to some classical function spaces such as the Sobolev space \citep{stone1994,Wasserman2006} commonly used in the nonparametric regression literature.\footnote{The minimax optimal rate in root-mean squared error norm for estimating a function in the standard Sobolev ball is known, and no nonparametric estimator can avoid the ``curse of dimensionality" for the standard Sobolev ball.} Moreover, it is still unclear how the
moment of the Fourier transform appeared in the ANN approximation error bounds depends on the dimension of the covariates. This moment is implicitly treated as a constant in the existing works on ANNs %
\citep{barron1993universal,chen1999improved,klusowski2018approximation} as
they consider fixed dimensions.

In this paper, we introduce a mixed smoothness class, and show that it is a subset of the Barron class. For any function in the mixed smoothness class, we derive an upper bound for the moment of its Fourier transform in terms of the dimension of the covariates. Functions in this mixed smoothness class need to be at least one order smoother in each coordinate than those in the standard Sobolev ball. We show that the nonlinear ANN estimators for functions in the mixed smoothness class have fast convergence rates. We also show that the conventional linear sieve approximators still suffer the ``curse of dimensionality'' when the target function belongs to the mixed smoothness class.
% \citep{Schmidt2020}.
%We also show that the
%conventional linear sieve approximators still have the dimensionality
%problem when the target function belongs to the mixed smoothness ball.
%
%Neither this connection nor the upper bound of the Fourier transform moment
%has been investigated in the literature.
Our new theoretical results enhance readers' understanding why single hidden layer ANNs perform better than nonparametric linear smoothers when estimating functions in a mixed smoothness class with increasing dimensional covariates.

%Moreover, different from \cite%
%{chen1999improved}, our ANNs class no longer requires the $L_{1}$ norm
%constraint on the weights. This greatly facilitate the computation of the
%ANNs estimates.

While the development of credible inferential
theories for the ANN-based estimator of TEs is essential to test the
significance of the various causal effects, it is also a daunting task because of the complex nonlinear structure of the ANNs. In this paper,
we establish the root-$n$ asymptotic normality of our ANN-based TE estimator when the number of the confounders is allowed to grow with the sample size.
%without the
%need of assuming that the PS function is bounded below by a constant. This is a strong
%assumption, especially in the settings with a large number of confounders,
%albeit still used in the literature, see for example %
%\citep{athey2018approximate}.
%and find that the number of the confounders is
%allowed to grow with the sample size with a rate no greater than $\{\log
%(n)\}^{1/2}$ to ensure the desirable asymptotic property of the ANNs-based
%TE estimator for conducting inference.
Different from the earlier works on semiparametric efficient estimation and inference for TEs (see, e.g.,
\cite{robins1994estimation,Hirano03,chen2008semiparametric,ai2018unified}), our semiparametric inferential theory allows for settings with
diverging dimensional confounders. To the best of our knowledge, our paper is the first to provide a thorough theoretical justification for the ANN-based inferential procedures for general TEs when the dimension of the confounders can grow with the
sample size.

Our ANN-based TE estimator is obtained directly from a generalized optimization procedure without estimating the Efficient Influence Function (EIF). The estimated EIF approach requires estimating two nuisance functions nonparametrically, while our optimization based procedure involves estimating one nuisance function only. Recently, \cite{farrell2018deep} proposed an ANN-based Doubly Robust (DR) (or EIF based) estimator of average TEs, which involves estimating both OR and PS nuisance functions via ANNs. They assume that both nuisance functions belong to the standard Sobolev (or H\"{o}lder) ball and the dimension of the confounders is fixed. Although the EIF estimation-based method is commonly used for estimating average TEs, see for example \cite{CTD09, T10, VR11, RLSR12,KMMS17,chernozhukov2018double}, it can be more difficult to apply in quantile, asymmetric and other complex TE settings, as these TE parameters can enter the estimated EIF equation in a nonlinear and non-separable fashion. When the nuisance functions are trained by nonlinear machine learning algorithms, it becomes even more computationally challenging to estimate the EIF in these complex settings. Our TE estimator is obtained directly from optimizing an objective function with a plug-in ANN-based estimator of the PS nuisance function, so a weighted bootstrap procedure can be conveniently applied for conducting inference without estimating the EIF nor the asymptotic variance function. To better illustrate our ANN-based TE estimation and inference procedures, we focus on using the ANNs with one hidden layer in the main text, and discuss the extension to ANNs with multiple hidden layers in the online supplement.

Finally, for those readers who care about efficient estimation of the averaged treatment effect (ATE) only, we also propose an
ANN-based efficient estimator of the ATE obtained from the ANN estimated OR nuisance function. Under standard regularity conditions, our proposed ANN-PS and ANN-OR estimators for ATE have the same asymptotic distribution, and are both asymptotic efficient when the number of covariates is fixed. We show that, unlike the estimators using IPW and DR methods, the ANN-OR based ATE estimator can achieve the root-$n$ asymptotic normality without imposing the strict overlap condition on the PS function. Consequently, the ANN-OR based ATE estimator is more robust than the ANN-PS based estimator when the true unknown PS function is close to zero. In our Monte Carlo simulations, we also observe that ANN-OR based ATE estimator performs slightly better than ANN-DR ATE estimator, which in turn performs slightly better than ANN-IPW ATE estimator. However, it is difficult to apply the ANN-OR based procedure to estimate other types of multi-valued TEs such as quantile TEs.

The rest of the paper is organized as follows. Section \ref{sec:ANN} provides a new approximation error rate result for ANNs to a mixed smoothness class of functions with diverging dimension. Section \ref{sec:framework} introduces the general multi-valued TEs and our proposed ANN-based estimators for TEs. Section \ref{sec:asymptotics} establishes the large sample properties and Section \ref{sec:variance} presents the inferential procedures. Section \ref{sec:GTT} extends the optimization procedures and the asymptotic properties to general multi-valued treatment effects for the treated subgroups. Section \ref{sec:simulation} reports simulation studies and
Section \ref{sec:application} contains a real data application. Section \ref{sec:conclusion} briefly concludes. All the technical proofs and additional simulation results are provided in the Appendix and the on-line
Supplemental Materials.

\section{ANNs approximation for functions in the Barron class}\label{sec:ANN}

For nonparametric estimation of a target function of high dimensional covariates, in addition to specifying the approximation basis, identifying a ``good'' target function space is also crucial in machine learning literature, as \cite{e2022barron} write:
\begin{quote}
		Sobolev/Besov type spaces are not the right function spaces for studying machine learning models that can potentially address the curse of dimensionality problem.
\end{quote}
In this section we present ANNs with one hidden layer and the related approximation results for a target function in the Barron class.

Let $\mathcal{X}$ denote the  support of a random vector $\boldsymbol{X}$ which is compact in $\mathbb{R}^p$. Without loss of generality, we assume $\mathcal{X}=[0,1]^p$. Let $F_X$ be the cumulative distribution function (CDF) of $\boldsymbol{X}$. Denote the $L^2(dF_X)$-norm of any function $f(\cdot)$ by $\|f\|_{L^2(dF_X)}:=\left\{\int_{\mathcal{X}}|f(x)|^2dF_X(x)\right\}^{1/2}$. Let $\widetilde{f}(\bs{a})$ be the Fourier transform of $f(x)$ defined by
\begin{align*}
	\widetilde{f}(\bs{a}):=\frac{1}{(2\pi)^p}\int_{\mathbb{R}^p}\exp\left(-i \bs{a}^\top x \right)f(x)dx,
\end{align*}
where $\bs{a}=(a_1,...,a_p)^{\top}\in\mathbb{R}^p$. We define the $m^{th}$ moment of the Fourier transform of $f(x)$ as $v_{f,m}:=\int_{\mathbb{R}^p} |\bs{a}|_1^m|\widetilde{f}(a)|da$, where $|\bs{a}|_1:=\sum_{i=1}^p|a_i|$.
 \cite{barron1993universal}, \cite{hornik1994degree}, \cite{chen1999improved}, and \cite{klusowski2018approximation} considered the target function belonging to the Barron  class  $	\mathcal{F}_p^m$:
\begin{align}\label{def:F_p^m+1}
	\mathcal{F}^m_p:=\bigg\{ f:\mathcal{X}\to \mathbb{R}: f(x)=\int_{\mathbb{R}^p} \exp\left(i \bs{a}^\top x \right)\widetilde{f}(\bs{a}) d\bs{a}, \ v_{f,m}<\infty \bigg\}.
\end{align}
$\mathcal{F}^m_p$ contains a class of functions of $p$ dimension that admit Fourier representations  with the finite $m^{th}$ moment. The input variables of functions in $\mathcal{F}^m_p$ have dimension $p$, which is allowed to grow to infinity as the sample size $n$ increases.  It is worth noting that $v_{f,m}$ depends on the dimension $p$, and its value can increase with $p$. In the nonparametric regression literature, spaces with certain smoothness constraints such as the H\"{o}lder or Sobolev space are instead more commonly used \citep{stone1994,Wasserman2006,chen2007large}.  We will build a connection between the function class $\mathcal{F}_p^m$ given in (\ref{def:F_p^m+1}) and a mixed smoothness ball, and will establish an upper bound for $v_{f,m}$ in terms of the dimension $p$ in Theorem \ref{thm:vfm_finite}. To the best of our knowledge, our paper is the first one that builds such a connection between the Fourier function class used for ANNs and a mixed smoothness ball and establishes an upper bound for $v_{f,m}$  which appears in the approximation error bounds.

Consider to approximate a target function $f\in \mathcal{F}^m_p$ using the ANNs, belonging to the class
\begin{align}
	\mathcal{G}(\psi,B,&r,p)
	=\bigg\{g:g(x)=g_0(x;\boldsymbol{\gamma}_0)+\frac{B}{r}\sum_{j=1}^{r} \gamma_j\psi(\bs{a}_j^\top x),\ \bs{a}_j=(a_{j1},...,a_{jp})^\top\in\mathbb{R}^p, \notag\\ & \ \|\bs{a}_{j}\|_2= 1, \ |\gamma_j|\leq 1, \ j\in\{1,...,r\}, \ B\in \mathbb{R}^+   \bigg\},\label{def:Gp}
\end{align}
where  $g_0(x;\boldsymbol{\gamma}_0)$ is a parametric function indexed by an unknown parameter vector $\boldsymbol{\gamma}_0$, and $\|\bs{a}_{j}\|_2:=\{|a_{j1}|^2+...+|a_{jp}|^2\}^{1/2}$. The structure of $g_0(x;\boldsymbol{\gamma}_0)$ depends on the type of the activation function that is used. For example, if the ReLU activation function is used, then $g_0(x;\boldsymbol{\gamma}_0)=\boldsymbol{\gamma}_0^\top x$. $\mathcal{G}(\psi,B,r,p)$ is the collection of output functions for neural networks with $p$-dimensional input feature $x$,  a single hidden layer with $r$ hidden units and an activation function $\psi$ and real-valued input-to-hidden unit weights $(\bs{a}_j)$, and hidden-to-output weights $(\gamma_j^*:=B\gamma_j)$. Note that for any outer parameter $\gamma_j^*\in \mathbb{R}$, it can be written as $\gamma_j^*=B\gamma_j$ with $\ |\gamma_j|\leq 1$, and $B$ is a scale factor of all  $\gamma_j^*$'s. Both $r$ and $B$ are allowed to increase with the sample size $n$, which will be discussed in Section \ref{IPW-TE}. The ANN class in (\ref{def:Gp}) comes from the class given in \cite{klusowski2018approximation}. The difference is that we change their $\ell_1$ constraint on the inner parameters $\bs{a}_j$ to a $\ell_2$ normalization $\|\bs{a}_{j}\|_2= 1$. This normalization has been commonly used in semiparametric index models, see   \cite{mahesingle2016}.

The approximation error for a target function depends on the smoothness of  the approximand, the dimension of the covariates, and the type of approximation basis.  We first present the approximation results based on some popularly used neural networks, which have been  established in the existing literature:
\begin{itemize}
	\item  (Sigmoid type activation function)  Suppose that  the function $f\in\mathcal{F}_p^1$,  $g_0(x;\boldsymbol{\gamma}_0)\equiv 0$, the activation function $\psi(\cdot)$ is compactly supported, bounded, and uniformly Lipschitz  continuous. If  $B\leq 2 v_{f,1}<\infty$,  then \citet[Theorem 2.1]{chen1999improved} show that the $L^2(dF_X)$-approximation rate of $f$ based on ANN is
	\begin{align}\label{rate:ANN}
		\inf_{g\in	\mathcal{G}(\psi,B,r,p)}\left\{\int_{\mathcal{X}}|f(x)-g(x)|^2dF_X(x)\right\}^{1/2}\leq \text{const}\times v_{f,1}\cdot r^{-\frac{1}{2}-\frac{1}{p}}.
	\end{align}
	The activation functions $\psi$ include the Heaviside, logistic, tanh, cosine squasher, and other sigmoid functions \citep{hornik1994degree,makovoz1996random}, but do not include the ReLU and squared ReLU ridge functions stated below.	
	\item  (ReLU activation function) Suppose that the function $f\in\mathcal{F}_p^2$,  $g_0(x;\boldsymbol{\gamma}_0)=\boldsymbol{\gamma}_0^\top x$ for $\boldsymbol{\gamma}_0\in \mathbb{R}^p$, and $\psi(\bs{a}^\top x)=(\bs{a}^\top x)_{+}$.  If $B\leq 2 v_{f,2}<\infty $, then \citet[Theorem 2 and its discussion on page 7651]{klusowski2018approximation} show that the $L^{2}(dF_X)$-approximation rate  based on ReLU ridge functions is
	\begin{align}\label{rate:ReLU}
		\inf_{g\in	\mathcal{G}(\psi,B,r,p)}\left\{\int_{\mathcal{X}}|f(x)-g(x)|^2dF_X(x)\right\}^{1/2}\leq \text{const}\times v_{f,2}\cdot r^{-\frac{1}{2}-\frac{1}{p}}.
	\end{align}
	\item  (Squared ReLU activation function) Suppose that the function $f\in\mathcal{F}^3_p$, $g_0(x;\boldsymbol{\gamma}_0)=\gamma_{01}^\top x+x^\top \gamma_{02} \cdot x$ for $\boldsymbol{\gamma}_0=\{\gamma_{01},\gamma_{02}\}\in \mathbb{R}^p\times \mathbb{R}^{p\times p}$, and $\psi(\bs{a}^\top x)=(\bs{a}^\top x)^2_{+}$. If  $B\leq 2 v_{f,3}<\infty$, then \citet[Theorem 3 and its discussion on page 7651]{klusowski2018approximation} show that the $L^2(dF_X)$-approximation rate  based on squared ReLU ridge functions is
	\begin{align}\label{rate:sqrReLU}
		\inf_{g\in	\mathcal{G}(\psi,B,r,p)}\left\{\int_{\mathcal{X}}|f(x)-g(x)|^2dF_X(x)\right\}^{1/2}\leq \text{const}\times v_{f,3}\cdot r^{-\frac{1}{2}-\frac{1}{p}}.
	\end{align}
\end{itemize}
If the target function $f(x)$ is in a Barron class with a finite moment given in (\ref{def:F_p^m+1}), then \eqref{rate:ANN}, \eqref{rate:ReLU} and \eqref{rate:sqrReLU} show that the $L^{2}(dF_{X})$-approximation rates of ANNs are $O(v_{f,m}\cdot r^{-1/2-1/p})=o(v_{f,m}\cdot r^{-1/2})$ for $m=1,2,3$, in which $r^{-1/2}$ no longer depends on the dimension $p$. Thus, the resulting ANNs estimator
can break the \textquotedblleft curse of dimensionality\textquotedblright that typically arises in the nonparametric kernel and linear sieve estimation.

\begin{rk}
Recently, \cite{devore2023weighted} propose weighted variation spaces that enlarge the second order Barron space $\mathcal{F}_p^2$. They show that the $L^2(dF_X)$-approximation rate based on the shallow ReLU neural networks  for a function $f(\cdot)$ belonging to their weighted variation spaces achieves   $\text{const}\times\|f\|_{\mathcal{V}_w}\cdot r^{-\frac{1}{2}-\frac{3}{2p}}$, where $\|\cdot\|_{\mathcal{V}_w}$ is the weighted variation norm defined in \citet[Section 4, page 8]{devore2023weighted}. Our theoretical results for the classic Barron space presented in this article  can also be adapted to the weighted variation spaces. We focus on the Barron space  for easing the presentation.
\end{rk}

\subsection{The mixed smoothness ball}\label{sec:mixH}
To the best of our knowledge, there are two questions that remain unanswered in the literature, including (1) how restrictive this Barron
 class $\mathcal{F}_p^m$ is compared to the conventional smoothness spaces such as the Sobolev ball typically assumed in
the multivariate nonparametric regression literature, and (2) how the moment of the Fourier transform $v_{f,m}$ depends on the dimension $p$. To address
these questions, we build a connection between the Barron class $\mathcal{F}_p^m$ and a mixed smoothness ball, and establish an upper bound
for $v_{f,m}$.

Given a $p$-tuple $\bs{\alpha}=(\alpha_1,...,\alpha_p)$ of nonnegative integers, set $|\boldsymbol{\alpha}|_{1}:=\sum_{j=1}^p\alpha_j$ and let $D^{\bs{\alpha}}$ denote the differential operator defined by
\begin{align*}
	D^{\bs{\alpha}}:=\frac{\partial^{|\bs{\alpha}|_1}}{\partial x_1^{\alpha_1}\cdots \partial x_p^{\alpha_p}}.
\end{align*}
For an integer $s\in\mathbb{N}$, denote the class of all $s$-times continuously differentiable real-valued functions on $\mathcal{X}$ by $\mathcal{C}^{s,\infty}(\mathcal{X})$:
\begin{align}\label{def:C^s}
\mathcal{C}^{s,\infty}(\mathcal{X}):=\left\{f(\cdot):\sup_{|\bs{\alpha}|_1\leq s}|D^{\bs{\alpha}}f(x)|\leq 1\right\}.
\end{align}
The upper bound ``$1$" used in the definition of $\mathcal{C}^{s,\infty}(\mathcal{X})$  is only for notational simplicity and it can be replaced by any finite positive constant. The function class $\mathcal{C}^{s,\infty}(\mathcal{X})$ is called the  Sobolev ball (in $L^{\infty}$-norm) of order $s$.

Let $\partial_{x_1}\cdots \partial_{x_p}$ be the  partial derivative with respect to $x=(x_1,...,x_p)^{\top}$ defined by
\begin{align*}
\partial_{x_1}\cdots \partial_{x_p}:=\frac{\partial^{p}}{\partial x_1\cdots \partial x_p}.
\end{align*}
For any  $\delta>0$, define $\Delta_{x_i}^{\delta}$ to be the difference operator by
\begin{align*}
	\Delta_{x_i}^{\delta}f\left(x_1,...,x_p\right):=f\left(x_1,...,x_{i-1},x_i,x_{i+1},...,x_p\right)-f\left(x_1,...,x_{i-1},x_i-\delta,x_{i+1},...,x_p\right),
\end{align*}
for $i\in\{1,2,...,p\}$. We have
\begin{align*}
	\lim_{\delta_1,...,\delta_p\to 0}\frac{\Delta_{x_p}^{\delta_p}\cdots \Delta_{x_1}^{\delta_1}f\left(x\right)}{\delta_1\cdots \delta_p}=\partial_{x_1}\cdots \partial_{x_p} f(x),
\end{align*}
provided that  the partial derivative $\partial_{x_1}\cdots \partial_{x_p} f(x)$ exists.

Being motivated by the application of sparse grids in dealing with high-dimensional partial differential equations (PDEs) \citep{bungartz2004sparse}, we define $\mathcal{W}^{m,1+\epsilon,\infty }(\mathcal{X})$ for $m\in\mathbb{N}$ and $\epsilon\in (0,1]$ to be the \emph{mixed smoothness ball} of order $(m,1+\epsilon)$:
\begin{align}\label{def:mix}
\mathcal{W}^{m,1+\epsilon,\infty}&(\mathcal{X}):=\bigg\{f:f(\cdot)\in\mathcal{C}^{m,\infty}(\mathcal{X}),\sup_{\{\forall \boldsymbol{\alpha}: |\boldsymbol{\alpha}|_{1}= m\}}\sup_{x\in \mathcal{X}}\left|\partial_{x_1}\cdots \partial_{x_p}D^{\boldsymbol{\alpha}}f(x)\right|\leq 1, \notag \\
& \quad  \sup_{\{\forall \boldsymbol{\alpha}: |\boldsymbol{\alpha}|_{1}= m\}}\sup_{\{x\in \mathcal{X},\delta_1>0,...,\delta_p>0\}}\frac{\left|\Delta^{\delta_p}_{x_p}\cdots\Delta^{\delta_1}_{x_1}\partial_{x_1}\cdots \partial_{x_p}D^{\boldsymbol{\alpha}}f(x)\right|}{\delta_1^{\epsilon}\cdots \delta_p^{\epsilon}}\leq 1\bigg\}.
\end{align}

  The following result states that $\mathcal{W}^{m,1+\epsilon,\infty}(\mathcal{X})$ with $\epsilon\in (0,1]$  is a subspace of $\mathcal{F}_p^m$. The proof is relegated to Appendix \ref{appendix_ANN_approx}.
\begin{thm}\label{thm:vfm_finite}
	Let $f\in \mathcal{W}^{m,1+\epsilon,\infty}(\mathcal{X})$ for some $\epsilon\in (0,1]$, then  $f\in\mathcal{F}_p^m$ and $$v_{f,m}\leq 2\cdot \left(M_0\cdot \frac{\pi}{2}\right)^m \cdot M^{p}$$ for some universal constant $M$ defined by
	$M:=M_0\cdot \left(\frac{\pi}{2}\right)\cdot \left(\frac{1}{2\epsilon}+\frac{1}{2}\right) ,$
	where the definition of $M_0$ is given in \eqref{def:M0}.
\end{thm}
To the best of our knowledge, Theorem \ref{thm:vfm_finite} is the first result in the literature that provides a connection between the mixed smoothness ball $\mathcal{W}^{m,1+\epsilon,\infty }(\mathcal{X})$ and $\mathcal{F}_p^m$. Theorem \ref{thm:vfm_finite} shows that the mixed smoothness ball $\mathcal{W}^{m,1+\epsilon,\infty }(\mathcal{X})$ is a subspace of the Barron class $\mathcal{F}_p^m$. Thus, for any functions in $\mathcal{W}^{m,1+\epsilon,\infty }(\mathcal{X})$, their ANNs approximators enjoy the nice approximation rates given in \eqref{rate:ANN}-\eqref{rate:sqrReLU}.  Furthermore, Theorem \ref{thm:vfm_finite} explicitly provides an upper bound for the $m^{th}$ moment of the Fourier transform $v_{f,m}$, which enables us to evaluate the effect of the dimension $p$ on the approximation rates of
ANNs. This upper bound has not been provided in the literature, and the theories  in most existing works on ANNs are established by assuming that $p$ is fixed.

\begin{rk}
In particular, if $\epsilon=1$ in \eqref{def:mix}, we  obtain
\begin{align*}
	\mathcal{W}^{m,2,\infty}(\mathcal{X}):=&\bigg\{f(\cdot):f(\cdot)\in\mathcal{C}^{m,\infty}(\mathcal{X}) \ \text{and} \ \sup_{\{\forall \boldsymbol{\alpha}:  |\boldsymbol{\alpha}|_{1}= m\}}\sup_{x\in \mathcal{X}}\left|\partial^2_{x_1}\cdots \partial^2_{x_p}D^{\boldsymbol{\alpha}}f(x)\right|\leq 1\bigg\},
\end{align*}
where
\begin{align*}
\partial^2_{x_1}\cdots \partial^2_{x_p}:=\frac{\partial^{2p}}{\partial x^2_1\cdots \partial x^2_p}.
\end{align*}
 The functions in the mixed smoothness ball $\mathcal{W}^{m,2,\infty}(\mathcal{X})$ need to be $2$-order smoother in each coordinate of $\bs{X}$ than the functions in the regular Sobolev ball of order $s=m$ given in \eqref{def:C^s}, i.e.  $\mathcal{C}^{m,\infty}(\mathcal{X})$. It is worth noting that  $\mathcal{W}^{m,2,\infty}(\mathcal{X})$  is much broader than the Sobolev ball of order $s=m+2p$; indeed, we have the following inclusion relation:
\begin{align*}
	&\mathcal{C}^{m+2p,\infty}(\mathcal{X})\subsetneq \mathcal{W}^{m,2,\infty}(\mathcal{X})\subset \mathcal{W}^{m,1+\epsilon,\infty}(\mathcal{X}) \subset \mathcal{F}^m_p, \quad \text{for} \ \epsilon\in(0,1].
\end{align*}
The functions in this mixed smoothness ball do not need a compositional structure such as a hierarchical interaction structure considered in \cite{bauer2019deep} and \cite{Schmidt2020}. We should be mindful that breaking the curse of dimensionality happens at the cost of
sacrificing flexibility. If a function is assumed to be in the Sobolev ball of order $m$, the nonparametric optimal minimax rates suffer from the curse of dimensionality, i.e., no nonparametric estimator can avoid the dimensionality problem under this condition \citep{Schmidt2020}.

\end{rk}

\section{Parameters of interest and ANN-based estimators}\label{sec:framework}
In this section, we first define our general treatment effect parameters of interest, and then introduce our ANN optimization based estimators.

Let $D$ denote a treatment variable taking value in $\mathcal{D}=\{0,1,...,J\}$, where $J\geq 1$ is a positive integer. Let $
Y^*(d)$ denote the potential outcome when the treatment status $D=d$ is assigned. The probability density of $Y^*(d)$ exists, denoted by $f_{Y^*(d)}$, is continuously differentiable.   Let $\mathcal{L}(\cdot
)$ denote a nonnegative and strictly convex loss function satisfying $\mathcal{L}(0)=0$ and $\mathcal{L}(v)\geq 0$ for all $v\in\mathbb{R}$. The derivative of $\mathcal{L}(\cdot)$ exists almost everywhere and non-constant which is denoted by $
\mathcal{L}^{\prime }(\cdot )$. Let $\boldsymbol{\beta}^*=({\beta}^*_0,{\beta}^*_1,\ldots,{\beta}^*_J)^\top\in\mathbb{R}^{J+1}$ be the parameter of interest which is uniquely identified through the following optimization problem:
\begin{align}\label{id:beta*_potential}
	\boldsymbol{\beta}^*:=\arg\min_{\boldsymbol{\beta}}\sum_{d=0}^J\mathbb{E}\left[\mathcal{L}\left(Y^*(d)-\beta_d\right)\right],
\end{align}
where $\boldsymbol{\beta}=(\beta_0,\beta_1,...,\beta_J)^\top\in\mathbb{R}^{J+1}$ and $J\in\mathbb{N}$.  The formulation \eqref{id:beta*_potential} permits various definitions of treatment effect (TE) parameters, some of which have been considered in the literature. For example,
\begin{itemize}
	\item $\mathcal{L}(v)=v^2$ and $J=1$,  then $\beta^*_0=\mathbb{E}[Y^*(0)]$ and $\beta^*_1=\mathbb{E}[Y^*(1)]$, and $\beta^*_1-\beta^*_0$ is the average treatment effects (ATE) studied by \cite{hahn1998role}, \cite{Hirano03}, \cite{chan2016globally} and many others. When $J\geq 2$, then $\beta^*_d=\mathbb{E}[Y^*(d)]$ is the multi-valued ATE first studied by \cite{cattaneo2010efficient}.
	\item $\mathcal{L}(v)=v\cdot\{\tau-\mathds{1}(v\leq 0)\}$ for some $\tau\in (0,1)$ and $J= 1$, then $\beta^*_0=F_{Y^*(0)}^{-1}(\tau)$ and $\beta^*_1=F_{Y^*(1)}^{-1}(\tau)$, and $\beta^*_1-\beta^*_0$ is the quantile treatment effects \citep[QTE,][]{Firpo2007Efficient,chen2008semiparametric,HKZ2019}.
	\item $\mathcal{L}(v)=v^2\cdot |\tau-\mathds{1}(v\leq 0)|$ is the asymmetric least  square treatment effects \citep[ALSTE,][]{newey1987asymmetric}. {\color{black} ALSTE estimators have properties analogue to QTE estimators, but they are easier to compute. ALSTE has a variety of applications, such as the study of racial/ethnic disparities in health care, in which the data are often skewed.
		\iffalse
		When $\tau=0.5$, the ALSTE reduces to the ATE. When $\tau\neq 0.5$, due to the asymmetric nature and  smoothness of $\mathcal{L}(v)$, the ALS loss provides a convenient and computationally efficient way of summarizing the counterfactual distributions of the potential outcome $Y^*(d)$. Applications of the ALSTE include testing treatment effect based on the expected shortfall \citep{he2010detection} and the asymmetric decisions problems \citep{elliott2008economic}.
		\fi}
\end{itemize}

The problem with \eqref{id:beta*_potential} is that the potential outcomes $(Y^*(0),Y^*(1),...,Y^*(J))$ cannot all be
observed. The observed outcome is denoted by $Y:=Y^*(D)=\sum_{d=0}^{J+1}\mathds{1}(D=d)Y^*(d)$. One may attempt to solve the problem:
\begin{align*}
	\min_{\boldsymbol{\beta}}\sum_{d=0}^J\mathbb{E}\left[\mathcal{L}\left(Y-\beta_d\right)\right].
\end{align*}
However, due to the selection in treatment, the true value $\boldsymbol{\beta}^*$ is not the solution of the above problems. To address this problem, most literature imposes the following \emph{unconfoundedness} condition \citep{rosenbaum1983central}:
\begin{assumption}\label{as:CIA} For each $d\in \mathcal{D}$,
	$Y^*(d)\perp D|\boldsymbol{X}$.
\end{assumption}
This condition is also maintained in our work. Nevertheless, we depart from the classical semiparametric estimation and inference for various TEs by allowing the dimension $p$ of the confounders $\boldsymbol{X}$ to grow with sample size $n$. Specifically, we work with triangular array data $\{((D_{i,n},\boldsymbol{X}_{i,n},Y_{i,n}),i=1,...,n),n=1,2,...\}$  defined on some common probability space $(\Omega,\mathcal{A},\mathbb{P})$. Each $\boldsymbol{X}_{i,n}$ is a vector whose dimension $p_n$ may grow with $n$, the support of $\boldsymbol{X}_{i,n}$ is assumed to be $[0,1]^{p_n}$. For each given $n$, these vectors are independent across $i$, but not necessarily identically	distributed. The law $\mathbb{P}_n$ of $\{(D_{i,n},\boldsymbol{X}_{i,n},Y_{i,n}),i=1,...,n\}$ can change with $n$, though we do not make explicit use of $\mathbb{P}_n$. Thus, all parameters (including $p_n$) that characterize the distribution of $\{(D_{i,n},\boldsymbol{X}_{i,n},Y_{i,n}),i=1,...,n\}$  are implicitly indexed by the sample size $n$, but we omit the index $n$ in what follows to simplify notation.

\subsection{ANN-IPW estimator for general TEs}\label{IPW-TE}

Under Assumption \ref{as:CIA}, the causal parameters $\boldsymbol{\beta}^*$ can be identified by the minimizer of the following optimization problem:
\begin{align}\label{def:beta*_opt}
	\boldsymbol{\beta}^*=\arg\min_{\boldsymbol{\beta}}\sum_{d=0}^J\mathbb{E}\left[\frac{\mathds{1}(D_i=d)}{\pi^*_d(\boldsymbol{X}_{i})}\mathcal{L}\left(Y_i-\beta_d\right)\right],
\end{align}
where $\pi^*_d(\boldsymbol{X}_{i}):=\mathbb{P}(D_i=d|\boldsymbol{X}_{i})$ is the propensity score (PS) function which is unknown in practice.
\iffalse
To take high-dimensional data into account, we follow the conceptual considerations in \cite{farrell2015robust} and \cite{fan2022estimation} and treat the DGP (the
distribution of $(D,Y,\boldsymbol{X})$) depending on the sample size $n$, allowing, in particular, the dimension of $\boldsymbol{X}$ to grow with the sample size $n$. This implies that the nuisance functions, $\{\pi_d^*(\boldsymbol{X})\}_{d=0}^J$, may generally depend on $n$ as well.
\fi

Based on \eqref{def:beta*_opt}, existing approaches  rely on parametric or nonparametric estimation of the PS function $\pi^*_d(\cdot)$. Parametric methods suffer from model misspecification problems, while conventional nonparametric methods, such as linear sieve or kernel regression, fail to work if the dimension of covariates $p$ is large  which is known as the ``curse of dimensionality".
The goal of this article is to efficiently estimate $\boldsymbol{\beta}^*$ under this general framework when the dimension of covariates $p$ is large, and it possibly increases as the sample size $n$ grows. We propose to estimate the PS function $\pi^*_d(\cdot)$ using feedforward ANNs with one hidden layer described below.

All three ANNs described in Section \ref{sec:ANN} can be applied to estimate the  PS function $\pi_d^*(\cdot)$, and the resulting TE estimators have the same asymptotic properties based on the three ANNs. For convenience of presentation, we use the sigmoid type ANNs to present the theoretical results in this section.  To facilitate our subsequent statistical applications, we allow  $r=r_n$ and $B=B_n$ to depend on sample size $n$.  We denote the resulting ANN sieve space as
$$\mathcal{G}_n:=\mathcal{G}(\psi,B_n,r_n,p).$$
Denote $D_{di}:=\mathds{1}(D_i=d)$ for brevity.  Let $L(a):=\exp(a)/(1+\exp(a))$, for $a\in\mathbb{R}$, be the logistic function. The  inverse logistic transform of the true PS is defined by $$g_d^*(x):=L^{-1}\left(\pi_d^*(x)\right)=\log\left\{\pi_d^*(x)/(1-\pi_d^*(x))\right\},$$ and it satisfies $\mathbb{E}[\ell_d(D_{di},\boldsymbol{X}_{i};g^*_d)]\geq \mathbb{E}[\ell_d(D_{di},\boldsymbol{X}_{i};g_d)]$ for all $g_d\in\mathcal{G}_n$, where
\begin{align*}
	\ell_d(D_{di},\boldsymbol{X}_{i};g_d):=&D_{di}\log L\left(g_d(\boldsymbol{X}_{i})\right)+\{1-D_{di}\}\log \left(1-L\left(g_d(\boldsymbol{X}_{i})\right)\right)\\
	=&D_{di}g_d(\boldsymbol{X}_i)-\log\left[1+\exp(g_d(\boldsymbol{X}_i))\right].
\end{align*}		
Let $\widehat{g}_d$ be the ANN estimator of $g_d^*$ based on the space $\mathcal{G}_n$, i.e.
\begin{align}\label{as:pihat_1st}
	L_{d,n}(\widehat{g}_d)\geq \sup_{g_d\in\mathcal{G}_n}L_{d,n}(g_d)-O(\epsilon^2_n),
\end{align}
where
$L_{d,n}(g_d):=n^{-1}\sum_{i=1}^n\ell_d(D_{di},\boldsymbol{X}_{i};g_d)$
is the empirical criterion, and $\epsilon_n=o(n^{-1/2})$.

The ANN estimator of  $g^*_d$ depends on the sample size $n$. For notational simplicity, we omit the index $n$. \eqref{as:pihat_1st} states that the ANN estimator $\widehat{g}_{d}$ of $g^* _{d}$ does not need to be the global maximizer of the objective function $L_{d,n}(g_{d})$, which may not be obtained in practice. It can be any local solutions satisfying  \eqref{as:pihat_1st}, i.e., the values of the objective function evaluated at the local solutions and at the global maximizer cannot be far away from each other, and their difference needs to satisfy the order $O(\epsilon _{n}^{2})$. This assumption is also imposed for sieve extreme estimation; see \cite{shen1997methods}, \cite{chen1998sieve} and \cite{chen1999improved}. The estimator of $\pi_d^*$ is defined by $\widehat{\pi}_d:=L(\widehat{g}_d)$,  then we use the empirical version of \eqref{def:beta*_opt} to construct the estimator of $\boldsymbol{\beta }^{\ast }$, denoted by $\widehat{\boldsymbol{\beta}}=(\widehat{\beta}_{0},...,\widehat{\beta}_{J})^\top$ where
\begin{align}\label{def:betahat}
	\widehat{\beta}_d:= \arg\min_{\beta\in\Theta}\frac{1}{n}\sum_{i=1}^n\frac{D_{di}}{\widehat{\pi}_d(\boldsymbol{X}_{i})}\mathcal{L}\left(Y_i-{\beta}\right),
\end{align}
for every $d\in \mathcal{D}=\{0,1,...,J\}$. $\widehat{\boldsymbol{\beta}}$ is called the artificial neural networks-based inverse probability weighting (ANN-IPW) estimator.

\subsection{ANN-OR estimator for ATE}\label{sec:OR-ATE}
In this subsection, we consider an alternative estimator for a particularly  important parameter of interest, ATE, which corresponds to a loss function $\mathcal{L}(v)=v^2$. Using Assumption \ref{as:CIA} and the property of conditional expectation, we can rewrite \eqref{id:beta*_potential} as follows:
\begin{align}
	\label{eq:beta_regression}
	\boldsymbol{\beta}^*=\arg\min_{\boldsymbol{\beta}}\sum_{d=0}^J \mathbb{E}\left[ \mathbb{E}[\mathcal{L}(Y_i-\beta_d)|\boldsymbol{X}_{i},D_i=d]\right].
\end{align}
Based on the above expression, an alternative estimation strategy for $\boldsymbol{\beta}^*$ is to first estimate the conditional expectation $ \mathbb{E}[\mathcal{L}(Y_i-\beta_d)|\boldsymbol{X}_{i},D_i=d]$ (with $\beta_d$ being fixed), and then estimate $\boldsymbol{\beta}^*$ by minimizing the empirical version of \eqref{eq:beta_regression} with $ \mathbb{E}[\mathcal{L}(Y_i-\beta_d)|\boldsymbol{X}_{i},D_i=d]$ replaced by its estimate. Unlike the estimator $\widehat{\mathcal{E}}_d(x)$ in \eqref{eq:Ed_hat} of the supplement, where $\widehat{\beta}_d$ involved in $\widehat{\mathcal{E}}_d(x)$ is separately obtained through \eqref{def:betahat}, solving an empirical version of \eqref{eq:beta_regression} is difficult for a general $\mathcal{L}(\cdot)$, since  $\boldsymbol{\beta}^*$ is involved in the ANN estimator of $\mathbb{E}[\mathcal{L}(Y_i-\beta_d)|\boldsymbol{X}_{i},D=d]$ which may not have a closed-form expression.

In this case, $\beta_d^*=\mathbb{E}[Y^*(d)]=\mathbb{E}[z_d^*(\boldsymbol{X}_{i})]$, where  $z_d^*(\boldsymbol{X}_{i}):=\mathbb{E}[Y_i|\boldsymbol{X}_{i},D_i=d]$ is the outcome regression (OR) function and satisfies $\mathbb{E}[\ell^{OR}_d(D_{di},\boldsymbol{X}_{i},Y_i;z^*_d)]\geq \mathbb{E}[\ell^{OR}_d(D_{di},\boldsymbol{X}_{i},Y_i;z_d)]$ for all $z_d\in\mathcal{G}_n$, where
\begin{align*}
	\ell_d^{OR}(D_{di},\boldsymbol{X}_{i},Y_i;z_d):=-D_{di}\left\{Y_i-z_d(\boldsymbol{X}_{i})\right\}^2.
\end{align*}		
Let $\widehat{z}_d$ be the ANN estimator of $z_d^*$ based on the space $\mathcal{G}_n$, i.e.
\begin{align}\label{as:Rhat_1st}
	L^{OR}_{d,n}(\widehat{z}_d)\geq \sup_{z_d\in\mathcal{G}_n}L^{OR}_{d,n}(z_d)-O(\epsilon^2_n),
\end{align}
where
$L^{OR}_{d,n}(z_d):=n^{-1}\sum_{i=1}^n\ell^{OR}_d(D_{di},\boldsymbol{X}_{i},Y_i;z_d)$
is the empirical criterion, and $\epsilon_n=o(n^{-1/2})$. Then the ANN-OR estimator of $\beta_d^*$ is defined to be
\begin{align}\label{def:tauhat_OR}
	\widehat{\beta}^{OR}_d=\frac{1}{n}\sum_{i=1}^n\widehat{z}_d(\boldsymbol{X}_{i}).
\end{align}

\section{Large sample properties of estimators}\label{sec:asymptotics}

\subsection{Properties of the ANN-IPW estimator for general TEs}\label{IPW-property}

We first introduce sufficient conditions for the convergence rates of our ANN estimators $\{\widehat{\pi}_d\}_{d=0}^J$ for the unknown PS nuisance functions.

	\begin{assumption}\label{as:parameter} 	For every $d\in \mathcal{D}=\{0,1,...,J\}$ and  $m\geq 1$, we assume $g _{d}^{\ast }(\cdot )\in \mathcal{F}_p^m$.
	\end{assumption}

\begin{assumption}\label{as:rate_rp}  (i) The dimension of $\boldsymbol{X}_{i}$ is  denoted by $p\in\mathbb{N}$ and the number of hidden units is denoted by $r_n\in\mathbb{N}$. They satisfy
	\begin{align*}
\max\left\{v_{g_d^*,m}\cdot r_n^{-\frac{1}{2}-\frac{1}{p}},\sqrt{\frac{r_n\cdot p\cdot \log n}{n}}\right\}=o\left(n^{-\frac{1}{4}}\right).
	\end{align*}
 (ii) The bound of the hidden-to-output weights, $B_n$, specified in \eqref{def:Gp} satisfies  $B_n\leq 2v_{g_d^*,m}$.
\end{assumption}
Assumption  \ref{as:parameter}  is a smoothness condition imposed on the transformed PS functions. Assumption \ref{as:rate_rp} (i) allows the dimension of covariates going to infinity as the sample size grows, while it imposes restrictions on the growth rate of the dimension of covariates and that of the number of hidden units to ensure that the $L^2(dF_X)$-convergence rate of estimated PS attains $o_P(n^{-1/4})$, which is needed to establish the $\sqrt{n}$-asymptotic normality for the proposed TE estimator.
	
\begin{rk}
 As shown in Theorem \ref{thm:vfm_finite}, the mixed smoothness ball $\mathcal{W}^{m,1+\epsilon,\infty}(\mathcal{X})$ belongs to the Barron space $\mathcal{F}_p^m$.  Assumption \ref{as:rate_rp} (i) can be implied by the following  primitive condition:
	\begin{quote} 	\noindent \textbf{Assumption \ref{as:rate_rp}$'$}. (i) Suppose $g_d^*\in\mathcal{W}^{m,1+\epsilon,\infty}(\mathcal{X})$,  $p$ and $r_n$ are allowed to grow to infinity as the sample size $n$ increases, with the rates
		\begin{align*}
			p\leq a_n\cdot\left(\log n\right)^{\frac{1}{2}}\ \text{and} \  C_1\cdot n^{\frac{p+1}{2(p+2)}}\leq r_n \leq C_2\cdot (\log n)^{-\frac{3}{2}}\cdot n^{\frac{1}{2}},
		\end{align*}
		where $a_n\to 0$ can be arbitrarily slow, and $C_1$ and $C_2$ are two positive constants.
		%The bound of the hidden-to-output weights, $B_n$, specified in \eqref{def:Gp} satisfies  $B_n\leq 2v_{\pi_d^*,m}$.
	\end{quote}
\end{rk}

The following result establishes the convergence rates of ${g}^*_d$ and ${\pi}^*_d$.
\begin{thm}\label{thm:rate_ANN} Suppose   Assumptions \ref{as:parameter} and \ref{as:rate_rp}   hold. Then
	\begin{align*}
		&\|\widehat{g}_d-g_d^*\|_{L^2(dF_X)}=O_P\left(\max\left\{v_{g_d^*,m}\cdot r_n^{-\frac{1}{2}-\frac{1}{p}},\sqrt{\frac{r_n\cdot p\cdot \log n}{n}}\right\}\right)=o_P\left(n^{-1/4}\right),
	\end{align*}
	and	
	\begin{align*}
		&\|\widehat{\pi}_d-\pi_d^*\|_{L^2(dF_X)}=O_P\left(\max\left\{v_{g_d^*,m}\cdot r_n^{-\frac{1}{2}-\frac{1}{p}},\sqrt{\frac{r_n\cdot p\cdot \log n}{n}}\right\}\right)=o_P\left(n^{-1/4}\right),
	\end{align*}
	where the constants hiding inside $O_P$ and $o_P$ do not depend on
	$p$ and $n$.	
\end{thm}

The proof of Theorem \ref{thm:rate_ANN} is provided in Supplement \ref{app:rate_ANN}. Theorem \ref{thm:rate_ANN} shows that under a suitable smoothness condition, the $M$-estimates based on ANNs with a single hidden layer circumvent the curse of dimensionality and achieve a desirable rate for establishing the asymptotic normality of plug-in estimators \citep{chen2003estimation}.   \cite{bauer2019deep} showed that their least squares estimator based on multilayer neural networks with a smooth activation function can achieve the convergence rate of $n^{-2s /(2s +d^{\ast })}$ (up to a log factor), if
the regression function satisfies a $s$-smooth generalized hierarchical
interaction model of order $d^{\ast }$, where $d^{\ast }$ is fixed. \cite{Schmidt2020} established a similar rate for the ReLU activation function.  However,  the target function class considered in \cite{bauer2019deep} and \cite{Schmidt2020} is different from that used in our paper. The extension of our results  for multilayer neural networks is beyond the scope of the current article.  We refer to the Supplement for more discussion.

Let $\mathcal{E}_d(x,\beta_d^*):=\mathbb{E}[\mathcal{L}'(Y_i^*(d)-{\beta}^*_d)|\boldsymbol{X}_{i}=x]$, $u_d^*(x):=\mathcal{E}_d(x;\beta_d^*)/\pi_d^*(x)$, and
$\overline{g}(g_d,\epsilon_n):=(1-\epsilon_n)\cdot g_d+\epsilon_n\cdot \{u_d^*+g_d^*\}$ be a local alternative value around $g_d\in\mathcal{G}_n$. The directional derivative of $\ell_d(D_{di},\boldsymbol{X}_{i};g_d)$ is given by \begin{align*}
	\frac{\partial}{\partial g_d}\ell_d(D_{di},\boldsymbol{X}_{i};g_d)[u]:=&\lim_{t\to 0}\frac{\ell_d(D_{di},\boldsymbol{X}_{i};g_d+t\cdot u)-\ell_d(D_{di},\boldsymbol{X}_{i};g_d)}{t}\\
	=&\left\{D_{di}-L(g_d(\boldsymbol{X}_i))\right\} u(\boldsymbol{X}_{i}), \ \text{for}\ u\in L^2(dF_X).
\end{align*}
We now introduce sufficient conditions and additional notation for the asymptotic normality of our ANN-IPW estimators $\widehat{\boldsymbol{\beta}}$ for the general TE parameters.

%Assumption \ref{as:density_smooth} imposes smoothness on data %distributions which are needed for asymptotic analysis and ensuring finite %asymptotic variance.

\begin{assumption}\label{as:parameter_beta}
	(i)	Let $\Theta$ be a compact set of $\mathbb{R}^{J+1}$ containing  the true parameters $\boldsymbol{\beta}^*$. (ii)  The propensity scores are uniformly bounded away from zero, i.e., there exists  a constant $\underline{c}$ such that  $0<\underline{c}\leq \pi^*_d(x)$ for all $x\in\mathcal{X}$ and $d\in\{0,1,...,J\}$. (iii) For every  $d\in\{0,1,...,J\}$ and $m\geq 1$, we assume the function $\mathcal{E}_d(\cdot,\beta_d^*)$  is uniformly bounded.
\end{assumption}
\begin{assumption}\label{as:app_error} (Approximation error) We assume the following conditions hold:
	\begin{align*}
		\sup_{\{g_d\in \mathcal{G}_n:\|g_d-g_d^*\|_{L^2(dF_X)}\leq \delta_n\} }\|\text{Proj}_{\mathcal{G}_{n}}\overline{g}({g}_d,\epsilon_n)-\overline{g}({g}_d,\epsilon_n) \|_{L^2(dF_X)} =O\left(\frac{\epsilon_n^2}{\delta_n}\right),
	\end{align*}
	and
	\begin{align*}
		\sup_{\{g_d\in \mathcal{G}_n:\|g_d-g_d^*\|_{L^2(dF_X)}\leq \delta_n\} }	\frac{1}{n}\sum_{i=1}^n\left(\frac{\partial}{\partial g_d}\ell_d(D_{di},\boldsymbol{X}_{i};g^*_d)[\overline{g}({g}_d,\epsilon_n)-\text{Proj}_{\mathcal{G}_{n}}\overline{g}({g}_d,\epsilon_n)]\right)=O_P(\epsilon_n^2),
	\end{align*}
	where $\text{Proj}_{\mathcal{G}_n}\overline{g}(g_d,\epsilon_n)$ denotes the $L^2(dF_X)$-projection of $\overline{g}(g_d,\epsilon_n)$ on the ANN space $\mathcal{G}_n$ and $\delta_n$ is a  sequence of positive real numbers satisfying $\|\widehat{g}_d-g_d^*\|_{L^2(dF_X)}=o_P(\delta_n)$.
\end{assumption}

\begin{assumption}\label{as:L2_continuous} \
	\begin{enumerate}
		\item  There exists a finite positive constant $\kappa\geq 1/2$ such that  for any $\beta\in \Theta$ and any $\delta>0$ in a neighborhood of zero,
		\begin{align*}
			\left\{\mathbb{E}\left[\sup_{\widetilde{\beta}:|\widetilde{\beta}-\beta|<\delta}\left\{\mathcal{L}'(Y-\widetilde{\beta})-\mathcal{L}'(Y-{\beta})\right\}^2\right]\right\}^{1/2}\leq \text{const}\times \delta^{\kappa};
		\end{align*}
		\item $\sup_{\beta\in \Theta}\mathbb{E}\left[|\mathcal{L}'(Y-{\beta})|^2\right]<\infty$ and $\mathbb{E}\left[\sup_{\beta\in \Theta}|\mathcal{L}'(Y-{\beta})|\right]<\infty$;
		\item $\sup_{x\in\mathcal{X}}\mathbb{E}\left[|\mathcal{L}'(Y-{\beta}^*_d)||\boldsymbol{X}=x\right]<C<\infty$ for some finite constant $C>0$;
		\item $H_d:=-\partial_{\beta_d}\mathbb{E}[\mathcal{L}'(Y^*(d)-\beta_d^*)]>0$.
	\end{enumerate}	
\end{assumption}
Assumption \ref{as:parameter_beta} (i) is a standard condition for the parameter space. Assumption \ref{as:parameter_beta} (ii) is a strict overlap condition ensuring the existence of participants at all treatment levels, which is commonly assumed in the literature. \cite{DDFLS21} discussed the applicability of the strict overlap condition with high-dimensional covariates, and provided a variety of circumstances under which this condition holds. They also argued that the strict overlap condition may not be necessary if other smoothness conditions are imposed on the potential outcomes, or it can be technically relaxed with some non-standard
asymptotic analyses \citep[e.g.][]{hong2020inference,ma2020robust} and the sacrifice of  uniform
inference on ATE. Assumption \ref{as:parameter_beta} (iii) is a smoothness condition for approximation. The functions $\{\pi_{d}^*(\cdot),\mathcal{E}_d(\cdot,\beta_d^*)\}_{d=0}^J$ generally depend on the sample size $n$.   Assumption \ref{as:app_error} specifies both approximation error and stochastic equicontinuity in neural network space, which is needed for establishing Lemma \ref{lemma:projection} in the supplemental material. Such a condition is also imposed in  \citet[Condition (C)]{shen1997methods},  \citet[Condition (B.3)]{chen1998sieve},  and \citet[Assumption 3.3 (ii)]{chen2015sieve}.   Assumption \ref{as:L2_continuous} concerns $L^2$ continuity and envelope conditions, which are needed for the applicability of the uniform
law of large numbers,
establishing stochastic equicontinuity and weak convergence, see \cite{chen2008semiparametric}. Again, they are
satisfied by widely used loss functions such as $\mathcal{L}(v)=v^2$, $\mathcal{L}(v)=v\{\tau-\mathds{1}(v\leq 0) \}$, and $\mathcal{L}(v)=v^2\cdot |\tau-\mathds{1}(v\leq 0)|$ discussed in Section \ref{sec:framework}. Assumption \ref{as:L2_continuous} (3) implies $\sup_{x\in\mathcal{X}}|\mathcal{E}(x;\beta_d^*)|<C<\infty$ by Jensen's inequality.

The following theorem shows the asymptotic distribution of the proposed estimator $\widehat{\boldsymbol{\beta}}$, whose proof is presented in Appendix \ref{app:thm:eff_IPW_shorter} and Supplement \ref{app:thm:eff_IPW}.
\begin{theorem}\label{thm:eff_IPW}
	Under Assumptions \ref{as:CIA}-\ref{as:L2_continuous}, for any $d\in\{0,1,..,J\}$, we have $\widehat{\beta}_d\xrightarrow{p}\beta_d^*$ and
	\begin{align}\label{eq:betahat-beta*}
		\sqrt{n}(\widehat{\beta}_d-\beta^*_d)=H_d^{-1}\cdot\frac{1}{\sqrt{n}}\sum_{i=1}^nS_d (Y_i,D_{di},\boldsymbol{X}_{i};\beta_d^*)+o_P(1),
	\end{align}
	where  $H_d=-\partial_{\beta_d}\mathbb{E}[\mathcal{L}'(Y^*(d)-\beta_d^*)]$ and
	\begin{align*}
		S_d=S_d (Y_i,D_{di},\boldsymbol{X}_{i};\beta_d^*):=	\frac{D_{di}}{\pi_d^*(\boldsymbol{X}_{i})}\mathcal{L}'\{Y_i-\beta_d^*\}-\left\{\frac{D_{di}-\pi_d^*(\boldsymbol{X}_{i})}{\pi_d^*(\boldsymbol{X}_{i})}\right\}\mathcal{E}_d(\boldsymbol{X}_{i},\beta_d^*).
	\end{align*}
	Consequently,  $$\boldsymbol{V}^{-1/2}\cdot \sqrt{n}\left\{\widehat{\boldsymbol{\beta}}-\boldsymbol{\beta}^*\right\}\xrightarrow{d}\mathcal{N}\left(0,I_{(J+1)\times (J+1)}\right),$$ where $I_{(J+1)\times (J+1)}$ is the $(J+1)\times(J+1)$  identity matrix, $\boldsymbol{V}=\boldsymbol{H}^{-1} \mathbb{E}[\boldsymbol{S} \boldsymbol{S}^\top ]\boldsymbol{H}^{-1}$,  $\boldsymbol{H}=\text{Diag}\{H_0,...,H_J\}$ and $\boldsymbol{S}=(S_{0},...,S_J)^\top$.
\end{theorem}
Based on the strict overlap condition and the integrability of the outcome, Assumption \ref{as:parameter_beta} (ii) and Assumption \ref{as:L2_continuous} (ii), we have that  the asymptotic variance is finite, which implies that the proposed estimator $\widehat{\boldsymbol{\beta}}$ is {\color{black}$\sqrt{n}$-consistent.} In addition, when $p$ is a fixed number, our estimator attains the semiparametric efficiency bound given in \cite{ai2018unified}.
\iffalse
{\color{red}The estimation of $\beta_d^*$ in \eqref{def:betahat} can be casted as an $M$-estimation problem with a plug-in nuisance parameter. If the linear sieve space is used, the general asymptotic results for $M$-estimators  have been established in existing literature, see \cite{chen2014sieveinference} and \cite{chen2015sieve}. The proof of our Theorem \ref{thm:eff_IPW} involves checking some high level conditions in \cite{chen2014sieveinference} and \cite{chen2015sieve}; for example, our Lemma \ref{lemma:projection} in the supplemental material is similar to the high level condition imposed in  \citet[Assumption A.1]{chen2015sieve}, whose  proof depends on the approximation and statistical properties of neural networks described in Assumption \ref{as:app_error}. }
\fi

\subsection{Property of the ANN-OR estimator for ATE}\label{sec:OR-property}
The asymptotic normality of the ANN-IPW estimator requires the strict overlap condition, i.e. Assumption \ref{as:parameter_beta} (ii). In this section, we prove that such a condition can be possibly relaxed for the ANN-OR estimator of ATE defined in \eqref{def:tauhat_OR}. From both the theoretical analysis and the numerical comparison in Section \ref{sec:simulation}, we recommend the use of ANN-OR estimator for estimating ATE in practice and the use of ANN-IPW estimator for estimating other types of causal effects such as QTE.

Let $w_d^*(x):=f_X(x)/f_{X|D}(x|d)$, $z_d^*(x)=\mathbb{E}[Y|\bs{X}=x,D=d]$, and
$\overline{z}(z_d,\epsilon_n):=(1-\epsilon_n)\cdot z_d+\epsilon_n\cdot \{w_d^*+z_d^*\}$ be a local alternative value around $z_d\in\mathcal{G}_n$.

\begin{assumption}\label{as:z_d*}
		For every $d\in \mathcal{D}=\{0,1,...,J\}$ and  $m\geq 1$, we assume $z _{d}^{\ast }(\cdot )\in \mathcal{F}_p^m$.
\end{assumption}
\begin{assumption}\label{as:rate_rp_or}  (i) We assume
	\begin{align*}
		\max\left\{v_{z_d^*,m}\cdot r_n^{-\frac{1}{2}-\frac{1}{p}},\sqrt{\frac{r_n\cdot p\cdot \log n}{n}}\right\}=o\left(n^{-\frac{1}{4}}\right).
	\end{align*}
 (ii) The bound of the hidden-to-output weights, $B_n$, specified in \eqref{def:Gp} satisfies  $B_n\leq 2v_{z_d^*,m}$.
\end{assumption}

\begin{assumption}\label{as:parameter_beta_OR}  (i) $\mathbb{P}(D_{di}=1)\in(0,1)$ and $\pi_d^*(\boldsymbol{X})\in (0,1)$;
	(ii) There exists a constant $\bar{c}$ such that
	\begin{align*}
		\mathbb{E}\left[\left\{w_d^*(\boldsymbol{X})\right\}^2\right]=\mathbb{E}\left[\left\{\frac{f_X(\boldsymbol{X})}{f_{X|D}(\boldsymbol{X}|d)}\right\}^2\right]<\overline{c}<\infty.
	\end{align*}
\end{assumption}

\begin{assumption}\label{as:app_error_OR} (Approximation error) We assume the following conditions hold:
	$$\sup_{\{z_d\in \mathcal{G}_n:\|z_d-z_d^*\|_{L^2(dF_X)}\leq \delta_n\} }\|\text{Proj}_{\mathcal{G}_{n}}\overline{z}({z}_d,\epsilon_n)-\overline{z}(z_d,\epsilon_n) \|_{L^2(dF_X)} =O\left(\frac{\epsilon_n^2}{\delta_n}\right), and$$
	\vskip -0.5cm
	$$\sup_{\{z_d\in \mathcal{G}_n:\|z_d-z_d^*\|_{L^2(dF_X)}\leq \delta_n\} }	\frac{1}{n}\sum_{i=1}^n\left(\overline{z}({z}_d,\epsilon_n)(\boldsymbol{X}_i)-\text{Proj}_{\mathcal{G}_{n}}\overline{z}({z}_d,\epsilon_n)(\boldsymbol{X}_i)\right)=O_P(\epsilon_n^2),$$
	where $\text{Proj}_{\mathcal{G}_n}\overline{z}(z_d,\epsilon_n)$ denotes the $L^2(dF_X)$-projection of $\overline{z}(z_d,\epsilon_n)$ on the ANN space $\mathcal{G}_n$.
\end{assumption}

\begin{assumption}\label{as:z_dmoment}
	$\sup_{x\in\mathcal{X}}\mathbb{E}[\{Y^*(d)\}^2|\boldsymbol{X}=x]<C<\infty$ for some finite constant $C>0$.
\end{assumption}
Assumptions \ref{as:z_d*}-\ref{as:z_dmoment} are comparable to Assumptions \ref{as:parameter}-\ref{as:L2_continuous}. It's worth noting that Assumption \ref{as:parameter_beta_OR} does not restrict the propensity score $\pi_d^*(\boldsymbol{X})$ to be uniformly bounded  below by a constant.

\begin{theorem}\label{thm:regression}
	Under Assumptions \ref{as:CIA}, \ref{as:z_d*}-\ref{as:z_dmoment}, for every $d\in\{0,1,...,J\}$, we have $\widehat{\beta}^{OR}_d\xrightarrow{p}\beta^*_d$ and
	\begin{align*}
		\sqrt{n}(\widehat{\beta}^{OR}_d-\beta^*_d)=&\frac{1}{\sqrt{n}}\sum_{i=1}^nS_d^{OR}(Y_i,D_{di},\boldsymbol{X}_i;\beta_d^*)+o_P(1),
	\end{align*}
	where $$S_d^{OR}=S_d^{OR}(Y_i,D_{di},\boldsymbol{X}_i;\beta_d^*)=\frac{D_{di}}{\pi_d^*(\boldsymbol{X}_{i})}Y_i-\left\{\frac{D_{di}-\pi_d^*(\boldsymbol{X}_{i})}{\pi_d^*(\boldsymbol{X}_{i})}\right\}\cdot z_d^*(\boldsymbol{X}_{i})-\mathbb{E}[z_d^*(\boldsymbol{X}_{i})].$$
	Consequently,  $$\left\{\boldsymbol{V}^{OR}\right\}^{-1/2}\cdot \sqrt{n}\left\{\widehat{\boldsymbol{\beta}}^{OR}-\boldsymbol{\beta}^*\right\}\xrightarrow{d}\mathcal{N}\left(0,I_{(J+1)\times (J+1)}\right),$$ where $I_{(J+1)\times (J+1)}$ is the $(J+1)\times(J+1)$  identity matrix, $\boldsymbol{V}^{OR}= \mathbb{E}[\boldsymbol{S}^{OR}\cdot (\boldsymbol{S}^{OR})^\top ]$,  and $\boldsymbol{S}^{OR}=(S^{OR}_{0},...,S^{OR}_J)^\top$.
\end{theorem}
The proof of Theorem \ref{thm:regression} is provided in Supplement \ref{proof:thm:regression}.

Note that $1/\pi_d^*(\boldsymbol{X})=w^*_d(\boldsymbol{X})/\mathbb{P}(D_{di}=1)$, with Assumptions \ref{as:parameter_beta_OR} and  \ref{as:z_dmoment}, we  have that the asymptotic variance is finite, which implies the proposed estimator $\widehat{\boldsymbol{\beta}}^{OR}$ is $\sqrt{n}$-consistent.  Moreover, the ANN-OR estimator $\widehat{\boldsymbol{\beta}}^{OR}$ has the same asymptotic variance as the ANN-IPW estimator  $\widehat{\boldsymbol{\beta}}$ when $\mathcal{L}(v)=v^2$ for ATE. We can take the same inferential strategies as given in Section \ref{sec:variance} to conduct inference based on the ANN-OR estimator.

\section{Statistical inference}\label{sec:variance}
This section presents a weighted bootstrap procedure  to conduct statistical inference for $\boldsymbol{\beta}^*$. Our TE estimator is obtained from directly optimizing an objective function, so a weighted bootstrap procedure can be performed to conduct inference without the need of estimating the asymptotic variance function. Estimation of the variance function can be challenging in the quantile TE setting. In Supplement \ref{app:variance}, we discuss a possible method for the estimation of the asymptotic variance based on the asymptotic formula given in Theorem \ref{thm:eff_IPW}.

Let $\{\omega_{d1},...,\omega_{dn}\}$ be $i.i.d.$ positive random weights that are independent of data satisfying $\mathbb{E}[\omega_{di}]=1$ and $Var(\omega_{di})=1$, where $d\in\{0,1,...,J\}$. The weighted bootstrap estimator of the inverse logistic PS $g_d^*$  is defined by satisfying
\begin{align*}
	L^{B}_{d,n}(\widehat{g}^{B}_d)\geq \sup_{g_d\in\mathcal{G}_n}L^{B}_{d,n}(g_d)-O(\epsilon^2_n),
\end{align*}
where $
L^{B}_{d,n}(g_d):=n^{-1}\sum_{i=1}^n\omega_{di}\ell_d(D_{di},\boldsymbol{X}_{i};g_d(\cdot))$
is the bootstrapped empirical criterion, and $\epsilon_n=o(n^{-1/2})$. Let $\widehat{\pi}^{B}_d:=L(\widehat{g}^{B}_d)$. Then the weighted bootstrap IPW estimator of $\beta_d^*$ is given by
\begin{align*}
	\widehat{\beta}^{B}_d=\arg\min_{\beta\in\Theta}\frac{1}{n}
	\sum_{i=1}^n\frac{\omega_{di}D_{di}}{\widehat{\pi}^{B}_d(\boldsymbol{X}_{i})}\mathcal{L}\left(Y_i-\beta\right),\ d\in \{0,1,...,J\}.
\end{align*}
{\color{black}The weighted bootstrap OR estimator of ATE can be derived similarly. The weighted bootstrap estimator of the OR function $z_d^*$  is defined by satisfying
	\begin{align*}
		L^{OR,B}_{d,n}(\widehat{z}^{B}_d)\geq \sup_{z_d\in\mathcal{G}_n}L^{OR,B}_{d,n}(z_d)-O(\epsilon^2_n),
	\end{align*}
	where $
	L^{OR,B}_{d,n}(z_d):=n^{-1}\sum_{i=1}^n\omega_{di}\ell^{OR}_d(D_{di},\boldsymbol{X}_{i},Y_i;z_d(\cdot))$.
	Then the weighted bootstrap OR estimator of $\mathbb{E}[Y^*(d)]$ is given by
	\begin{align*}
		\widehat{\beta}^{OR,B}_d=\frac{1}{n}
		\sum_{i=1}^n\omega_{d,i}\widehat{z}^{B}_d(\boldsymbol{X}_i),\ d\in \{0,1,...,J\}.
	\end{align*}
	Let $\widehat{\boldsymbol{\beta}}^{B}:=(\widehat{\beta}^{B}_0,...,\widehat{\beta}^{B}_{J+1})^\top$ and $\widehat{\boldsymbol{\beta}}^{OR,B}:=(\widehat{\beta}^{OR,B}_0,...,\widehat{\beta}^{OR,B}_{J+1})^\top$. The following theorem justifies the validation of the proposed bootstrap inference.
	\begin{theorem}\label{thm:eff_IPW_bootsrap}
		(i) Under Assumptions \ref{as:CIA}-\ref{as:L2_continuous}, for any $d\in\{0,1,..,J\}$, then conditionally on the data we have
		\begin{align*}
			\boldsymbol{V}^{-1/2}\cdot \sqrt{n}\left(\widehat{\boldsymbol{\beta}}^{B}-\widehat{\boldsymbol{\beta}}\right)\xrightarrow{d}\mathcal{N}(0,I_{(J+1)\times (J+1)}).
		\end{align*}	
		(ii) 	Under Assumptions \ref{as:CIA}, \ref{as:rate_rp}, \ref{as:z_d*}-\ref{as:z_dmoment}, for any $d\in\{0,1,..,J\}$, then conditionally on the data we have
		\begin{align*}
			\{\boldsymbol{V}^{OR}\}^{-1/2}\cdot \sqrt{n}\left(\widehat{\boldsymbol{\beta}}^{OR,B}-\widehat{\boldsymbol{\beta}}^{OR}\right)\xrightarrow{d}\mathcal{N}(0,I_{(J+1)\times (J+1)}).
		\end{align*}	
\end{theorem}}
The proof of Theorem \ref{thm:eff_IPW_bootsrap} is presented in Supplement \ref{appendix:eff_IPW_bootsrap}.

\subsection{Possible challenge of applying the EIF based method to quantile TE estimation}\label{sec:EIFestimator}

The EIF can be applied to different loss functions. When EIF is given, the estimator of $\beta
_{d}^{\ast }$ can be obtained by solving the estimated efficient score
function \citep{tsiatis2007semiparametric}. For example, when the loss
function $\mathcal{L}(v)=v^{2}$ corresponding to ATE, the EIF of $\beta
_{d}^{\ast }=\mathbb{E}[Y^{\ast }(d)]$ is
\begin{equation}
\frac{D_{di}}{\pi _{d}^{\ast }(\boldsymbol{X}_{i})}Y_{i}-\left\{ \frac{D_{di}%
}{\pi _{d}^{\ast }(\boldsymbol{X}_{i})}-1\right\} \mathbb{E}[Y_{i}|D_{di}=1,%
\boldsymbol{X}_{i}]-\beta _{d}^{\ast }.  \label{eq:ate_IF}
\end{equation}%
It involves the PS function $\pi _{d}^{\ast }(x)$ and the OR function $%
\mathbb{E}[Y_{i}|D_{di}=1,\boldsymbol{X}_{i}=x]$ that can be estimated separately. As a result, the ATE of $\beta
_{d}^{\ast }$ can be obtained with the estimated PS and OR functions directly plug into the function given in (\ref{eq:ate_IF}).

When the loss function $\mathcal{L}(v)=v\cdot \{\tau -\mathds{1}(v\leq 0)\}$
corresponding to the $\tau ^{th}$-quantile TE, the specific form of EIF for $%
\beta _{d}^{\ast }=F_{Y^{\ast }(d)}^{-1}(\tau )$ can also be derived from $%
H_{d}^{-1}S_{d}(Y_{i},D_{di},\boldsymbol{X}_{i};\beta
_{d}^{\ast })$. As a result, its estimator can be obtained from solving
the estimated efficient score equation%
\begin{equation}
\sum_{i=1}^{n}\left[ \frac{D_{di}}{\widehat{\pi }_{d}(\boldsymbol{X}
_{i})}\left\{ \tau -\mathds{1}(Y_{i}\leq \beta )\right\} -\left\{ \frac{%
D_{di}}{\widehat{\pi }_{d}(\boldsymbol{X}_{i})}-1\right\} \left\{
\tau -\widehat{\mathbb{E}}[\mathds{1}(Y_{i}\leq \beta )|D_{di}=1,\boldsymbol{%
X}_{i}]\right\} \right] =0,  \label{eq:QTE}
\end{equation}%
where  $\widehat{\pi }_{d}(x)$ and $\widehat{\mathbb{E}}[\mathds{1}%
(Y_{i}\leq \beta )|D_{di}=1,\boldsymbol{X}_{i}=x]$ are estimates of {$\pi
_{d}^{\ast }(x)$ and }$\mathbb{E}[\mathds{1}(Y_{i}\leq \beta )|D_{di}=1,%
\boldsymbol{X}_{i}=x]$, respectively. We can see that the estimation of quantile TEs from (%
\ref{eq:QTE}) is challenging when ANNs or other nonlinear machine
learning methods are employed to obtain $\widehat{\mathbb{E}}[\mathds{1}%
(Y_{i}\leq \beta )|D_{di}=1,\boldsymbol{X}_{i}=x]$, as it intertwines with
	the unknown quantile TE parameter $\beta$ nonlinearly.

Different from the aforementioned estimators constructed based on the
estimated EIF, our TE estimators are directly obtained from optimizing an
objective function that only involves the ANN-based estimated PS function.
This approach greatly facilitates the computation of obtaining TE estimates
and conducting causal inference without the need to estimate the EIF. %Computational convenience is of critical importance
%when we have large-scale observational data and ANNs approximation is involved.

%
%\section{Outcome regression (OR) estimation}\label{sec:OR}
%

%--------------------------------------------

\section{Extension to the general treatment effect on the treated}\label{sec:GTT}

The above results can be easily extended to other multi-valued causal parameters defined on the treated subgroup. Let
\begin{align}\label{id:beta*_potential_att}
	\boldsymbol{\beta}_{d'}^*:=\arg\min_{\boldsymbol{\beta}}\sum_{d=0}^J\mathbb{E}\left[\mathcal{L}\left(Y^*(d)-\beta_d\right)|D=d'\right],
\end{align}
for some fixed $d'\in\{0,1,...,J\}$, where $\boldsymbol{\beta}=(\beta_0,\beta_1,...,\beta_J)$ and $\boldsymbol{\beta}_{d'}^*=(\beta^*_{0,d'},\beta^*_{1,d'},...,\beta^*_{J,d'})$. The formulation \eqref{id:beta*_potential_att} includes the following important cases discussed in \cite{lee2018efficient}:
\begin{itemize}
	\item $\mathcal{L}(v)=v^2$, then $\beta^*_{d,d'}=\mathbb{E}[Y^*(d)|D=d']$ is the average treatment effects on the treated.
	\item $\mathcal{L}(v)=v\{\tau-\mathds{1}(v\leq 0)\}$, then $\beta^*_{d,d'}=F_{Y^*(d)|D}^{-1}(\tau|d')$ is the $\tau^{th}$ quantile of $Y^*(d)$ conditioned on the treated group $\{D=d'\}$.
\end{itemize}
Under Assumption \ref{as:CIA}, using the property of conditional expectation, the parameter of interest $\boldsymbol{\beta}_{d'}^*$ is identified by
\begin{align*}
	\boldsymbol{\beta}_{d'}^*:=&\arg\min_{\boldsymbol{\beta}}\sum_{d=0}^J\frac{1}{p_{d'}}\mathbb{E}\left[\mathds{1}(D=d')\mathcal{L}\left(Y^*(d)-\beta_d\right)\right]\\
	=&\arg\min_{\boldsymbol{\beta}}\sum_{d=0}^J\frac{1}{p_{d'}}\mathbb{E}\left[\pi^*_{d'}(\boldsymbol{X})\cdot \mathbb{E}[\mathcal{L}\left(Y^*(d)-\beta_d\right)|\boldsymbol{X}]\right]\\
	=&\arg\min_{\boldsymbol{\beta}}\sum_{d=0}^J\frac{1}{p_{d'}}\mathbb{E}\left[ \pi^*_{d'}(\boldsymbol{X})\cdot \mathbb{E}\left[\mathcal{L}\left(Y^*(d)-\beta_d\right)|\boldsymbol{X}\right]\cdot \mathbb{E}\left[\frac{\mathds{1}(D=d)}{\pi^*_{d}(X)}\bigg|\boldsymbol{X}\right]\right]\\
	=&\arg\min_{\boldsymbol{\beta}}\sum_{d=0}^J\frac{1}{p_{d'}}\cdot \mathbb{E}\left[\mathds{1}(D=d)\cdot \frac{\pi^*_{d'}(\boldsymbol{X})}{\pi^*_{d}(\boldsymbol{X})}\cdot \mathcal{L}\left(Y-\beta_d\right)\right],
\end{align*}
where $p_{d'}:=\mathbb{P}(D=d')$. The estimator of $\boldsymbol{\beta}_{d'}^*$ is obtained by minimizing the empirical analogue of the above equation:
\begin{align*}	\widehat{\boldsymbol{\beta}}_{d'}=\arg\min_{\boldsymbol{\beta}}\sum_{d=0}^J\frac{\sum_{i=1}^nD_{di}\widehat{\pi}_{d'}(\boldsymbol{X}_{i}) \mathcal{L}\left(Y_i-\beta_d\right)/\widehat{\pi}_{d}(\boldsymbol{X}_{i})}{\sum_{i=1}^nD_{d'i}},
\end{align*}
where $\widehat{\pi}_{d}$ is the ANN estimator of $\pi^*_{d}$. The estimator of $\beta^*_{d,d'}$ for $d\in\{0,1,...,J\}$ can be defined as
\begin{align*}	\widehat{\beta}_{d,d'}=\arg\min_{\beta\in\Theta}\frac{1}{n}\sum_{i=1}^n\frac{D_{di}}{\widehat{\pi}_{d}(\boldsymbol{X}_{i})}\widehat{\pi}_{d'}(\boldsymbol{X}_{i}) \mathcal{L}\left(Y_i-\beta\right).
\end{align*}

Similar to the proof of Theorem \ref{thm:eff_IPW} we obtain the following result for $\boldsymbol{\beta}_{d'}^*$.
\begin{theorem}\label{thm:att}
	Under Assumptions \ref{as:CIA}-\ref{as:L2_continuous}, for any $d,d'\in\{0,1,..,J\}$, we have that
	\begin{align*}
		\sqrt{n}(\widehat{\beta}_{d,d'}-\beta^*_{d,d'})=H_{d,d'}^{-1}\cdot\frac{1}{\sqrt{n}}\sum_{i=1}^nS_{d,d'}(\boldsymbol{X}_{i},D_{di},Y_i;\beta_{d,d'}^*)+o_P(1),
	\end{align*}
	where $H_{d,d'}=-\partial_{\beta_d}\mathbb{E}[\pi_{d'}^*(\boldsymbol{X}_{i})\mathcal{L}'(Y_i^*(d)-\beta_d^*)]$ and
	\begin{align*}
		S_{d,d'}(Y_i,D_{di},\boldsymbol{X}_{i};\beta_{d,d'}^*):= \frac{D_{di}}{{\pi}^*_d(\boldsymbol{X}_{i})}{\pi}^*_{d'}(\boldsymbol{X}_{i})
\mathcal{L}'(Y_i-\beta_{d,d'}^*)-\left\{\frac{D_{di}}
{\pi_d^*(\boldsymbol{X}_{i})}\pi^*_{d'}(\boldsymbol{X}_{i})-D_{d'i}\right\}
\mathcal{E}_{d}(\boldsymbol{X}_{i},\beta_{d,d'}^*).
	\end{align*}
\end{theorem}

\section{Simulation studies}\label{sec:simulation}
\subsection{Background and methods used}
In this section, we illustrate the finite sample performance of our proposed methods via simulations in which we generate data from models in Section \ref{subsec:simu_models}. Our proposed IPW estimator can be applied to various types of treatment effects. We use ATE, ATT (average treatment effects on the treated), QTE and QTT (quantile treatment effects on the treated) for illustration of the performance of the IPW estimator. For QTE and QTT, we consider the 25th (Q1), 50th (Q2) and 75th (Q3) percentiles. We also illustrate the performance of the OR estimator for ATE and ATT. To obtain the IPW and OR estimators, we estimate the PS and OR functions by using our proposed ANN method as well as five other popular methods, including the generalized linear models (GLM), the generalized additive models (GAM), the random forests (RF), the gradient boosted machines (GBM) and the deep neural networks with three hidden layers (DNN). We make a comparison of the performance of the resulting TE estimators with the nuisance functions estimated by the aforementioned six methods. Moreover, we compare our IPW and OR estimators with the doubly robust (DR) estimator \citep{farrell2018deep} and the Oracle estimator for ATE. For the DR estimator, the IPW and OR functions are also approximated by ANNs. The Oracle estimator is constructed based on the efficient influence function with the true PS and OR functions plugged in, see \cite{hahn1998role}. The Oracle estimators are infeasible in practice, but they are expected to perform the best for the estimation of ATE, and serve as a benchmark to compare with. In the quantile TE settings, both DR (EIF-based) and OR estimators are difficult to obtain, so we only show the performance of the IPW estimator.

We use the Rectified Linear Unit (ReLU) as the activation function for both ANN and DNN. We use cubic regression spline basis functions for GAM. We apply grid search with 5-fold cross-validation to select hyperparameters for all methods, including the number of neurons for ANN DNN, the number of trees and max depths of trees for RF and GBM, and the learning rate for GBM. All the simulation studies are implemented in Python 3.9. The DNN, GLM, GAM, RF and GBM methods are implemented using the packages tensorflow, statsmodel, pyGAM and scikit-learn, respectively.

\subsection{Data generating process}\label{subsec:simu_models} We generate the treatment and outcome variables from a nonlinear model and a linear model, respectively, given as follows.

Model 1 (nonlinear model) :
\begin{align*}
	\text{logit} \{\mathbb{E}(D_i|\boldsymbol{X}_i) \}&=0.5(X_{i1}^*X_{i2}^*-0.7sin((X_{i3}^*+X_{i4}^*)(X_{i5}^*-0.2))-0.1),\\
	\mathbb{E}(Y_i^*(1)|\bs{X}_i)&=\mathbb{E}(Y_i|\bs{X}_i, D_i=1)= 0.3(X_{i1}^*-0.9)^2+0.1(X_{i2}^*-0.5)^2\\&-0.6X_{i2}^*X_{i3}^*+sin(-1.7(X_{i1}^*+X_{i3}^*-1.1)+X_{i4}^*X_{i5}^*)+1,\\
	\mathbb{E}(Y_i^*(0)|\bs{X}_i)&=\mathbb{E}(Y_i|\bs{X}_i, D_i=0)= 0.64(X_{i1}^*-0.9)^2+0.16(X_{i2}^*+0.2)^2\\&-0.6X_{i2}^*X_{i3}^*+sin(-1.7(X_{i1}^*+X_{i3}^*-1.1)+X_{i4}^*X_{i5}^*)-1;
\end{align*}

Model 2 (linear model) :
\begin{align*}
	\text{logit} \{ \mathbb{E}(D_i|\bs{X}_i) \}&=0.1(X_{i1}^*+X_{i2}^*-2X_{i3}^*+3X_{i4}^*-3X_{i5}^*),\\
	\mathbb{E}(Y^*_i(1)|\bs{X}_i)&=\mathbb{E}(Y_i|\bs{X}_i,D_i=1)= 4X_{i1}^*+3X_{i2}^*-X_{i3}^*-5X_{i4}^*+7X_{i5}^*+1,\\
	\mathbb{E}(Y^*_i(0)|\bs{X}_i)&=\mathbb{E}(Y_i|\bs{X}_i,D_i=0)= 4X_{i1}^*+3X_{i2}^*-X_{i3}^*-5X_{i4}^*+7X_{i5}^*-1,
\end{align*}
where $X^*_{ij'} = c_p\frac{5}{p}\sum_{j=p(j'-1)/5+1}^{pj'/5}X_{ij}$ for $1\leq j' \leq 5, 1\leq i \leq n$, and $Y^*_{i}(d)=\mathbb{E}(Y^*_{i}(d)\mid  \bs{X}_{i})+\epsilon _{i}$, $d=\{0,1\}$,  $\epsilon _{i} \stackrel{i.i.d.}{\sim} \mathcal{N}\left( 0, 1 \right)$ for $1\leq i \leq n$.

We generate the confounders from $X_{ij} = 2(F(Z_{ij})-0.5)$, where $Z_{i}=(Z_{i1}, ...,Z_{ip})^\top \stackrel{i.i.d.}{\sim} \mathcal{N}\left( 0, \mathbf{\Sigma} \right)$, $\mathbf{\Sigma} = \{\sigma_{kk'} \}$, ${\sigma_{kk'}=0.2^{|k-k'|}}$ for $1\leq k, k' \leq p$, and $F(\cdot)$ is the cumulative distribution function of the standard normal for $1\leq i \leq n, 1\leq j \leq p$. Let $c_p$=1. We partition the confounders into 5 subgroups, and $X^*_{ij'}$ is the average of the $p/5$ confounders in the j'-th subgroup for $j'=1, ...,5$, so that every confounder is included in our models. We consider $p$ = 5, 10 and $n$ = 1000, 2000, 5000. All simulation results are based on 400 realizations.

We also use the nonlinear model (Model 1) to illustrate the performance of our proposed methods for $p=100$ and $n=2000$, with the confounders $X_{ij} = 2(F(Z_{ij}/\sigma_j)-0.5)$, where $\sigma_j$ is the standard deviation of $Z_{ij}$, and $Z_{i}=(Z_{i1}, ...,Z_{ip})^\top$ are generated from two designs.
\begin{itemize}
	\item Design 1 (factor model):  $Z_{ij} = F_i^{\top}L_j + \eta_{ij}$, where $F_i \stackrel{i.i.d.}{\sim} \mathcal{N}\left( 0, \mathbf{\Sigma^*} \right)$, $\mathbf{\Sigma^*} = \{\sigma_{kk'} \}$, in which ${\sigma_{kk'}=0.2^{|k-k'|}}$ for $1\leq k, k' \leq 10$, $L_j$ is a constant vector kept fixed for each realization and is generated from $L_j \stackrel{i.i.d.}{\sim} \mathcal{N}\left( 0, \mathbf{\Sigma^*} \right)$, and $\eta_{ij} \stackrel{i.i.d.}{\sim} \mathcal{N}\left( 0, 1 \right)$. Let $c_p$=7.
	\item Design 2 (multivariate normal): $Z_{i} \stackrel{i.i.d.}{\sim} \mathcal{N}\left( 0, \mathbf{\Sigma} \right)$, $\mathbf{\Sigma} = \{\sigma_{kk'} \}$, ${\sigma_{kk'}=0.2^{m(|k-k'|)}}$ for $1\leq k, k' \leq p$, where $m(x) = \lceil x/10 \rceil$, and $\lceil a \rceil$ denotes the smallest integer no less than $a$.  Let $c_p$=4.
\end{itemize}

\subsection{Simulation results} \label{simulationresults}

To compare the performance of different methods for estimating the TEs, we report the following statistics: the absolute value of bias (bias), the empirical standard deviation (emp\_sd), the average value of the estimated  standard deviations based on the asymptotic formula (est\_sd) and obtained from the weighted bootstrapping (est\_sd\_boot), and the empirical coverage rates of the 95\% confidence intervals based on the estimated asymptotic standard deviations (cover\_rate) and the weighted bootstrapping method (cover\_rate\_boot). The 95\% confidence intervals based on bootstrapping are obtained from the 2.5th percentile and 97.5th percentile of the weighted bootstrapping estimates. The bootstrap confidence intervals and the estimated standard deviations (est\_sd\_boot) are obtained based on 400 bootstrap replicates for each simulation sample. The bootstrap weights are randomly generated from the exponential distribution with mean 1 according to \cite{makosorok2005bootstrap}.

Tables \ref{tab:simu_average_ate_p5} -  \ref{tab:simu_average_ate_p10} report the numerical results for different estimators of ATE for Model 1 with $p$ = 5, 10,  respectively. We see that
as $n$ increases, the empirical coverage rates (cover\_rate and cover\_rate\_boot) based on our proposed ANN-based IPW and OR estimates become closer to the nominal level 95\%. The biases are close to zero, and the values of emp\_sd, est\_sd and est\_sd\_boot decrease as $n$ increases. These results corroborate our asymptotic theories.  We observe that our ANN-based IPW and OR estimators have comparable performance to the DR and the Oracle estimators when estimating ATE. The proposed ANN-based OR estimator slightly outperforms the ANN-based IPW and DR estimators in the sense that it has the smallest emp\_sd value. It is possible that the estimated PS functions have a few values close to zero. This can affect the emp\_sd value of the IPW estimate for ATE. The DR estimator which is constructed based on the estimates of both IPW and OR functions has larger emp\_sd values than the OR estimator, but it yields smaller emp\_sd values than the IPW estimator. Our numerical results suggest that the proposed ANN-based OR estimator is preferred for the estimation of ATE. However, in practice, it can be difficult to construct OR and DR estimators for other types of TEs, such as quantile TEs. Then the proposed ANN-based IPW estimator becomes a more appealing tool. Moreover, our numerical results given in Tables \ref{tab:simu_quantile_qte_p5} -  \ref{tab:simu_quantile_qte_p10} show that the performance of the ANN-based IPW estimators for quantile TEs is less influenced by the small values of the estimated PS functions because of the robustness of the quantile objective functions. For our proposed ANN-based IPW and OR estimators, it is convenient to apply the proposed weighted bootstrap procedure for conducting inference. We find that the empirical coverage rates of $95\%$ confidence intervals obtained from the weighted bootstrapping are closer to the nominal level than those obtained from the estimated asymptotic standard deviations.

Next, we compare the performance of different machine learning (ML) methods for the estimation of ATE. We see that the GLM and GAM methods yield large estimation biases due to the model misspecification problem. Our numerical results show that the proposed ANN method outperforms the other two ML methods, RF and GBM, for the estimation of TEs. The empirical coverage rates based on the ANN method are closer to the nominal level in all cases than the rates obtained from RF and GBM. It is worth noting that our ANN-based TE estimators enjoy the properties of root-n consistency and semiparametric efficiency. In general, our numerical results corroborate those theoretical properties.  Moreover, for RF and GBM, the OR estimator also performs better than the IPW estimator for ATE estimation. The empirical coverage rates of the $95\%$ confidence intervals obtained from the weighted bootstrapping are improved compared to the rates obtained from the estimated asymptotic standard deviation. The DNN method has comparable performance to the ANN method.

\begin{figure}
	\centering
	\includegraphics[width = 0.4\textwidth]{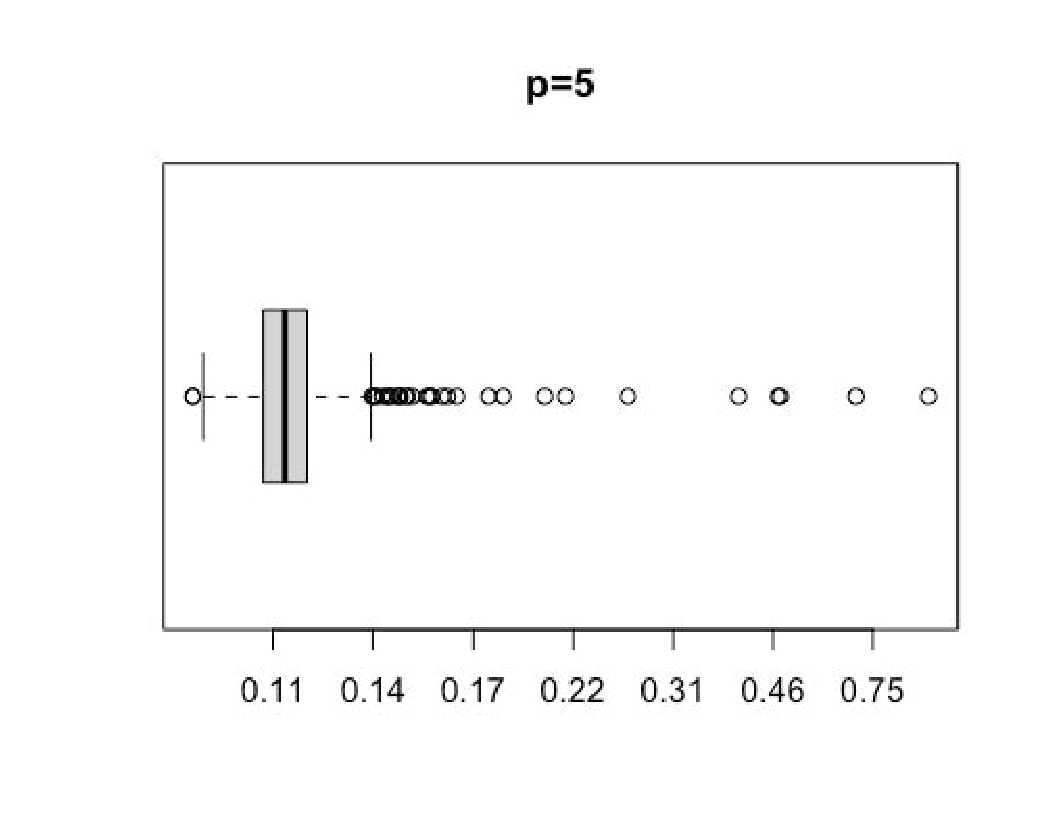}
	\includegraphics[width = 0.4\textwidth]{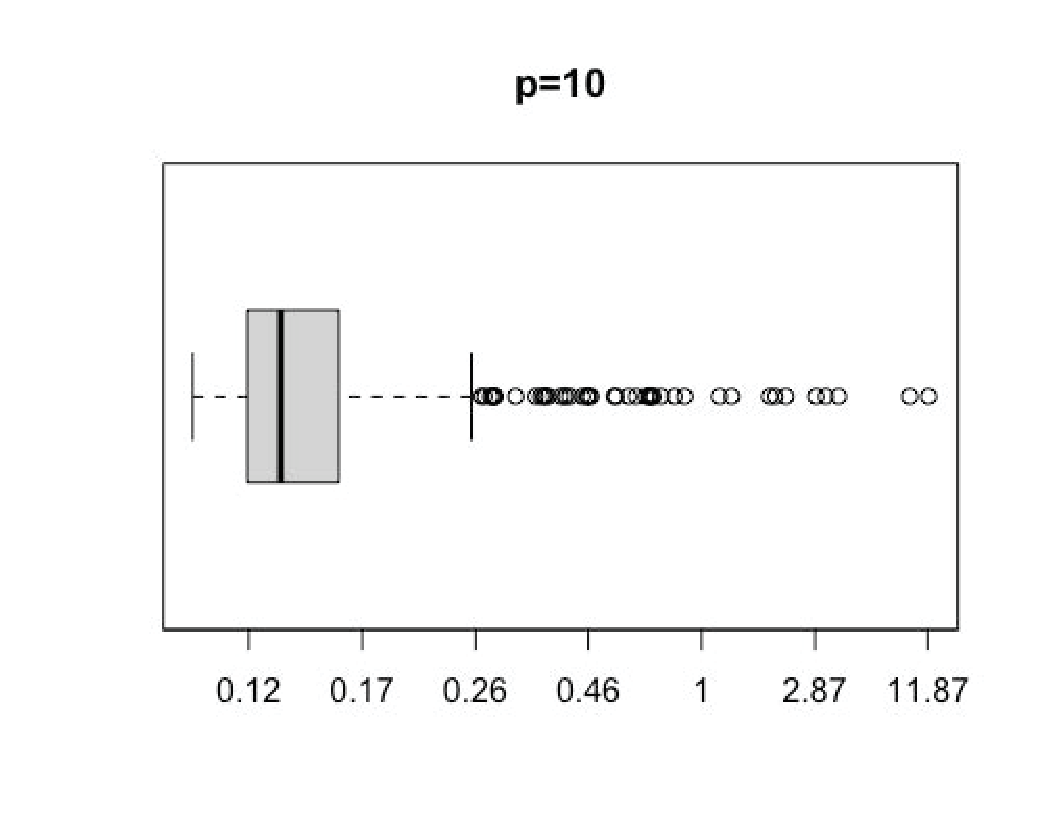}
	\caption{Boxplots of the estimated asymptotic standard deviations of QTE(Q1) for Model 1, $n=1000$.}
	\label{fig:sigma_boxplot}
\end{figure}

Tables \ref{tab:simu_quantile_qte_p5} -  \ref{tab:simu_quantile_qte_p10} show the numerical results of different methods for the estimation of QTEs for Model 1 with $p$ = 5, 10,  respectively. It is difficult to construct OR and DR estimators for QTEs, so we only report the results for the IPW estimators, which are very convenient to be obtained in this context. The PS functions are estimated by different ML methods, and the  numerical results of the resulting IPW estimates are summarized in Tables \ref{tab:simu_quantile_qte_p5} -  \ref{tab:simu_quantile_qte_p10}. In general, we observe similar patterns of numerical performance of different methods as shown in Table \ref{tab:simu_average_ate_p5} -  \ref{tab:simu_average_ate_p10}. It is worth noting that the proposed ANN-based IPW method has very stable performance for the estimation of QTEs. The resulting emp\_sd values are not influenced by possibly small values of the estimated PS functions because of the robustness nature of the quantile objective function. Moreover, in the QTE settings, estimation of the asymptotic standard deviations can involve a complicated procedure, and several approximations are needed. As a result, the estimation is not guaranteed to perform well. Figure \ref{fig:sigma_boxplot} shows the boxplots of the estimated asymptotic standard deviations of QTE (Q1) for Model 1 with $p=5,10$, $n=1000$. We see that the estimated values are large for some simulation replicates. In contrast, the estimated standard deviations obtained from the weighted bootstrapping have more reliable performance. In complex TE settings such as QTEs, the proposed weighted bootstrap method that avoids the estimation of the asymptotic variance provides a robust way to conduct statistical inference, and thus it is recommended in practice.  It is convenient to apply the weighted bootstrap method in our proposed TE estimation procedure, as the TE estimators are obtained from optimizing a general objective function. We apply different ML methods to estimate the PS function. The numerical results show that the ANN and DNN methods have comparable performance, and they still outperform other methods for the estimation and inference of QTEs.

\iffalse
The ANN and DNN estimates outperform other estimates, the bias are close to zero, and coverage rates  (cover\_rate and cover\_rate\_boot) are closer to the nominal 95\% as $n$ increases. We also observe that the est\_sd values are quite large when sample when $n$ is 1000, while the est\_sd\_boot values are comparable with the emp\_sd values, indicating bootstrapping provides a reliable estimate of standard deviations for QTE. In summary, ANN estimate performs well in estimating QTE.

Estimating standard deviation of QTE from asymptotic formula is more challenging than that of ATE, the former requires correct estimation of marginal density of the potential outcomes and an unknown conditional expectation term which also relies on correct estimation of PS function. Figure \ref{fig:sigma_boxplot} shows the boxplot of the standard deviation estimator of QTE of Q1 for Model 1 with $p=10$, $n=1000$. There are some outliers that are quite large, resulting in an inaccurate estimate of the variance. In summary, bootstrapping provides a robust way to estimate the standard deviation, especially when $n$ is small. However, we can't apply bootstrapping for DR estimate which shows another limitation of this method.
\fi

To save space, the numerical results of different methods for ATTs and QTTs for Model 1 and all the numerical results for Model 2 are presented in Tables \ref{tab:simu_average_att_p5} - \ref{tab:simu_quantile_qtt_p10_2} of Section \ref{S6section} in the Supplementary Materials.  Tables \ref{tab:simu_average_att_p5} - \ref{tab:simu_quantile_qtt_p10} show that the numerical results of different methods for estimating ATTs and QTTs have similar patterns as those given in Tables  \ref{tab:simu_average_ate_p5} -  \ref{tab:simu_quantile_qte_p10} for ATEs and QTEs. In Model 2, both PS and OR functions are generated from linear models, so the GLM and GAM methods no longer have the model misspecification problem, and GLM is expected to have the best performance. However, we can see from Tables \ref{tab:simu_average_ate_p5_2} - \ref{tab:simu_quantile_qtt_p10_2} that the ANN and DNN methods have comparable performance to GLM for the estimation of TEs in all cases. It is worth noting that  our proposed method can also be applied to the estimation of  asymmetric least squares TEs and other types of TEs, and it has similar patterns of numerical performance as the estimation of ATEs and QTEs. The numerical results are not presented due to space limitations.

At last, we evaluate the performance of our proposed TE estimators in the settings with $p= 100$ and $n=2000$. In this scenario, the number of confounders is very large compared to the sample size, and it does not satisfy the order requirement given in Assumption 4. Note that when dealing with high-dimensional covariates, one often assumes a parametric structure on the regression model and imposes a sparsity condition such that a small number of covariates are useful for the prediction %\citep{belloni2014inference,belloni2017program}.
The sparsity assumption and the parametric structure are not required in our setting. For the purpose of dimensionality reduction, we apply Principal Component Analysis (PCA) to extract the first 20 leading principal components, and use them to estimate the PS and OR functions via ANNs. For comparison, we also use the original covariates matrix without PCA to fit the nuisance models via ANNs. The resulting TE estimators with and without the PCA procedure are called ANN-PCA and ANN, respectively. Tables \ref{tab:simu_average_p100} - \ref{tab:simu_quantile_p100}  report the summary statistics of the ANN-based TE estimators for ATE, ATT, QTE and QTT for Model 1 with $p= 100$  and $n= 2000$ , based on 400 simulation realizations, when the confounders are generated from Designs 1 \& 2 given in Section \ref{subsec:simu_models}. For QTE and QTT, we only report the estimated standard deviations and empirical coverage rates from the weighted bootstrapping, as it is difficult to estimate the asymptotic standard deviations in the quantile settings. The ATE and ATT are estimated by the IPW, OR and DR methods, respectively, while the QTE and QTT are only estimated by the IPW method.

From Table \ref{tab:simu_average_p100}, for the estimation of ATE and ATT, we see that the empirical coverage rates obtained from all of the three methods, IPW, OR and DR, are smaller than the nominal level $0.95$, and the values of bias and emp\_sd are larger than those values given in Tables \ref{tab:simu_average_ate_p5} -  \ref{tab:simu_average_ate_p10} for $p=5,10$. The ANN-PCA method yields larger biases but smaller emp\_sd than the ANN method. The empirical coverage rates from the ANN-PCA method are closer to the nominal level than those from the ANN method for both designs, but they still cannot reach the nominal level. It is expected that these ANN-based methods have inferior performance for $p=100$ compared to the $p=5,10$ settings, as the order assumption on the dimension $p$ required for ANN approximations does not hold anymore when $p=100$. As a result, the ANN-based estimators of the nuisance functions (OR and PS functions) are not guaranteed to be consistent estimators, yielding deteriorated performance, and those estimates further affect the estimation of ATE and ATT. The formula of est\_sd involves the estimates of both OR and PS functions, so it is not surprising that its value is also affected. From Table \ref{tab:simu_quantile_p100} for the estimation of QTE and QTT, we can observe similar patterns as the results in Table \ref{tab:simu_average_p100}, except that the ANN method has slightly larger empirical coverage rates than the ANN-PCA method. In sum, the TE estimation using ANNs in the context of ultra-high dimensional covariates is a challenging task. A sparse model assumption may be needed for high-dimensional settings.   The investigation of its methodology and theories is beyond the scope of this paper, and it can be an interesting topic to pursue in the future.

\begin{table}[]
	\centering
	\caption{The summary statistics of the estimated ATEs for Model 1 with p=5}
	\renewcommand{\arraystretch}{1}
	\resizebox{\textwidth}{!}{%
\begin{tabular}{lcccccccccccccc}
			\hline
			& \multicolumn{6}{c}{IPW} & \multicolumn{6}{c}{OR} & \multirow{2}{*}{DR} & \multirow{2}{*}{Oracle} \\ \cmidrule(r){2-7} \cmidrule(r){8-13}
			& \textbf{ANN} & GLM & GAM & RF & GBM & DNN & \textbf{ANN} & GLM & GAM & RF & GBM & DNN &  &  \\ \hline
			\\
			n=1000 &  &  &  &  &  &  &  &  &  &  &  &  &  &  \\
			\hline
			bias & \textbf{0.0008} & 0.0584 & 0.0594 & 0.0523 & 0.0562 & 0.0015 & \textbf{0.0010} & 0.0586 & 0.0574 & 0.0151 & 0.0144 & 0.0012 & 0.0026 & 0.0006 \\
			emp\_sd & \textbf{0.0758} & 0.0907 & 0.0924 & 0.0830 & 0.0844 & 0.0789 & \textbf{0.0713} & 0.0906 & 0.0930 & 0.0769 & 0.0828 & 0.0726 & 0.0752 & 0.0668 \\
			est\_sd & \textbf{0.0701} & 0.0923 & 0.0903 & 0.0511 & 0.0483 & 0.0715 & \textbf{0.0701} & 0.0923 & 0.0903 & 0.0511 & 0.0483 & 0.0715 & 0.0701 & 0.0686 \\
			cover\_rate & \textbf{0.9275} & 0.9025 & 0.8750 & 0.6700 & 0.6400 & 0.9250 & \textbf{0.9425} & 0.9050 & 0.8900 & 0.8000 & 0.7450 & 0.9425 & 0.9325 & 0.9600 \\
\hline
		\end{tabular}
		\label{tab:simu_average_ate_p5}}
\end{table}

\begin{table}[]
	\centering
	\caption{The summary statistics of the estimated ATEs for Model 1 with p=10}
	\renewcommand{\arraystretch}{1}
	\resizebox{\textwidth}{!}{%
		\begin{tabular}{lcccccccccccccc}
			\hline
			& \multicolumn{6}{c}{IPW} & \multicolumn{6}{c}{OR} & \multirow{2}{*}{DR} & \multirow{2}{*}{Oracle} \\ \cmidrule(r){2-7} \cmidrule(r){8-13}
			& \textbf{ANN} & GLM & GAM & RF & GBM & DNN & \textbf{ANN} & GLM & GAM & RF & GBM & DNN &  &  \\ \hline
			\\
			n=1000 &  &  &  &  &  &  &  &  &  &  &  &  &  &  \\
			\hline
			bias & \textbf{0.0010} & 0.0635 & 0.0655 & 0.0574 & 0.0536 & 0.0010 & \textbf{0.0064} & 0.0637 & 0.0640 & 0.0264 & 0.0279 & 0.0037 & 0. 0073 & 0.0077 \\
			emp\_sd & \textbf{0.0855} & 0.0900 & 0.1127 & 0.0887 & 0.0915 & 0.0859 & \textbf{0.0793} & 0.0901 & 0.1041 & 0.0825 & 0.0832 & 0.0785 & 0.0823 & 0.0666 \\
			est\_sd & \textbf{0.0789} & 0.0943 & 0.0982 & 0.0547 & 0.0512 & 0.0784 & \textbf{0.0789} & 0.0943 & 0.0982 & 0.0547 & 0.0512 & 0.0784 & 0.0789 & 0.0695 \\
			cover\_rate & \textbf{0.9250} & 0.9075 & 0.8675 & 0.6725 & 0.6500 & 0.9275 & \textbf{0.9400} & 0.9075 & 0.8800 & 0.7775 & 0.7375 & 0.9450 & 0.9325 & 0.9550 \\
\hline
		\end{tabular}
		\label{tab:simu_average_ate_p10}
	}
	
\end{table}

\begin{table}[]
	\centering
	\caption{The summary statistics of the estimated QTEs by the IPW method for Model 1 with p=5}
	\renewcommand{\arraystretch}{1}
	\resizebox{\textwidth}{!}{%
		\begin{tabular}{lcccccccccccccccccc}
			\hline
			& \multicolumn{6}{c}{Q1} & \multicolumn{6}{c}{Q2} & \multicolumn{6}{c}{Q3} \\ \cmidrule(r){2-7} \cmidrule(r){8-13} \cmidrule(r){14-19}
			& \textbf{ANN} & GLM & GAM & RF & GBM & DNN & \textbf{ANN} & GLM & GAM & RF & GBM & DNN & \textbf{ANN} & GLM & GAM & RF & GBM & DNN \\ \hline
			\\
			n=1000 &  &  &  &  &  &  &  &  &  &  &  &  &  &  &  &  &  &  \\
			\hline
			bias & \textbf{0.0045} & 0.0601 & 0.0589 & 0.0630 & 0.0684 & 0.0036 & \textbf{0.0012} & 0.0534 & 0.0574 & 0.0507 & 0.0539 & 0.0025 & \textbf{0.0098} & 0.0363 & 0.0407 & 0.0286 & 0.0302 & 0.0067 \\
			emp\_sd & \textbf{0.1266} & 0.1358 & 0.1418 & 0.1274 & 0.1301 & 0.1235 & \textbf{0.1075} & 0.1116 & 0.1166 & 0.1055 & 0.1074 & 0.1099 & \textbf{0.1196} & 0.1151 & 0.1201 & 0.1121 & 0.1130 & 0.1206 \\
			est\_sd & \textbf{0.1414} & 0.1403 & 0.1596 & 0.1398 & 0.1405 & 0.1404 & \textbf{0.1238} & 0.1240 & 0.1447 & 0.1241 & 0.1249 & 0.1222 & \textbf{0.1251} & 0.1232 & 0.1412 & 0.1237 & 0.1247 & 0.1243 \\
			cover\_rate & \textbf{0.9500} & 0.9275 & 0.9375 & 0.9350 & 0.9250 & 0.9525 & \textbf{0.9675} & 0.9550 & 0.9500 & 0.9675 & 0.9650 & 0.9575 & \textbf{0.9600} & 0.9650 & 0.9650 & 0.9650 & 0.9625 & 0.9625 \\
\hline
		\end{tabular}
		\label{tab:simu_quantile_qte_p5}
	}
	
\end{table}

\begin{table}[]
	\centering
	\caption{The summary statistics of the estimated QTEs by the IPW method for Model 1 with p=10}
	\renewcommand{\arraystretch}{1}
	\resizebox{\textwidth}{!}{%
		\begin{tabular}{lcccccccccccccccccc}
			\hline
			& \multicolumn{6}{c}{Q1} & \multicolumn{6}{c}{Q2} & \multicolumn{6}{c}{Q3} \\ \cmidrule(r){2-7} \cmidrule(r){8-13} \cmidrule(r){14-19}
			& \textbf{ANN} & GLM & GAM & RF & GBM & DNN & \textbf{ANN} & GLM & GAM & RF & GBM & DNN & \textbf{ANN} & GLM & GAM & RF & GBM & DNN \\ \hline
			\\
			n=1000 &  &  &  &  &  &  &  &  &  &  &  &  &  &  &  &  &  &  \\
			\hline
			bias & \textbf{0.0019} & 0.0717 & 0.0750 & 0.0700 & 0.0696 & 0.0023 & \textbf{0.0047} & 0.0544 & 0.0541 & 0.0507 & 0.0472 & 0.0033 & \textbf{0.0030} & 0.0488 & 0.0476 & 0.0394 & 0.0367 & 0.0036 \\
			emp\_sd & \textbf{0.1585} & 0.1345 & 0.1634 & 0.1325 & 0.1364 & 0.1592 & \textbf{0.1437} & 0.1195 & 0.1485 & 0.1179 & 0.1206 & 0.1437 & \textbf{0.1332} & 0.1185 & 0.1529 & 0.1193 & 0.1208 & 0.1331 \\
			est\_sd & \textbf{0.2530} & 0.1444 & 0.4770 & 0.1442 & 0.1487 & 0.2630 & \textbf{0.2639} & 0.1269 & 0.4708 & 0.1269 & 0.1312 & 0.2684 & \textbf{0.2034} & 0.1253 & 0.4458 & 0.1255 & 0.1294 & 0.2109 \\
			cover\_rate & \textbf{0.9675} & 0.9200 & 0.9950 & 0.9350 & 0.9375 & 0.9675 & \textbf{0.9425} & 0.9450 & 1.0000 & 0.9475 & 0.9525 & 0.9450 & \textbf{0.9575} & 0.9325 & 1.0000 & 0.9350 & 0.9500 & 0.9575 \\
\hline
		\end{tabular}
		\label{tab:simu_quantile_qte_p10}
	}
\end{table}

\begin{table}[]
	\centering
	\caption{The summary statistics of the estimated ATEs and ATTs for Model 1 with p=100 and n=2000}
	\renewcommand{\arraystretch}{1}
	\resizebox{0.7\textwidth}{!}{%
		\begin{tabular}{llcccccccc}
			\hline
			&  & \multicolumn{4}{c}{Design 1} & \multicolumn{4}{c}{Design 2} \\
			\cmidrule(r){3-6} \cmidrule(r){7-10}
			&  & \multicolumn{2}{c}{ATE} & \multicolumn{2}{c}{ATT} & \multicolumn{2}{c}{ATE} & \multicolumn{2}{c}{ATT} \\
			\cmidrule(r){3-4} \cmidrule(r){5-6}  \cmidrule(r){7-8} \cmidrule(r){9-10}
			&  & ANN & ANN-PCA & ANN & ANN-PCA & ANN & ANN-PCA & ANN & ANN-PCA \\ \hline
			\\
			\multirow{6}{*}{IPW} & bias & 0.0104 & 0.0444 & 0.0205 & 0.0516 & 0.0163 & 0.0446 & 0.0197 & 0.0578 \\
			& emp\_sd & 0.0983 & 0.0719 & 0.1293 & 0.0790 & 0.1080 & 0.0776 & 0.1673 & 0.1005 \\
			& est\_sd & 0.0512 & 0.0651 & 0.0764 & 0.0789 & 0.0618 & 0.0733 & 0.0782 & 0.0865 \\
			& cover\_rate & 0.8225 & 0.8650 & 0.8025 & 0.8875 & 0.8100 & 0.8800 & 0.7650 & 0.8500 \\
			& est\_sd\_boot & 0.0694 & 0.0668 & 0.0899 & 0.0796 & 0.0798 & 0.0737 & 0.1036 & 0.0934 \\
			& cover\_rate\_boot & 0.8650 & 0.8675 & 0.8350 & 0.8900 & 0.8550 & 0.8925 & 0.8100 & 0.8825 \\
			\\
			\multirow{6}{*}{OR} & bias & 0.0097 & 0.0379 & 0.0132 & 0.0356 & 0.0222 & 0.0388 & 0.0193 & 0.0511 \\
			& emp\_sd & 0.0892 & 0.0722 & 0.1163 & 0.0870 & 0.0988 & 0.0760 & 0.1356 & 0.0961 \\
			& est\_sd & 0.0512 & 0.0651 & 0.0764 & 0.0789 & 0.0618 & 0.0733 & 0.0782 & 0.0865 \\
			& cover\_rate & 0.8350 & 0.8900 & 0.8475 & 0.8950 & 0.7975 & 0.9150 & 0.8150 & 0.8400 \\
			& est\_sd\_boot & 0.0625 & 0.0639 & 0.0871 & 0.0847 & 0.0714 & 0.0698 & 0.0939 & 0.0942 \\
			& cover\_rate\_boot & 0.8775 & 0.8850 & 0.8575 & 0.9150 & 0.8450 & 0.8950 & 0.8575 & 0.8850 \\
			\\
			\multirow{6}{*}{DR}  & bias & 0.0101 & 0.0382 & 0.0158 & 0.0457 & 0.0195 & 0.0390 & 0.0204 & 0.0548 \\
			& emp\_sd & 0.0894 & 0.0723 & 0.1207 & 0.0862 & 0.0992 & 0.0765 & 0.1397 & 0.0964 \\
			& est\_sd & 0.0521 & 0.0669 & 0.0785 & 0.0807 & 0.0633 & 0.0724 & 0.0801 & 0.0849 \\
			& cover\_rate & 0.8375 & 0.8850 & 0.8500 & 0.9050 & 0.8025 & 0.8975 & 0.8275 & 0.8375 \\
			\hline
		\end{tabular}
		\label{tab:simu_average_p100}
	}
\end{table}

\begin{table}[]
	\centering
	\caption{The summary statistics of the estimated QTEs and QTTs by the IPW method for Model 1 with p=100 and n=2000}
	\renewcommand{\arraystretch}{1}
	\resizebox{0.7\textwidth}{!}{%
		\begin{tabular}{llcccccccc}
			\hline
			&  & \multicolumn{4}{c}{Design 1} & \multicolumn{4}{c}{Design 2} \\
			\cmidrule(r){3-6} \cmidrule(r){7-10}
			&  & \multicolumn{2}{c}{QTE} & \multicolumn{2}{c}{QTT} & \multicolumn{2}{c}{QTE} & \multicolumn{2}{c}{QTT} \\
			\cmidrule(r){3-4} \cmidrule(r){5-6}  \cmidrule(r){7-8} \cmidrule(r){9-10}
			&  & ANN & ANN-PCA & ANN & ANN-PCA & ANN & ANN-PCA & ANN & ANN-PCA \\ \hline
			\\
			\multirow{4}{*}{Q1} & bias & 0.0157 & 0.0259 & 0.0139 & 0.0313 & 0.0132 & 0.0337 & 0.0111 & 0.0418 \\
			& emp\_sd & 0.1379 & 0.1101 & 0.1917 & 0.1257 & 0.1988 & 0.1210 & 0.2551 & 0.1542 \\
			& est\_sd\_boot & 0.1305 & 0.1004 & 0.1812 & 0.1234 & 0.1753 & 0.1190 & 0.2236 & 0.1497 \\
			& cover\_rate\_boot & 0.9050 & 0.8850 & 0.9125 & 0.8875 & 0.9125 & 0.8775 & 0.9050 & 0.8575 \\
			\\
			\multirow{4}{*}{Q2} & bias & 0.0131 & 0.0283 & 0.0249 & 0.0302 & 0.0154 & 0.0326 & 0.0055 & 0.0386 \\
			& emp\_sd & 0.1256 & 0.1017 & 0.1687 & 0.1123 & 0.1680 & 0.1101 & 0.2056 & 0.1315 \\
			& est\_sd\_boot & 0.1104 & 0.0977 & 0.1447 & 0.1026 & 0.1544 & 0.1041 & 0.1869 & 0.1278 \\
			& cover\_rate\_boot & 0.9225 & 0.9150 & 0.9075 & 0.9125 & 0.9300 & 0.9025 & 0.9225 & 0.9050 \\
			\\
			\multirow{4}{*}{Q3} & bias & 0.0265 & 0.0324 & 0.0446 & 0.0301 & 0.0248 & 0.0354 & 0.0272 & 0.0314 \\
			& emp\_sd & 0.1250 & 0.0966 & 0.1539 & 0.1041 & 0.1784 & 0.1077 & 0.2196 & 0.1192 \\
			& est\_sd\_boot & 0.1213 & 0.0935 & 0.1495 & 0.1066 & 0.1647 & 0.1003 & 0.1967 & 0.1138 \\
			& cover\_rate\_boot & 0.9175 & 0.9100 & 0.9025 & 0.8925 & 0.9200 & 0.9050 & 0.9175 & 0.8975 \\
			
			\hline
		\end{tabular}
		\label{tab:simu_quantile_p100}
	}
\end{table}

\section{Application}\label{sec:application}
In this section, we apply the proposed methods to the data from the National Health and Nutrition Examination Survey (NHANES) to investigate the causal effect of smoking on body mass index (BMI). The collected data consist of 6647 subjects, including 3359 smokers and 3288 nonsmokers. The confounding variables include four continuous variables: age, family poverty income ratio (Family PIR), systolic blood pressure (SBP), and diastolic blood pressure (DBP); six binary variables: gender, marital status, education, alcohol use, vigorous activity over past 30 days (PHSVIG), and moderate activity over past 30 days (PHSMOD). Table \ref{tab:nhanes_groupcompare} presents the group comparisons of all confounding variables in the full dataset. Mean and standard deviation (SD) are presented for continuous variables, while the count and percentage (\%) of observations for each group are presented for categorical variables. Standardized difference(Std. Dif.) is calculated as $(\bar{x}_{ns}-\bar{x}_s)/\sqrt{s_{ns}^2/n_{ns}+s_s^2/n_s}$ for continuous variables, and $(p_{ns}-p_s)/\sqrt{pq/n_{ns}+pq/n_s}$ for categorical variables, where $\bar{x}$, $s^2$ and $p$ denote sample mean, sample variance and sample proportion, and the subscripts $ns$ and $s$ refer to nonsmokers and smokers respectively, and $p,q$ are the overall proportions. The last column shows the p-value of group comparison for each covariate. We notice that the smoking group and nonsmoking group differ greatly in their group characteristics. A naive comparison of the sample mean between smoking and nonsmoking groups will lead to a biased estimation of the smoking effects on BMI.

\begin{table}[]
	\centering
	\centering
	\caption{Group comparisons}
	\resizebox{0.7\textwidth}{!}{%
		\begin{tabular}{llllllll}
			\hline
			Covariates &  & \multicolumn{2}{l}{Non-smoker ($N_{ns}$=3288)}  & \multicolumn{2}{l}{Smoker ($N_{ns}$=3359)} & Std. Dif. & p-value \\
			\hline
			Gender & 1 = Male & 1404 & (41.8\%) & 2019 & (61.41\%) & -15.99 & \textless{}0.001 \\
			& 0 = Female & 1955 & (58.2\%) & 1269 & (38.59\%) &  &  \\
			Age & Mean(SD) & 48.97 & (19) & 51.73 & (17.57) & -6.14 & \textless{}0.001 \\
			Marital & 1 = Yes & 1989 & (59.21\%) & 1867 & (56.78\%) & 2.01 & 0.0446 \\
			& 0 = No & 1370 & (40.79\%) & 1421 & (43.22\%) &  &  \\
			Education & 1 = College or above & 1626 & (48.41\%) & 1297 & (39.45\%) & 7.36 & \textless{}0.001 \\
			& 0 = Less than college & 1733 & (51.59\%) & 1991 & (60.55\%) &  &  \\
			Family PIR & Mean(SD) & 2.79 & (1.63) & 2.57 & (1.6) & 5.62 & \textless{}0.001 \\
			Alcohol & 1 = Yes & 1897 & (56.48\%) & 2708 & (82.36\%) & -22.87 & \textless{}0.001 \\
			& 0 = No & 1462 & (43.52\%) & 580 & (17.64\%) &  &  \\
			PHSVIG & 1 = Yes & 1102 & (32.81\%) & 908 & (27.62\%) & 4.61 & \textless{}0.001 \\
			& 0 = No & 2257 & (67.19\%) & 2380 & (72.38\%) &  &  \\
			PHSMOD & 1 = Yes & 1491 & (44.39\%) & 1376 & (41.85\%) & 2.09 & 0.0366 \\
			& 0 = No & 1868 & (55.61\%) & 1912 & (58.15\%) &  &  \\
			SBP & Mean(SD) & 126.42 & (21.04) & 126.63 & (19.98) & -0.43 & 0.6684 \\
			DBP & Mean(SD) & 72.1 & (13.56) & 71.61 & (14.1) & 1.44 & 0.15\\
			\hline
		\end{tabular}%
	}
	\label{tab:nhanes_groupcompare}
\end{table}

We apply our proposed ANN methods to estimate the PS and OR functions, respectively. We estimate ATE by the proposed IPW and OR methods, and estimate QTE by the IPW method only. The number of neurons is selected using grid search with 5-fold cross-validation.
\iffalse
Hyperparameters for neural networks including number of hidden units, learning rates, batches and number of epochs are selected using grid search with 5-fold cross-validation, and all the other hyperparameters are set to be the default values in the Python package tensorflow.
\fi
Table \ref{tab:nhanes_res} reports the estimates of ATE and QTE, the estimated standard deviations
based on the asymptotic formula (est\_sd) and obtained from the weighted bootstrapping
(est\_sd\_boot), and the corresponding z-values and p-values for testing ATE and QTE. The negative values of the estimates indicate that smoking has adverse effects on BMI. From the numerical results based on the estimated asymptotic standard deviations, we see that the p-values of testing ATE are 0.073 and 0.058 by the IPW and OR methods, respectively. We also notice that the p-value for testing QTE at the $25\%$ quantile is very small, which is 0.005. However, the p-value increases to 0.071 at the $50\%$ quantile (median), and further to 0.436 at the $75\%$ quantile.  This indicates that smoking has a more prominent effect on the population with smaller BMI, and its effect diminishes as BMI increases; i.e., the effect of smoking becomes less significant as the value of BMI becomes larger. This interesting pattern cannot be reflected in ATE. We can draw the same inferential conclusions as above when the weighted bootstrap method is applied.

\begin{table}[]
	\centering
	\caption{The estimates and standard errors of ATE and QTE}
	\resizebox{0.5\textwidth}{!}{%
		\begin{tabular}{lccccc}
			\hline
			& \multicolumn{2}{c}{ATE} & \multicolumn{3}{c}{QTE} \\
			\cmidrule(r){2-3}
			\cmidrule(r){4-6}
			& IPW & OR & Q1 & Q2 & Q3 \\
			\hline
			estimate & -0.224 & -0.241  & -0.400 & -0.269 & -0.040 \\
			\\
			est\_sd & 0.154 & 0.154 & 0.157 & 0.184 & 0.247 \\
			z-value &  -1.454 & -1.564 & -2.547&-1.467&-0.162\\
			p-value & 0.073 &0.058 & 0.005& 0.071&0.436\\
			\\
			est\_sd\_boot & 0.162 & 0.149 & 0.156 & 0.187 & 0.254 \\
			z-value\_boot &  -1.383 & -1.617 & -2.564&-1.443&-0.157\\
			p-value\_boot & 0.083 &0.053 & 0.005& 0.074&0.437\\
			\hline
		\end{tabular}%
	}
	\label{tab:nhanes_res}
\end{table}

\begin{figure}[]
	\centering
	\caption{The plots of $\tau_1(\cdot)$ and $\tau_0(\cdot)$ versus two continuous variables for the smoking and nonsmoking groups, and for males and females, respectively, where the blue solid curves represent nonsmoking group and red dashed line represent smoking group.}
	\includegraphics[width = 0.8\textwidth, height = 5cm]{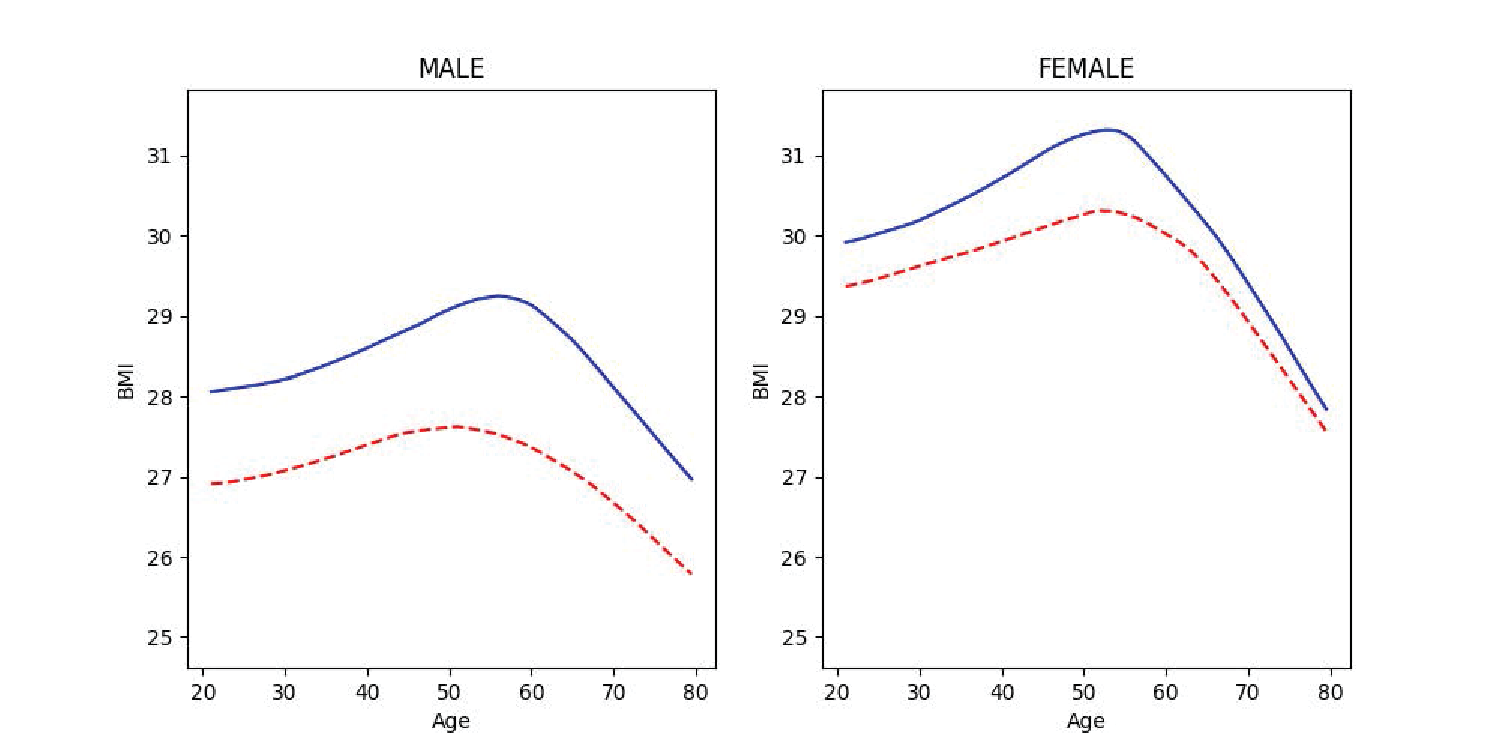}
	\includegraphics[width = 0.8\textwidth, height = 5cm]{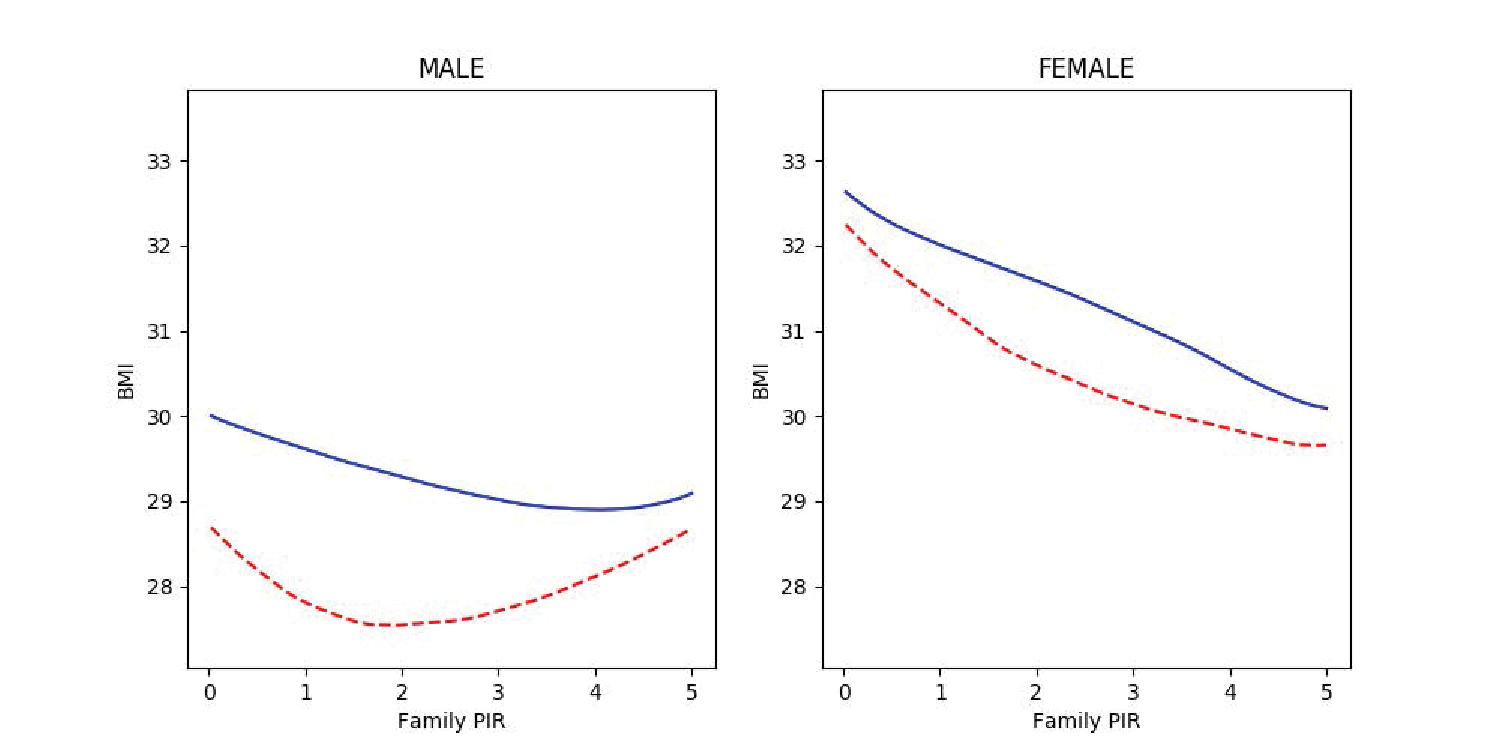}
	\label{fig:nhanes_compare}
\end{figure}

We also examine the relationship between BMI and two continuous confounding variables, age and family poverty income ratio (Family PIR). Figure \ref{fig:nhanes_compare} depicts the estimated  conditional mean functions (OR functions) $\tau_1(\cdot)$ and $\tau_0(\cdot)$ versus the two continuous variables for the smoking and nonsmoking groups, and for males and females, respectively. For each comparison, all the other confounding variables are fixed as constants: the continuous variables take the values of their means while the categorical variables are kept as married, college or above, drinks alcohol, no vigorous activity and no moderate activity. It is interesting to notice that for the same age or Family PIR, the estimated conditional mean in the smoking group is smaller than that in the nonsmoking group for both males and female, and the estimated conditional mean in the male group is also smaller than that in the female group for both smoker and nonsmoker. We can clearly see nonlinear relationships between age and BMI as well as between Family PIR and BMI. Age is positively associated with BMI when it is less than 50, and the association between age and BMI becomes more negative as people get older. We also see that the smoking effects on BMI are very different between  the male group and the female group. Smoking has a more significant effect on BMI for  males than for females at the same age. In the male group, the BMI decreases as family income increases until it reaches the poverty threshold, and then the BMI increases with family income for smokers. For  nonsmokers, it shows a relatively flatter trend. In the female group, the BMI keeps decreasing as family income increases for both smokers and nonsmokers.

\section{Conclusion}\label{sec:conclusion}
In this paper, we provide a unified framework for
efficient estimation of various types of TEs in observational data using ANNs with a
diverging number of covariates/confounders. Our framework allows for settings
with binary or multi-valued treatment variables, and includes the
average, quantile, and asymmetric least squares TEs as special cases. We
estimate the TEs through a generalized optimization, which involves an ANN estimation of one nuisance function only. When the unknown nuisance function is approximated by ANNs
with one hidden layer, we show that the number of confounders is
allowed to increase with the sample size. We further investigate how fast
the number of confounders can grow with the sample size ($n$) to ensure root-$n$ consistency, asymptotic normality and efficiency of the
resulting TE estimator. These statistical properties
are essential for inferring causations. We also show that a simple weighted bootstrap provides consistent confidence sets for the general TEs without the need to estimate the asymptotic variance. Compared to other approaches based on efficient influence functions, our general optimization-based estimation and inference methods are especially attractive for efficient estimation of complex TEs such as quantile and asymmetric TEs.
Practically, we illustrate our proposed method through simulation studies and a real data example. The numerical studies support our theoretical findings.

We have shown that the ANNs with one hidden layer can circumvent the ``curse of dimensionality'' and the resulting TE estimators enjoy root-n consistency under the condition that the target function is in a mixed smoothness class. Our new results advance the understanding of the required conditions and the statistical properties for ANNs in causal inference, and lay a theoretical foundation to demonstrate that ANNs are promising tools for causality analysis when the dimension is allowed to diverge, whereas most existing works on ANNs estimation still assume the dimension of co to be fixed. In the online supplemental materials, we discuss the extension of our method for efficient estimation of and inference on general TEs when the nuisance function is approximated by fully-connected ANNs with multiple
hidden layers. Finally, our optimization-based method can be also extended to causal
analysis with continuous treatment variables and with longitudinal data
designs. Thorough investigations are needed to develop the computational
algorithms and establish the theoretical properties of the resulting
estimators in these settings.

%\begin{supplement}[id=suppA]
	%\sname{Supplement A}\label{suppA}
%	\stitle{Supplement to ``Efficient Estimation of General Treatment Effects using Neural Networks with A Diverging Number of Confounders"}
	%\slink[doi]{COMPLETED BY THE TYPESETTER}
	% \sdatatype{.pdf}
%	\sdescription{The supplement contains the technical proofs of Theorems 2-5, some related extensions and additional simulation results. }
%\end{supplement}

\section*{Acknowledgement}
The authors sincerely thank the editor Elie Tamer and the referees for their
constructive suggestions and comments. The research of Chen is partially supported by Cowles Foundation.
 The research of Liu and Ma is supported in part by the U.S. NSF grants DMS-17-12558, DMS-20-14221 and DMS-23-10288 and the UCR Academic Senate CoR Grant.  Zheng Zhang is
supported by the fund from National Key R\&D Program of China [grant number 2022YFA1008300], the fund from the National
Natural Science Foundation of China [grant number 12371284], and the fund for building world-class universities (disciplines) of Renmin University of China [project number KYGJC2023011].
\clearpage
\appendix
\section*{Appendix}
\section{Polynomial approximation and curse of dimensionality}\label{appendix:polynomial}
Let $\mathcal{X}=[0,1]^p$ and $s_0\in\mathbb{N}$, \citet[Theorem 8]{lorentz1966approximation} states that  for any $s_0$-times continuously differentiable function $f(\cdot):\mathcal{X}\to\mathbb{R}$,  i.e.
 $\sup_{|\boldsymbol{\alpha}|_1\leq s_0 }\sup_{x\in\mathcal{X}}|D^{\boldsymbol{\alpha}}f(x)|\leq 1$, there exist polynomials $P_{n_1,...,n_p}(x)$, of degree $n_i$ in $x_i$, such that
\begin{align*}
	\sup_{x\in\mathcal{X}}|f(x)-	P_{n_1,...,n_p}(x)|\leq C_p\cdot \sum_{i=1}^p \frac{1}{n_i^{s_0}},
\end{align*}
where $C_p$ is a constant depending on $p$.

  Consider a $K$-dimensional polynomial sieve $\{u_K(x)\}$ of the form:
\begin{align*}
	u_1(x)=1,\ u_2(x)=(
	1, x_1)^\top,\ldots, u_{p+1}(x)=(
	1, x_1,...,x_p
	)^\top, \ u_{p+2}(x)=(
	1, x_1,...,x_p,x_1^2)^\top,....
\end{align*}
To ensure all degrees of $(x_1,...,x_p)$ get up to some order $n_0\in\mathbb{N}$, i.e. $\min\{n_1,...,n_p\}\geq n_0$, we require $K=(n_0+1)^p$. Therefore, for any function $f(\cdot)$ satisfying $\sup_{|\boldsymbol{\alpha}|_1\leq s_0 }\sup_{x\in\mathcal{X}}|D^{\boldsymbol{\alpha}}f(x)|\leq 1$, the approximation rate based on the polynomial sieve $\{u_K(x)\}$ is
\begin{align*}
	\inf_{\lambda_K\in\mathbb{R}^K}\sup_{x\in \mathcal{X}}|	f(x)-\lambda^\top_{K}u_K(x)|\leq  C_p\cdot p\cdot K^{-\frac{s_0}{p}}.
\end{align*}

Note that for any function in the mixed smoothness ball defined in Section \ref{sec:mixH}, i.e. $f(\cdot)\in\mathcal{W}^{m,1+\epsilon,\infty}(\mathcal{X})$ for $\epsilon\in (0,1)$, we  only have
$\sup_{|\boldsymbol{\alpha}|_1\leq m+1 }\sup_{x\in\mathcal{X}}|D^{\boldsymbol{\alpha}}f(x)|\leq 1$. In light of the compactness of $\mathcal{X}$, the $L^2(dF_X)$-approximation error based on the polynomial sieve $\{u_K(x)\}$ is $C_p\cdot p\cdot K^{-\frac{m+1}{p}}$, which severely suffers from the  curse of dimensionality.

\section{Proof of Theorem \ref{thm:vfm_finite}}\label{appendix_ANN_approx}
For a regular function $f(\cdot):\mathcal{X}\to \mathbb{R}$ whose Fourier transform is denoted by $\widetilde{f}(\cdot)$, by using the identity $e^{-\pi i}=-1$ and a change of variables, we have
\begin{align*}
	\widetilde{f}&(t_1,t_2,...,t_p)=\frac{1}{\{2\pi\}^{p}}\int_{\mathbb{R}^p}f(x_1,x_2,...,x_p)e^{-it_1x_1}\cdot e^{-it_2x_2}\cdots e^{-it_px_p}dx_1dx_2\cdots dx_p\\
	=&-\frac{1}{\{2\pi\}^{p}}\int_{\mathbb{R}^p}f(x_1,x_2,...,x_p)e^{-it_1x_1-i\pi}\cdot e^{-it_2x_2}\cdots \cdot e^{-it_px_p}dx_1dx_2\cdots dx_p\\
	=&-\frac{1}{\{2\pi\}^{p}}\int_{\mathbb{R}^p}f(x_1,x_2,...,x_p)e^{-it_1(x_1+\frac{\pi}{t_1})}\cdot e^{-it_2 x_2}\cdots e^{-it_px_p}dx_1dx_2\cdots dx_p\\
	=&-\frac{1}{\{2\pi\}^{p}}\int_{\mathbb{R}^p}f\left(x_1-\frac{\pi}{t_1},x_2,\cdots,x_p\right)e^{-it_1x_1}\cdot e^{-it_2x_2}\cdots e^{-it_px_p}dx_1dx_2\cdots dx_p.
\end{align*}
Then
\begin{align*}
	&2\widetilde{f}(t_1,t_2,...,t_p)\\
	=&\frac{1}{\{2\pi\}^{p}}\int_{\mathbb{R}^p}\left[f(x_1,x_2,\cdots,x_p)-f\left(x_1-\frac{\pi}{t_1},x_2,\cdots,x_p\right)\right]e^{-it_1x_1}\cdots  e^{-it_px_p}dx_1dx_2\cdots dx_p\\
	=&\frac{1}{\{2\pi\}^{p}}\int_{\mathbb{R}^p}\frac{\Delta_{x_1}^{\frac{\pi}{t_1}}f\left(x_1,x_2,...,x_p\right)}{\left(\frac{\pi}{t_1}\right)} \cdot \left(\frac{\pi}{t_1}\right) e^{-it_1x_1}\cdot e^{-it_2x_2}\cdots e^{-it_px_p}dx_1dx_2\cdots dx_p,
\end{align*}	
where $\Delta_{x_i}^{\delta}$ is the finite difference operator defined by
\begin{align*}
	\Delta_{x_i}^{\delta}f\left(x_1,...,x_p\right):=f\left(x_1,...,x_{i-1},x_i,x_{i+1},...,x_p\right)-f\left(x_1,...,x_{i-1},x_i-\delta,x_{i+1},...,x_p\right),
\end{align*}
for $i\in\{1,2,...,p\}$ and $\delta>0$.
Inductively, we have that for any nonnegative integer $s_1\in\mathbb{N}$ and a constant $\epsilon\in(0,1]$:
\begin{align*}
	&\widetilde{f}(t_1,t_2,...,t_p)\\
	=&\frac{1}{2}\cdot\frac{1}{\{2\pi\}^{p}}\int_{\mathbb{R}^p}\frac{\Delta_{x_1}^{\frac{\pi}{t_1}}f\left(\boldsymbol{x}\right)}{\left(\frac{\pi}{t_1}\right)}\cdot \left(\frac{\pi}{t_1}\right) e^{-it_1x_1}\cdot e^{-it_2x_2}\cdots e^{-it_px_p}dx_1dx_2\cdots dx_p\\
	=&\frac{1}{2^2}\cdot\frac{1}{\{2\pi\}^{p}}\int_{\mathbb{R}^p}\frac{\left(\Delta_{x_1}^{\frac{\pi}{t_1}}\right)^2f(\boldsymbol{x})}{\left(\frac{\pi}{t_1}\right)^2} \cdot \left(\frac{\pi}{t_1}\right)^2 e^{-it_1x_1}\cdot e^{-it_2x_2}\cdots e^{-it_px_p}dx_1dx_2\cdots dx_p\\
	\vdots &\\
	=&\frac{1}{2^{s_1}}\cdot\frac{1}{\{2\pi\}^{p}}\int_{\mathbb{R}^p}\frac{\left(\Delta_{x_1}^{\frac{\pi}{t_1}}\right)^{s_1}f(\boldsymbol{x})}{\left(\frac{\pi}{t_1}\right)^{s_1}} \cdot \left(\frac{\pi}{t_1}\right)^{s_1} e^{-it_1x_1}\cdot e^{-it_2x_2}\cdots e^{-it_px_p}dx_1dx_2\cdots dx_p\\
		=&\frac{1}{2^{s_1+1}}\cdot\frac{1}{\{2\pi\}^{p}}\int_{\mathbb{R}^p}\frac{\left(\Delta_{x_1}^{\frac{\pi}{t_1}}\right)^{s_1+1}f(\boldsymbol{x})}{\left(\frac{\pi}{t_1}\right)^{s_1+1}} \cdot \left(\frac{\pi}{t_1}\right)^{s_1+1} e^{-it_1x_1}\cdot e^{-it_2x_2}\cdots e^{-it_px_p}dx_1dx_2\cdots dx_p\\
		=&\frac{1}{2^{s_1+2}}\cdot\frac{1}{\{2\pi\}^{p}}\int_{\mathbb{R}^p}\frac{\left(\Delta_{x_1}^{\frac{\pi}{t_1}}\right)^{s_1+2}f(\boldsymbol{x})}{\left(\frac{\pi}{t_1}\right)^{s_1+1+\epsilon}} \cdot \left(\frac{\pi}{t_1}\right)^{s_1+1+\epsilon} e^{-it_1x_1}\cdot e^{-it_2x_2}\cdots e^{-it_px_p}dx_1dx_2\cdots dx_p.
\end{align*}
%Without loss of generality, for every $i\in \{1,2,...,p\}$, we have
%\begin{align*}
%	\widetilde{f}(t_1,t_2,...,t_p)= &\frac{1}{2^{s_i}}\cdot\frac{1}{\{2\pi\}^{p}}\int_{\mathbb{R}^p}\left(\Delta_{x_i}^{\frac{\pi}{t_i}}\right)^{s_i}f(\boldsymbol{x}) \cdot \left(\frac{\pi}{t_i}\right)^{s_i} e^{-it_1x_1}\cdots e^{-it_px_p}dx_1dx_2\cdots dx_p.
%\end{align*}
For a vector of nonnegative integers $(s_1,...,s_p)$, by applying the same argument to $\{t_{2},...,t_p\}$ gives:
\begin{align*}
	&\widetilde{f}(t_1,t_2,...,t_p)\\
	=&\left(\frac{1}{2}\right)^{\sum_{j=1}^p(s_j+2)}\cdot\frac{1}{\{2\pi\}^{p}}\int_{\mathbb{R}^p}\frac{\left(\Delta_{x_p}^{\frac{\pi}{t_p}}\right)^{s_p+2}\cdots \left(\Delta_{x_{1}}^{\frac{\pi}{t_{1}}}\right)^{s_{1}+2}f(\boldsymbol{x})}{\left(\frac{\pi}{t_p}\right)^{s_p+1+\epsilon}\cdots \left(\frac{\pi}{t_1}\right)^{s_1+1+\epsilon}} \cdot \left(\frac{\pi}{t_{1}}\right)^{s_{1}+1+\epsilon}\cdots \left(\frac{\pi}{t_p}\right)^{s_p+1+\epsilon} \\
	&\qquad\qquad \qquad\qquad  \qquad\qquad \times e^{-it_1x_1}\cdot e^{-it_2x_2}\cdots e^{-it_px_p}dx_1dx_2\cdots dx_p\\
	=&\left(\frac{\pi}{2}\right)^{|\boldsymbol{s}|_1+2p}\cdot\frac{1}{\{2\pi\}^{p}}\int_{[0,1]^p}\frac{\left(\Delta_{x_p}^{\frac{\pi}{t_p}}\right)^{s_p+2}\cdots \left(\Delta_{x_{1}}^{\frac{\pi}{t_{1}}}\right)^{s_{1}+2}f(\boldsymbol{x})}{\left(\frac{\pi}{t_p}\right)^{s_p+1+\epsilon}\cdots \left(\frac{\pi}{t_1}\right)^{s_1+1+\epsilon}} \cdot \frac{1}{t^{s_{1}+1+\epsilon}_{1}\cdots t_p^{s_p+1+\epsilon}} \\
	&\qquad\qquad \qquad\qquad  \qquad\qquad \times e^{-it_1x_1}\cdot e^{-it_2x_2}\cdots e^{-it_px_p}dx_1dx_2\cdots dx_p.
\end{align*}

Then for a vector of nonnegative integers $(s_1,...,s_p)$ and a constant $\epsilon\in(0,1]$:
\begin{align}
	&|\widetilde{f}(t_1,t_2,...,t_p)|  \label{eq:ftilde_bound}\\
	\leq& \left(\frac{\pi}{2}\right)^{|\boldsymbol{s}|_1+2p}\cdot\frac{1}{\{2\pi\}^{p}}\int_{[0,1]^p}\left|\frac{\left(\Delta_{x_p}^{\frac{\pi}{t_p}}\right)^{s_p+2}\cdots \left(\Delta_{x_{1}}^{\frac{\pi}{t_{1}}}\right)^{s_{1}+2}f(\boldsymbol{x})}{\left(\frac{\pi}{t_p}\right)^{s_p+1+\epsilon}\cdots \left(\frac{\pi}{t_1}\right)^{s_1+1+\epsilon}}\right| dx_1dx_2\cdots dx_p\\
	&\times  \frac{1}{|t_{1}|^{s_{1}+1+\epsilon}\cdots |t_p|^{s_p+1+\epsilon}}.\notag
\end{align}
Note that
\begin{align*}
	\lim_{t_{1},...,t_p\to\infty}\frac{\left(\Delta_{x_p}^{\frac{\pi}{t_p}}\right)^{s_p+1}\cdots \left(\Delta_{x_{1}}^{\frac{\pi}{t_{1}}}\right)^{s_{1}+1}f(\boldsymbol{x})}{\left(\frac{\pi}{t_p}\right)^{s_p+1}\cdots \left(\frac{\pi}{t_1}\right)^{s_1+1}}=\partial_{x_p}^{s_p+1}\cdots\partial_{x_{1}}^{s_{1}+1}f(\boldsymbol{x}),
\end{align*}
provided that the limit exists.

For any $f(\cdot)\in \mathcal{W}^{m,1+\epsilon,\infty}([0,1]^p)$, by definition we have $$\sup_{\{\forall\boldsymbol{\alpha}:|\boldsymbol{\alpha}|_1\leq  m\}}\sup_{\{\boldsymbol{x}\in [0,1]^p,t_1>0,...,t_p>0\}}\left|\frac{\left(\Delta_{x_p}^{\frac{\pi}{t_p}}\right)\cdots\left(\Delta_{x_{1}}^{\frac{\pi}{t_{1}}}\right)\partial_{x_1}\cdots \partial_{x_p}D^{\boldsymbol{\alpha}}f(\boldsymbol{x})}{\left(\frac{\pi}{t_p}\right)^{\epsilon}\cdots \left(\frac{\pi}{t_1}\right)^{\epsilon}}\right|\leq 1.$$
Then we can find a large enough constant $M_0>0$ such that $\min_{\{j\in 1,...,p\}}|t_j|\geq M_0$, we have
\begin{align*}
\sup_{\{\forall\boldsymbol{\alpha}:|\boldsymbol{\alpha}|_1\leq  m\}}	\sup_{\boldsymbol{x}\in[0,1]^p}\left|\frac{\left(\Delta_{x_p}^{\frac{\pi}{t_p}}\right)^{\alpha_p+2}\cdots \left(\Delta_{x_{1}}^{\frac{\pi}{t_{1}}}\right)^{\alpha_{1}+2}f(\boldsymbol{x})}{\left(\frac{\pi}{t_p}\right)^{\alpha_p+1+\epsilon}\cdots \left(\frac{\pi}{t_1}\right)^{\alpha_1+1+\epsilon}}\right|\leq 2,
\end{align*}
namely,
\begin{align}\label{def:M0}
	M_0:=\inf\left\{C\in\mathbb{R}:\sup_{\{|t_j|\geq C\}_{j=1}^p }\sup_{\left\{\forall \boldsymbol{\alpha}:|\boldsymbol{\alpha}|_1\leq m\right\}}\sup_{x\in \mathcal{X}}\frac{\left|\left(\Delta_{x_p}^{\frac{\pi}{t_p}}\right)^{\alpha_p+2}\cdots\left(\Delta_{x_{1}}^{\frac{\pi}{t_{1}}}\right)^{\alpha_1+2}f(x)\right|}{\left(\frac{\pi}{t_p}\right)^{\alpha_p+1+\epsilon}\cdots\left(\frac{\pi}{t_1}\right)^{\alpha_1+1+\epsilon}}\leq 2\right\}.
\end{align}

Hence, by \eqref{eq:ftilde_bound}, for any $f(\cdot)\in \mathcal{W}^{m,1+\epsilon,\infty}([0,1]^p)$, we have that for any $k\in \{1,..,p\}$:
\begin{align}
	&\sup_{t_1,...,t_k}|\widetilde{f}(t_1,...,t_k,t_{k+1},...,t_p)|\cdot \prod_{j=k+1}^pI(|t_j|\geq M_0) \notag \\
	\leq& \left(\frac{\pi}{2}\right)^{m+2p}\cdot\frac{2}{\{2\pi\}^{p}}\cdot   \frac{1}{|t_{k+1}|^{\gamma_{k+1}+1+\epsilon}\cdots |t_p|^{\gamma_{p}+1+\epsilon}}, \ \text{where}\   \sum_{j=k+1}^p\gamma_{j}\leq m.\label{eq:bound_ftilde}
\end{align}
We emphasize that \eqref{eq:bound_ftilde} holds for an \emph{arbitrarily} fixed $k\in \{1,...,p\}$ and for \emph{all} $\{(\gamma_{k+1},...,\gamma_p):\sum_{j=k+1}^p\gamma_{j}\leq m\}$.

We next find the bound for $v_{f,m}$. Without loss of generality, we assume the function $\widetilde{f}(t_1,..,t_p)$ is symmetric in $\bs{t}=(t_1,...,t_p)$. Note that
\begin{align*}
	&v_{f,m}=\int_{\mathbb{R}^p}|\boldsymbol{t}|_1^m\cdot |\widetilde{f}(\boldsymbol{t})|d\boldsymbol{t}\\
	=&\left(\int_{|t_p|\in [0,M_0]}+\int_{|t_p|\in [M_0,\infty)}\right)\cdots \left(\int_{|t_1|\in [0,M_0]}+\int_{|t_1|\in [M_0,\infty)}\right)\left\{\sum_{k=1}^p|t_k|\right\}^m\cdot |\widetilde{f}(\bs{t})|dt_1\cdots dt_p\\
	=&  \sum_{i=0}^p \begin{pmatrix}
		p\\
		i
	\end{pmatrix} \int_{|t_p|\in [M_0,\infty)}\cdots\int_{|t_{i+1}|\in [M_0,\infty)}  \int_{|t_i|\in [0,M_0]}\cdots\int_{|t_1|\in [0,M_0]}\left\{\sum_{k=1}^p|t_k|\right\}^m\cdot |\widetilde{f}(\bs{t})|dt_1\cdots dt_p\\
	=&\sum_{\sum_{j=1}^p\alpha_j=m}\begin{pmatrix}
		m\\
		\alpha_1,...,\alpha_p
	\end{pmatrix}  \sum_{i=0}^p \begin{pmatrix}
		p\\
		i
	\end{pmatrix}\\
	&\quad \times \int_{|t_p|\in [M_0,\infty)}\cdots\int_{|t_{i+1}|\in [M_0,\infty)}  \int_{|t_i|\in [0,M_0]}\cdots\int_{|t_1|\in [0,M_0]}|t_1|^{\alpha_1}\cdots |t_p|^{\alpha_p}\cdot |\widetilde{f}(\bs{t})|dt_1\cdots dt_p\\
	\leq &\sum_{\sum_{j=1}^p\alpha_j=m}\begin{pmatrix}
		m\\
		\alpha_1,...,\alpha_p
	\end{pmatrix}  \sum_{i=0}^p \begin{pmatrix}
		p\\
		i
	\end{pmatrix}\cdot M_0^{\alpha_1+\cdots\alpha_i+i}\\
	&\quad \times \int_{|t_p|\in [M_0,\infty)}\cdots\int_{|t_{i+1}|\in [M_0,\infty)}  |t_{i+1}|^{\alpha_{i+1}}\cdots |t_p|^{\alpha_p}\cdot \sup_{t_1,...,t_i}|\widetilde{f}(t_1,...,t_{i},t_{i+1}...,t_p)|dt_{i+1}\cdots dt_p.
\end{align*}	
For every $i\in\{0,1,..p\}$ and every $(\alpha_{i+1},...,\alpha_{p})$ in the summand, since $\alpha_{i+1}+\cdots \alpha_p\leq m$, by applying \eqref{eq:bound_ftilde} we have
\begin{align*}
	&\sup_{t_1,...,t_i}|\widetilde{f}(t_1,...,t_i,t_{i+1},...,t_p)|\cdot \prod_{j=i+1}^pI(|t_j|\geq M_0)  \\
	\leq& \left(\frac{\pi}{2}\right)^{m+2p}\cdot\frac{2}{\{2\pi\}^{p}}\cdot   \frac{1}{|t_{i+1}|^{\alpha_{i+1}+1+\epsilon}\cdots |t_p|^{\alpha_{p}+1+\epsilon}}.
\end{align*}
Then we have
\begin{align*}
	v_{f,m}	
	\leq & \sum_{\sum_{j=1}^p\alpha_j=m}\begin{pmatrix}
		m\\
		\alpha_1,...,\alpha_p
	\end{pmatrix}  \sum_{i=0}^p \begin{pmatrix}
		p\\
		i
	\end{pmatrix} M_0^{\alpha_1+\cdots+\alpha_i+i}\cdot \left(\frac{\pi}{2}\right)^{m+2p}\cdot\frac{2}{\{2\pi\}^{p}} \\
	&\quad \times   \int_{|t_p|\in [M_0,\infty)}\cdots\int_{|t_{i+1}|\in [M_0,\infty)}   \frac{1}{|t_{i+1}|^{1+\epsilon}\cdots t_p^{1+\epsilon}}dt_{i+1}\cdots dt_p \\
	\leq &\sum_{\sum_{j=1}^p\alpha_j=m}\begin{pmatrix}
		m\\
		\alpha_1,...,\alpha_p
	\end{pmatrix}  \sum_{i=0}^p \begin{pmatrix}
		p\\
		i
	\end{pmatrix} M_0^{m+i}\cdot \left(\frac{\pi}{2}\right)^{m+2p}\cdot\frac{2\cdot 2^p}{\{2\pi\}^{p}}\cdot \left(\frac{M_0^{^{-\epsilon}}}{\epsilon}\right)^{p-i}\\
	=&\sum_{\sum_{j=1}^p\alpha_j=m}\begin{pmatrix}
		m\\
		\alpha_1,...,\alpha_p
	\end{pmatrix}  \sum_{i=0}^p \begin{pmatrix}
		p\\
		i
	\end{pmatrix} \left(M_0\cdot \frac{\pi}{2}\right)^{m}\cdot \left(M_0^{1+\epsilon}\right)^i\cdot\left(\frac{\pi}{2}\right)^{2p}\cdot \frac{2}{\pi^{p}}\cdot \frac{\left(M_0^{-\epsilon}\right)^p}{\epsilon^{p-i}} \\
	\leq&\sum_{\sum_{j=1}^p\alpha_j=m}\begin{pmatrix}
		m\\
		\alpha_1,...,\alpha_p
	\end{pmatrix}  \sum_{i=0}^p \begin{pmatrix}
		p\\
		i
	\end{pmatrix} \left(M_0\cdot  \frac{\pi}{2}\right)^{m}\cdot\left(\frac{\pi}{2}\right)^{2p}\cdot \frac{2}{\pi^{p}}\cdot \frac{M_0^p}{\epsilon^{p-i}} \\
	= & \left(\frac{\pi}{2}\right)^{2p}\cdot  \frac{2}{\pi^{p}}\cdot M_0^p\cdot \left(M_0\cdot \frac{\pi}{2}\right)^{m}  \sum_{\sum_{j=1}^p\alpha_j=m}\begin{pmatrix}
		m\\
		\alpha_1,...,\alpha_p
	\end{pmatrix}  \sum_{i=0}^p \begin{pmatrix}
		p\\
		i
	\end{pmatrix} \cdot\left[\frac{1}{\epsilon}\right]^{p-i}\\
	=& \frac{2}{2^{p}}\cdot M_0^{p+m}\cdot \left( \frac{\pi}{2}\right)^{m+p}  \cdot p^m   \cdot\left[\frac{1}{\epsilon}+1\right]^{p}\\
	\leq & 2\cdot \left(M_0\cdot \frac{\pi}{2}\right)^m \cdot \left[M_0\cdot \left(\frac{\pi}{2}\right)\cdot \left(\frac{1}{2\epsilon}+\frac{1}{2}\right) \right]^{p}=2\cdot \left(M_0\cdot \frac{\pi}{2}\right)^m \cdot M^p,
\end{align*}
for some universial large constant $M$ defined by
\begin{align*}
	M:=M_0\cdot \left(\frac{\pi}{2}\right)\cdot \left(\frac{1}{2\epsilon}+\frac{1}{2}\right).
\end{align*}

\section{Proof of Theorem \ref{thm:eff_IPW}}\label{app:thm:eff_IPW_shorter}	
We first show $\widehat{\beta}_d\xrightarrow{p}\beta_d^*$ for any $d\in\{0,1,...,J\}$. The details of the proof are given in the on-line supplemental materials. Next, we establish the asymptotic normality for $\sqrt{n}\{\widehat{\beta}_d-\beta_d^*\}$. Since the loss function $\mathcal{L}(\cdot)$ may not be smooth (e.g. $\mathcal{L}(v)=v\{\tau-\mathds{1}(v\leq 0)\}$ in quantile regression),  the Delta method for deriving the large sample property is not applicable in our case. To circumvent this problem, we apply the \emph{nearness of arg mins} argument.  Define
\begin{align}\label{def:Gd}
	{G}_{d,n}(\beta,\widehat{\pi}_d):=\frac{1}{n}\sum_{i=1}^n\frac{D_{di}}{\widehat{\pi}_d(\boldsymbol{X}_{i})}\mathcal{L}(Y_i-\beta).
\end{align}
By definition
\begin{align}\label{def:beta_d}
	\widehat{\beta}_d=\arg\min_{\beta\in\Theta} G_{d,n}(\beta,\widehat{\pi}_d)=\arg\min_{\beta\in\Theta}\frac{1}{n}\sum_{i=1}^n\frac{D_{di}}{\widehat{\pi}_d(\boldsymbol{X}_{i})}\mathcal{L}(Y_i-\beta),
\end{align}
then
\begin{align*}
	\widehat{\beta}_d=&\arg\min_{\beta\in\Theta} n\left\{G_{d,n}(\beta,\widehat{\pi}_d)-G_{d,n}(\beta_d^*,\widehat{\pi}_d) \right\}\\
	=&\arg\min_{\beta\in\Theta}\sum_{i=1}^n\frac{D_{di}}{\widehat{\pi}_d(\boldsymbol{X}_{i})}\left\{\mathcal{L}(Y_i-\beta)-\mathcal{L}(Y_i-\beta_d^*)\right\}\\
	=&\arg\min_{\beta\in\Theta}\sum_{i=1}^n\frac{D_{di}}{\widehat{\pi}_d(\boldsymbol{X}_{i})}\bigg[-\mathcal{L}'(Y_i-\beta_d^*)(\beta-\beta_d^*)\\
	&\qquad\qquad +\left\{\mathcal{L}(Y_i-\beta)-\mathcal{L}(Y_i-\beta_d^*)+\mathcal{L}'(Y_i-\beta_d^*)(\beta-\beta_d^*)\right\}\bigg].
\end{align*}
By using change of variables and defining the following functions:
\begin{align*}
	&\widehat{u}_d:=\sqrt{n}(\widehat{\beta}_d-{\beta}_d^*), \  u:=\sqrt{n}(\beta-{\beta}_d^*),\\
	&R_d(Y_i,u):=\mathcal{L}\left(Y_i-\left\{\beta_d^*+\frac{u}{\sqrt{n}}\right\}\right)-\mathcal{L}(Y_i-\beta_d^*)+\mathcal{L}'(Y_i-\beta_d^*)\cdot \frac{u}{\sqrt{n}},\\
	&Q_{d,n}(u,\widehat{\pi}_d):=\sum_{i=1}^n\frac{D_{di}}{\widehat{\pi}_d(\boldsymbol{X}_{i})}\left[-\mathcal{L}'(Y_i-\beta_d^*)\cdot \frac{u}{\sqrt{n}}+R_d(Y_i,u)\right]=n\cdot \left[{G}_{d,n}(\beta,\widehat{\pi}_d)-{G}_{d,n}(\beta_d^*,\widehat{\pi}_d)\right].
\end{align*}
Then we get
\begin{align*}
	\widehat{u}_d=\arg\min_{u} Q_{d,n}(u,\widehat{\pi}_d).
\end{align*}
Next, we define the following quadratic function
\begin{align*}
	\widetilde{Q}_{d,n}(u):=&\frac{u}{\sqrt{n}}\sum_{i=1}^n\left[-\frac{D_{di}}{{\pi}^*_d(\boldsymbol{X}_{i})}\mathcal{L}'(Y_i-\beta_d^*)+\left(\frac{D_{di}}{\pi_d^*(\boldsymbol{X}_{i})}-1\right)\mathcal{E}_{d}(\boldsymbol{X}_{i},\beta_d^*) \right]\\
	&-\partial_{\beta_d}\mathbb{E}[\mathcal{L}'(Y^*_i(d)-\beta_d^*)]\cdot \frac{u^2}{2},
\end{align*}
which does not depend on $\widehat{\pi}_d$, and its minimizer is defined by
$$\widetilde{u}_{d}:=\arg\min_{u} \widetilde{Q}_{d,n}(u).$$
Since $\widetilde{Q}_{d,n}(u)$ is strictly convex and $\partial_{\beta}\mathbb{E}[\mathcal{L}'(Y^*_i(d)-\beta_d^*)]<0$, then the minimizer $\widetilde{u}_d$ is equal to
\begin{align*}
	\widetilde{u}_{d}=\frac{1}{\sqrt{n}}\sum_{i=1}^nH_d^{-1}\cdot S_d(Y_i,D_{di},\boldsymbol{X}_{i};\beta_d^*),
\end{align*}
where
\begin{align*}
	S_d(Y_i,D_{di},\boldsymbol{X}_{i};\beta_d^*):=\frac{D_{di}}{{\pi}^*_d(\boldsymbol{X}_{i})}\mathcal{L}'(Y_i-\beta_d^*)-\left(\frac{D_{di}}{\pi_d^*(\boldsymbol{X}_{i})}-1\right)\mathcal{E}_{d}(\boldsymbol{X}_{i},\beta_d^*)
\end{align*}
is the influence function of $\beta_d^*$ and $H_d:=-\partial_{\beta_d}\mathbb{E}[\mathcal{L}'(Y^*_i(d)-\beta_d^*)]$.

The desired result can be obtained via the following steps:
\begin{itemize}
	\item Step I: showing $\xi_{d,n}(u,\widehat{\pi}_d):=\widetilde{Q}_{d,n}(u)-Q_{d,n}(u,\widehat{\pi}_d)=o_p(1)$ for every fixed $u$;
	\item Step II: showing $|\widehat{u}_{d}-\widetilde{u}_{d}|=o_P(1)$;
	\item Step III: obtaining the desired result: \newline
	$\sqrt{n}\{\widehat{\beta}_d-\beta_d^*\}=\widetilde{u}_{d}+\{\widehat{u}_{d}-\widetilde{u}_{d}\}=n^{-1/2}\sum_{i=1}^nH_d^{-1}\cdot S_d(Y_i,D_{di},\boldsymbol{X}_{i};\beta_d^*)+o_P(1)$.
\end{itemize}
Note that both objective functions $\widetilde{Q}_{d,n}(u)$ and $Q_{d,n}(u,\widehat{\pi}_d)$ are convex in $u$ with probability approaching to one, the pointwise  convergence in Step I is sufficient for establishing Step II. The technical proofs of Step I-Step III are provided in the on-line supplemental materials.

\section{Notations}

Let $\|f\|_{L^2(dF_X)}:=\left\{\int_{\mathcal{X}} f(x)^2dF_X(x)\right\}^{1/2}$ and $\|f\|_{n}:=\left\{n^{-1} \sum_{i=1}^nf(\boldsymbol{X}_i)^2\right\}^{1/2}$.
We use the $\|\cdot\|_{L^2(dF_X)}$-bracketing number to measure the size of a functional space with respect to the norm $L^2(dF_X)$. Let $\mathcal{H}=\{h(\eta,\cdot):\eta\in \Xi \}$ be a class of measurable functions indexed by $\Xi$, such that $\int h(\eta,x)^2dF_X(x)<\infty$ for all $\eta\in \Xi$. Let $\mathbb{H}$ be the completion of $\mathcal{H}$ under $\|\cdot\|_{L^2(dF_X)}$. For any $\epsilon>0$, if there exists $S(\epsilon,\|\cdot\|_{L^2(dF_X)},k)=\{h_1^l,h_1^u,...,h_k^l,h_k^u\}\subset \mathbb{H}$ such that $\max_{1\leq j\leq k}\|h_j^u-h_j^l\|_{L^2(dF_X)}\leq \epsilon$, and if for any $h\in \mathcal{H}$, there exists a $j\in\{1,...,k\}$ with $h_j^l\leq h\leq h_j^u$ a.e., then $$N_{[\ ]}(\epsilon,\mathcal{H},\|\cdot\|_{L^2(dF_X)}):=\min\{k:S(\epsilon,\|\cdot\|_{L^2(dF_X)},k) \}$$
is defined as the $\|\cdot\|_{L^2(dF_X)}$-bracketing number of the space $\mathcal{H}$.

Given a collection $\mathcal{F}$ of measurable functions $f:\mathcal{X}\to\mathbb{R}$. The  $\mathcal{F}$-indexed \emph{empirical process} is defined by
\begin{align*}
	f\to \mathbb{G}_n f:=\frac{1}{\sqrt{n}}\sum_{i=1}^n\left\{f(\bs{X}_i)-\mathbb{E}[f(\bs{X}_i)]\right\}.
\end{align*}
Let $\|\mathbb{G}_n\|_{\mathcal{F}}:=\sup_{f\in\mathcal{F}}|\mathbb{G}_n f|$ and  $\mathbb{E}^*[\cdot]$ denote the expectation taken with respect to the outer measure $\mathbb{P}^*$. The notation ``$a \lesssim b$" denotes that $a$ is bounded by a universal constant times $b$.

%\newpage
\section{Proof of Theorem \ref{thm:rate_ANN}}\label{app:rate_ANN}
The ANNs is defined by
\begin{align*}
	\mathcal{G}_n=\bigg\{&g:g(x)=g_0(x;\gamma_0)+\frac{B_n}{r_n}\sum_{j=1}^{r_n} \gamma_j\psi(a_j^\top x),  \notag\\ & \quad a_j=(a_{j1},...,a_{jp})^\top\in\mathbb{R}^p,\  \|a_{j}\|_2= 1, \ |\gamma_j|\leq 1   \bigg\}.
\end{align*}
Without loss of generality, we set $g_0(x;\gamma_0)\equiv 0$ in the proof.
By Assumption \ref{as:parameter_beta} (ii) and $g_d^*(x)=L^{-1}(\pi_d^*(x))=\log\left(\pi_d^*(x)/(1-\pi_d^*(x))\right)$, we have that  $g_d^*(x)\in [L^{-1}(\underline{c})-1,L^{-1}(\overline{c})+1]$ holds uniformly over $x\in\mathcal{X}$. We define the bounded ANNs by $$\widetilde{\mathcal{G}}_n:=\mathcal{G}_n\cap \{g: L^{-1}(\underline{c})-1\leq g(x)\leq  L^{-1}(\overline{c})+1,\  \forall x\in\mathcal{X}\}.$$
The elements in $\widetilde{G}_n$ are uniformly bounded.    Then by Lemma \ref{claim:uc} in Section C.1, we have  $$\widehat{g}_d\in \widetilde{G}_n \quad   \text{with probability approaching to one}.$$

By the triangle inequality and $\text{Proj}_{\mathcal{G}_n}g_d^*=\text{Proj}_{\widetilde{\mathcal{G}}_n}g_d^*$, we have
\begin{align}
	\|\widehat{g}_d-g_d^*\|_{L^2(dF_X)}\leq& \|\widehat{g}_d-\text{Proj}_{\mathcal{G}_n}g_d^*\|_{L^2(dF_X)}+\|\text{Proj}_{\mathcal{G}_n}g_d^*-g_d^*\|_{L^2(dF_X)}\notag \\
	=&\|\widehat{g}_d-\text{Proj}_{\widetilde{\mathcal{G}}_n}g_d^*\|_{L^2(dF_X)}+\|\text{Proj}_{\mathcal{G}_n}g_d^*-g_d^*\|_{L^2(dF_X)}.
	\label{eq:triangle}
\end{align}
By Assumption \ref{as:parameter} and the $L^2(dF_X)$-approximation results in Section \ref{sec:ANN}, we have
\begin{align}\label{eq:app}
	\|g_d^*-\text{Proj}_{\mathcal{G}_n}g_d^*\|_{L^2(dF_X)}  = \text{const} \times v_{g_d^*,m}\cdot r_n^{-\frac{1}{2}-\frac{1}{p}} .
\end{align}

We next apply \citet[Theorem 3.4.1]{van1996weak} to establish the rate of convergence for $\|\widehat{g}_d-\text{Proj}_{\widetilde{\mathcal{G}}_n}g_d^*\|_{L^2(dF_X)}$.  By the mean value theorem, we have that for any $\delta>0$,
\begin{align*}
	&\sup_{\{g_d\in\widetilde{\mathcal{G}}_n:\delta/2\leq \|g_d-\text{Proj}_{\widetilde{\mathcal{G}}_n}g_d^*\|_{L^2(dF_X)}\leq\delta\}}\mathbb{E}\left\{\ell_d(D_{di},\boldsymbol{X}_{i};g_d)-\ell_d(D_{di},\boldsymbol{X}_{i};\text{Proj}_{\widetilde{\mathcal{G}}_n}g_d^*)\right\}\\
	=&\sup_{\{g_d\in\widetilde{\mathcal{G}}_n:\delta/2\leq \|g_d-\text{Proj}_{\widetilde{\mathcal{G}}_n}g_d^*\|_{L^2(dF_X)}\leq\delta\}}\mathbb{E}\bigg\{D_{di}g_d(\boldsymbol{X}_i)-\log\left[1+\exp(g_d(\boldsymbol{X}_i))\right]\\
	&\qquad\qquad\qquad\qquad\qquad\qquad\qquad\qquad-D_{di}\text{Proj}_{\widetilde{\mathcal{G}}_n}g_d^*(\bs{X}_i)+\log\left[1+\exp(\text{Proj}_{\widetilde{\mathcal{G}}_n}g_d^*(\bs{X}_i))\right]\bigg\}\\
	= &\sup_{\{g_d\in\widetilde{\mathcal{G}}_n:\delta/2\leq \|g_d-\text{Proj}_{\widetilde{\mathcal{G}}_n}g_d^*\|_{L^2(dF_X)}\leq\delta\}}\mathbb{E}\bigg\{\left\{D_{di}-L(\text{Proj}_{\widetilde{\mathcal{G}}_n}g_d^*(\bs{X}_i))\right\}\{g_d(\boldsymbol{X}_i)-\text{Proj}_{\widetilde{\mathcal{G}}_n}g_d^*(\bs{X}_i)\}\\
	&\qquad\qquad\qquad-\frac{1}{2}L'\left(\theta_d(\bs{X}_i)\right)\{g_d(\boldsymbol{X}_i)-\text{Proj}_{\widetilde{\mathcal{G}}_n}g_d^*(\bs{X}_i)\}^2\bigg\}\\
	\leq  & -\text{const}\times \sup_{\{g_d\in\widetilde{\mathcal{G}}_n:\delta/2\leq \|g_d-\text{Proj}_{\widetilde{\mathcal{G}}_n}g_d^*\|_{L^2(dF_X)}\leq\delta\}}\|g_d-\text{Proj}_{\widetilde{\mathcal{G}}_n}g_d^*\|^2_{L^2(dF_X)}\leq  -\text{const}\times \delta^2,
\end{align*}
where $\theta_d(\bs{X}_i)$ is between $g_d(\boldsymbol{X}_i)\in \widetilde{\mathcal{G}}_n$ and $\text{Proj}_{\widetilde{\mathcal{G}}_n}g_d^*(\bs{X}_i)\in \widetilde{\mathcal{G}}_n$ (it is also uniformly bounded), and  $L(a)=\exp(a)/\{1+\exp(a)\}$ is the logistic function whose derivative satisfies $L'(a)=L(a)\{1-L(a)\}$; the first inequality holds because $\mathbb{E}[D_{di}-L(\text{Proj}_{\mathcal{G}_n}g_d^*(\bs{X}_i))|\bs{X}_i]\xrightarrow{L^2} \mathbb{E}[D_{di}-\pi_d^*(\bs{X}_i)|\bs{X}_i]=0$  and $L'\left(\theta_d(\bs{X}_i)\right)$ is uniformly bounded below from zero. Then the condition in  \citet[Theorem 3.4.1]{van1996weak} holds.
Then by  \citet[Theorems 3.4.1]{van1996weak}, the convergence rate of  $\|\widehat{g}_d-\text{Proj}_{\widetilde{\mathcal{G}}_n}g_d^*\|_{L^2(dF_X)}$ is bounded by $\delta_n$:
\begin{align}\label{def:delta_n}
	\delta_n=\inf \left\{ \delta>0:\frac{1}{\delta^{2}}\cdot \mathbb{E}^*\left[\|\mathbb{G}_n\|_{\widetilde{\mathcal{L}}_{\delta}}\right]\leq \text{const}\times n^{1/2} \right\},
\end{align}
where
\begin{align*}
	\widetilde{\mathcal{L}}_{\delta}:=&\left\{\ell_d(D_{di},\boldsymbol{X}_i;g_d):g_d\in\widetilde{\mathcal{G}}_n,\|g_d-g_d^*\|_{L^2(dF_X)}\leq \delta \right\}\\
	=&\left\{D_{di}g_d(\boldsymbol{X}_i)-\log\left[1+\exp(g_d(\bs{X}_i))\right]:g\in\widetilde{\mathcal{G}}_n,\|g_d-g_d^*\|_{L^2(dF_X)}\leq \delta \right\}.
\end{align*}	
We compute $\mathbb{E}^*\left[\|\mathbb{G}_n\|_{\widetilde{\mathcal{L}}_{\delta}}\right]$ to determine $\delta_n$. Note that the envelope function of $\widetilde{\mathcal{L}}_{\delta}$ is bounded by the universal constant $2\cdot \max\{|L^{-1}(\underline{c})|+1,|L^{-1}(\overline{c})|+1\}$, then by \citet[Lemma 3.4.2]{van1996weak}, we have
\begin{align*}
	&\mathbb{E}^*\left[\|\mathbb{G}_n\|_{\widetilde{\mathcal{L}}_{\delta}}\right]\\
	\lesssim& \widetilde{J}_{[\ ]}(\delta,\widetilde{\mathcal{L}}_{\delta},\|\cdot\|_{L^2(dF_{X})})\left(1+\frac{\widetilde{J}_{[\ ]}(\delta,\widetilde{\mathcal{L}}_{\delta},\|\cdot\|_{L^2(dF_{X})})}{\sqrt{n}\delta^2}\cdot 2\cdot \max\{|L^{-1}(\underline{c})|+1,|L^{-1}(\overline{c})|+1\}\right),
\end{align*}
where
\begin{align*}
	\widetilde{J}_{[\ ]}(\delta,\widetilde{\mathcal{L}}_{\delta},\|\cdot\|_{L^2(dF_{X})}):=&\int_0^{\delta}\sqrt{1+\log N_{[\ ]}(\epsilon,\widetilde{\mathcal{L}}_{\delta},\|\cdot\|_{L^2(dF_{X})})}d\epsilon.
\end{align*}
Hence, to calculate $\delta_n$, an upper bound of $\log N_{[\ ]}(\epsilon,\widetilde{\mathcal{L}}_{\delta},L^2(dF_X))$ is needed. Note that $\ell_d(D_{di},\boldsymbol{X}_i;g_d)$ is uniformly Lipschitz in $g_d$,  we can have
\begin{align*}
	\log N_{[\ ]}(\epsilon,\widetilde{\mathcal{L}}_{\delta},\|\cdot\|_{L^2(dF_{X})}) \lesssim 	\log N_{[\ ]}(\epsilon,\widetilde{\mathcal{G}}_{n},\|\cdot\|_{L^2(dF_{X})}) \leq 	\log N_{[\ ]}(\epsilon,\mathcal{G}_{n},\|\cdot\|_{L^2(dF_{X})}).
\end{align*}
Thus, to choose $\delta_n$ satisfying \eqref{def:delta_n}, it is sufficient to choose $\delta_n$ such that
\begin{align} \label{def:delta_n_suff}
	\delta_n=\inf \left\{ \delta>0:\frac{1}{\delta^{2}}\cdot \int_0^{\delta}\sqrt{1+\log N_{[\ ]}(\epsilon,{\mathcal{G}}_{n},\|\cdot\|_{L^2(dF_{X})})}d\epsilon\lesssim n^{1/2} \right\}.
\end{align}
We claim the following result:
\begin{align}\label{eq:Hbound1}
	\log N_{[\ ]}(\epsilon,\mathcal{G}_n,\|\cdot\|_{L^2(dF_X)})\leq \text{const}\times r_n(p+1)\log \left(\frac{B_n\cdot(p+1)}{\epsilon}\right),
\end{align}
whose proof will be given later. Then
\begin{align}
	&\frac{1}{\delta^2}\int_0^{\delta}\sqrt{1+\log N_{[\ ]}(\epsilon,{\mathcal{G}}_{n},\|\cdot\|_{L^2(dF_{X})})}d\epsilon \notag \\
	\lesssim&\frac{1}{\delta^2}\int_{0}^{\delta}	\left\{r_n\cdot (p+1)\cdot \log \left(\frac{ B_n\cdot (p+1)}{\epsilon}\right)\right\}^{1/2}d\epsilon\notag\\
	\lesssim &\frac{1}{\delta}\cdot r_n^{1/2}\cdot  (p+1)^{1/2}\cdot  \left\{  \log \left(  B_n\cdot (p+1)\right)\right\}^{1/2}. \label{eq:Hbound2}
\end{align}
Then we can choose $\delta_n$  as follows  so that  \eqref{def:delta_n_suff} holds:
\begin{align}
	\delta_n =&\text{const}\times\sqrt{\frac{ r_n\cdot (p+1)\cdot \log(B_n\cdot (p+1))}{n}} \notag\\
	\leq & \text{const}\times \sqrt{\frac{r_n\cdot p\cdot \log (n\cdot B_n)}{n}}\leq  \text{const}\times \sqrt{\frac{ r_n\cdot p\cdot (\log n+\log 2v_{g_d^*,m} )}{n}} \notag\\
	\leq & \text{const}\times \sqrt{\frac{ r_n\cdot p\cdot (\log n+p )}{n}}\leq \text{const}\times \sqrt{\frac{ r_n\cdot p\cdot \log n }{n}},\label{eq:delta}
\end{align}
where the second inequality holds by assumption that $B_n\leq 2v_{g_d^*,m}$,  the third inequality holds by Theorem \ref{thm:vfm_finite}, and the last inequality holds by noting $p\leq C_0\cdot \sqrt{\log n}$. Combining \eqref{eq:triangle}, \eqref{eq:app}, \eqref{def:delta_n}, \eqref{eq:delta}, and Assumption \ref{as:rate_rp}, we obtain
\begin{align*}
	&\|\widehat{g}_d-g_d^*\|_{L^2(dF_X)}=O_P\left(\max\left\{v_{g_d^*,m}\cdot r_n^{-\frac{1}{2}-\frac{1}{p}},\sqrt{\frac{r_n\cdot p\cdot \log n}{n}}\right\}\right)=o_P(n^{-1/4}),
\end{align*}
where the constants hiding inside $O_p$ and $o_p$ do not depend on
$p$ and $n$.

In light of that $L'(a)=L(a)\{1-L(a)\}\leq 1/4$ for all $a\in\mathbb{R}$, we further have
\begin{align*}
	&\|\widehat{\pi}_d-\pi_d^*\|_{L^2(dF_X)}=\|L\left(\widehat{g}_d\right)-L\left(g_d^*\right)\|_{L^2(dF_X)}\\
	\leq &\frac{1}{4}\cdot\|\widehat{g}_d-g_d^*\|_{L^2(dF_X)}= O_P\left(\max\left\{v_{g_d^*,m}\cdot r_n^{-\frac{1}{2}-\frac{1}{p}},\sqrt{\frac{r_n\cdot p\cdot \log n}{n}}\right\}\right)=o_P\left(n^{-1/4}\right).
\end{align*}
Similarly, we can also obtain the rates of convergence under the $\|\cdot\|_n$-norm:
\begin{align*}
	&\|\widehat{g}_d-g_d^*\|_{n}=O_P\left(\max\left\{v_{g_d^*,m}\cdot r_n^{-\frac{1}{2}-\frac{1}{p}},\sqrt{\frac{r_n\cdot p\cdot \log n}{n}}\right\}\right)=o_P\left(n^{-1/4}\right),\\
	&\|\widehat{\pi}_d-\pi_d^*\|_{n}=O_P\left(\max\left\{v_{g_d^*,m}\cdot r_n^{-\frac{1}{2}-\frac{1}{p}},\sqrt{\frac{r_n\cdot p\cdot \log n}{n}}\right\}\right)=o_P\left(n^{-1/4}\right).
\end{align*}

We now begin to prove \eqref{eq:Hbound1}. Denote $\boldsymbol{\theta}:=(\gamma_1,...,\gamma_{r_n},a_1^{\top},...,a_{r_n}^{\top})$, $\Theta:=\{\boldsymbol{\theta}:\|a_j\|_2= 1, \  |\gamma_j|\leq  1, \ j\in\{1,...,r_n\} \}$, and $g(x;\boldsymbol{\theta}):=(B_n/r_n)\sum_{j=1}^{r_n} \gamma_j\psi(a_j^\top x)$ for $\boldsymbol{\theta}\in\Theta$. Consider two elements $g(x;\boldsymbol{\theta})$ and $g(x;\widetilde{\boldsymbol{\theta}})$ in $\mathcal{G}_{n}$, we have
\begin{align*}
	&|g(x;\boldsymbol{\theta})-g(x;\widetilde{\boldsymbol{\theta}})|\\
	=&	\left|\frac{B_n}{r_n}\sum_{j=1}^{r_n} \gamma_j\psi(a_j^\top x)-\frac{B_n}{r_n}\sum_{j=1}^{r_n} \widetilde{\gamma}_j\psi(\widetilde{a}_j^\top x)\right|\\
	\leq & \frac{B_n}{r_n}\sum_{j=1}^{r_n}\left|\gamma_j\psi(a_j^\top x)-\widetilde{\gamma}_j\psi(\widetilde{a}_j^\top x)\right|\\
	\leq&   \frac{B_n}{r_n}\sum_{j=1}^{r_n}\left|\gamma_j-\widetilde{\gamma}_j\right|\cdot |\psi(a_j^\top x)|+ \frac{B_n}{r_n}\sum_{j=1}^{r_n}|\widetilde{\gamma}_j|\cdot\left|\psi(a_j^\top x)-\psi(\widetilde{a}_j^\top x)\right|\\
	\leq &  \frac{B_n}{r_n}\sum_{j=1}^{r_n}\left|\gamma_j-\widetilde{\gamma}_j\right|\cdot C_{\psi}\cdot \|x\|_{\infty}\cdot \|a_j\|_{1}+\frac{B_n}{r_n}\sum_{j=1}^{r_n}|\widetilde{\gamma}_j|\cdot C_{\psi}\cdot \|x\|_{\infty} \cdot \|a_j-\widetilde{a}_j\|_1\\
	\leq & \frac{B_n}{r_n} C_{\psi}\cdot \sqrt{p} \sum_{j=1}^{r_n}\left|\gamma_j-\widetilde{\gamma}_j\right| +\frac{B_n}{r_n}C_{\psi}\cdot \sum_{j=1}^{r_n}   \|a_j-\widetilde{a}_j\|_1\\
	\leq& C_{\psi}\cdot \frac{B_n}{r_n}\cdot \sqrt{p}\sum_{j=1}^{r_n}   \left\{\left|\gamma_j-\widetilde{\gamma}_j\right|+\|a_j-\widetilde{a}_j\|_1\right\}\\
	= &C_{\psi}\cdot \frac{B_n}{r_n}\cdot \sqrt{p}\cdot \|\boldsymbol{\theta}-\tilde{\boldsymbol{\theta}}\|_1\leq C_{\psi}\cdot \frac{B_n}{r_n}\cdot \sqrt{p} \sqrt{(p+1)r_n}\cdot \|\boldsymbol{\theta}-\tilde{\boldsymbol{\theta}}\|_2\\
	\leq& C_{\psi}\cdot \frac{B_n}{\sqrt{r_n}}\cdot (p+1) \cdot \|\boldsymbol{\theta}-\tilde{\boldsymbol{\theta}}\|_2,
\end{align*}
where $C_{\psi}$ is the Lipschitz coefficient of $\psi(\cdot)$, and the forth inequality holds by using $\|a_j\|_1\leq \sqrt{p}\cdot \|a_j\|_2\leq \sqrt{p}$.  We make the following claim:
\begin{align}\label{eq:cover<bracket}
	N_{[\ ]}\left(\epsilon,\mathcal{G}_{n},L^2(dF_X) \right)\leq N\left(\frac{\epsilon\cdot \sqrt{r_n}}{2C_{\psi}\cdot B_n\cdot (p+1)},\Theta,\|\cdot\|_2\right),
\end{align}
where $N\left(\frac{\epsilon\cdot \sqrt{r_n}}{2C_{\psi}\cdot B_n\cdot (p+1)},\Theta,\|\cdot\|_2\right)$ is the $\frac{\epsilon\cdot \sqrt{r_n}}{2C_{\psi}\cdot B_n\cdot (p+1)}$-covering number of $\Theta$ with respect to the metric $\|\cdot\|_2$.   We prove  \eqref{eq:cover<bracket}. Let $\{\boldsymbol{\theta}_i\}_{i=1}^N$ be an $\frac{\epsilon\cdot \sqrt{r_n}}{2C_{\psi}\cdot B_n\cdot (p+1)}$-covering of $\Theta$ under $\|\cdot\|_2$, then define
\begin{align*}
	u_{i}(x):=h(x;\boldsymbol{\theta}_i)+\frac{\epsilon}{2}\  \text{and} \	l_{i}(x):=h(x;\boldsymbol{\theta}_i)-\frac{\epsilon}{2}.
\end{align*}
We know that for any $\boldsymbol{\theta}\in\Theta$, $\exists \boldsymbol{\theta}_i$, s.t. $\|\boldsymbol{\theta}-\boldsymbol{\theta}_i\|_2\leq\frac{\epsilon\cdot \sqrt{r_n}}{2C_{\psi}\cdot B_n\cdot (p+1)}$, and from Lipschitz properties of $h(x;\boldsymbol{\theta})$, we have:
\begin{align*}
	|h(x;\boldsymbol{\theta})-h(x;\boldsymbol{\theta}_i)|\leq C_{\psi}\cdot \frac{B_n}{\sqrt{r_n}}\cdot (p+1)\cdot \|\boldsymbol{\theta}-\boldsymbol{\theta}_i\|_2\leq \frac{\epsilon}{2}.
\end{align*}
Thus for all $x\in \mathcal{X}$,
\begin{align*}
	l_{i}(x)	\leq h(x;\boldsymbol{\theta})\leq u_{i}(x).
\end{align*}
As  for all $1\leq i\leq N$,  $\left\{\int_{\mathcal{X}}\|u_{i}(x)-l_{i}(x)\|^2dF_X(x)\right\}^{1/2}\leq \epsilon$ (we have $\epsilon$-separation). This ends the proof of \eqref{eq:cover<bracket}.

Note that the dimension of $\Theta$ is of $r_n(p+1)$ and
\begin{align*}
	\sup_{\boldsymbol{\theta}\in\Theta}\|\boldsymbol{\theta}\|_2=\sup_{\boldsymbol{\theta}\in\Theta}\left\{ \sum_{j=1}^{r_n}|\gamma_j|^2 +\sum_{j=1}^{r_n}\|a_j\|_2^2 \right\}^{1/2}
	\leq \sup_{\boldsymbol{\theta}\in\Theta}\left\{ r_n  +r_n \right\}^{1/2}\leq \sqrt{2r_n}.
\end{align*}
Then an upper bound of $N\left(\frac{\epsilon\cdot \sqrt{r_n}}{2C_{\psi}\cdot B_n\cdot (p+1)},\Theta,\|\cdot\|_2\right)$ is
\begin{align*}
	&N\left(\frac{\epsilon\cdot \sqrt{r_n}}{2C_{\psi}\cdot B_n\cdot (p+1)},\Theta,\|\cdot\|_2\right)\\
	\leq & \text{const}\times \left(\frac{\sup_{\boldsymbol{\theta}\in\Theta}\|\boldsymbol{\theta}\|_2}{\epsilon}\cdot \frac{2C_{\psi}\cdot B_n\cdot  (p+1)}{\sqrt{r_n}}\right)^{r_n(p+1)} \\
	\leq & \text{const}\times \left(\frac{2\sqrt{2}}{\epsilon}\cdot C_{\psi}\cdot B_n\cdot  (p+1)\right)^{r_n(p+1)},
\end{align*}
which further gives the desired result \eqref{eq:Hbound1}:
\begin{align*}
	&\log N_{[\ ]}\left(\epsilon,\mathcal{G}_{n},L^2(dF_X)\right)
	\leq  \log N\left(\frac{\epsilon\cdot \sqrt{r_n}}{2C_{\psi}\cdot B_n\cdot (p+1)},\Theta,\|\cdot\|_2\right)\\
	\leq & \text{const}\times r_n(p+1)\left\{\log 2\sqrt{2}C_{\psi}+ \log B_n(p+1)-\log \epsilon \right\}\\
	\leq &  \text{const}\times r_n(p+1)\log \left(\frac{B_n\cdot(p+1)}{\epsilon}\right).
\end{align*}

{
	\subsection{Discussion on Assumption \ref{as:rate_rp}' (i)}
	This section shows that Assumption \ref{as:rate_rp}' (i) and implies Assumption \ref{as:rate_rp} (i), i.e. under Assumption \ref{as:rate_rp}' (i), we have
	\begin{align}\label{eq:rate-variance}
		&\sqrt{\frac{ r_n\cdot p\cdot \log n}{n}}=o\left(n^{-1/4}\right),\\
		&v_{g_d^*,m}\cdot r_n^{-\frac{1}{2}-\frac{1}{p}}=O\left( M^p\cdot r_n^{-\frac{1}{2}-\frac{1}{p}}\right)=o\left(n^{-1/4}\right). \label{eq:rate-bias}
	\end{align}
	By Assumption \ref{as:rate_rp}' (i), we have
	\begin{align}\label{eq:rate_rp}
		p\leq a_n\cdot\left(\log n\right)^{\frac{1}{2}}\ \text{and} \  C_1\cdot n^{\frac{p+1}{2(p+2)}}\leq r_n \leq C_2\cdot  (\log n)^{-\frac{3}{2}}\cdot n^{\frac{1}{2}},
	\end{align}
	where $a_n\to 0$ can be arbitrarily slow, and $C_1$ and $C_2$ are two positive constants. We show that the choice of $p$ and $r_n$ in \eqref{eq:rate_rp} implies \eqref{eq:rate-variance} and \eqref{eq:rate-bias} hold. Since
	\begin{align*}
		\sqrt{\frac{r_n\cdot p\cdot \log n}{n}}=n^{-1/4}\cdot n^{-1/4}\cdot r_n^{1/2}\cdot p^{1/2}\cdot \{\log n\}^{1/2},
	\end{align*}
	then  \eqref{eq:rate-variance} holds in light of the following fact:
	\begin{align*}
		& n^{-1/4}\cdot r_{n}^{1/2}\cdot p^{1/2}\cdot \{\log n\}^{1/2}\rightarrow 0\\
		\Leftarrow & -\frac{1}{4}\log n+\frac{1}{2}\log r_{n}+\frac{1}{4}\cdot \log\log n+\frac{\log a_{n}}{2}+\frac{1}{2}\log \log n\rightarrow -\infty  \\
		\Leftarrow & -\frac{1}{4}\log n+\frac{1}{4}\log n-\frac{3}{4}\log \log n+\frac{1}{2}\log C_2+\frac{3}{4}\cdot \log \log n+\frac{1}{2}\log a_{n}\rightarrow -\infty  \\
		\Leftrightarrow & \frac{1}{2}\log a_{n}+\frac{1}{2}\log C_2\rightarrow -\infty ,
	\end{align*}%
	where the last convergence holds since $a_{n}\rightarrow 0$.

	Note that
	\begin{align*}
		M^p\cdot r_n^{-\frac{1}{2}-\frac{1}{p}}= M^p\cdot O\left(n^{-\frac{p+1}{4p}}\right)=O\left(n^{-\frac{1}{4}}\cdot  M^p\cdot n^{-\frac{1}{4p}}\right),
	\end{align*}
	then \eqref{eq:rate-bias} holds in light of the following fact:
	\begin{align*}
		& M^p\cdot n^{-\frac{1}{4p}}\to 0\\
		\Leftrightarrow&p\cdot \log M-\frac{1}{4p}\log n\to -\infty\\
		\Leftarrow&\left\{a_n\cdot \log M-\frac{1}{4a_n}\right\}\left(\log n\right)^{\frac{1}{2}}\to -\infty,
	\end{align*}
	where the last convergence holds since $a_n\to 0$.

	Therefore, under Assumptions \ref{as:parameter}, \ref{as:rate_rp} (ii), and \ref{as:rate_rp}'(i), we have 	\begin{align*}
		&\|\widehat{g}_d-g_d^*\|_{L^2(dF_X)}=O_P\left(\max\left\{v_{g_d^*,m}\cdot r_n^{-\frac{1}{2}-\frac{1}{p}},\sqrt{\frac{r_n\cdot p\cdot \log n}{n}}\right\}\right)=o_P\left(n^{-1/4}\right),
	\end{align*}
	where the constants hiding inside $O_p$ and $o_p$ do not depend on
	$p$ and $n$.}

%\newpage

\section{Key Lemmas} \label{app:lemma:projection}
By Assumptions \ref{as:rate_rp}, \ref{as:parameter} and Theorem \ref{thm:rate_ANN}, we have $g_d^*\in\mathcal{F}^{m}_p$ and
\begin{align*}
	\|\widehat{g}_d-{g}^*_d\|_{L^2(dF_X)}=O_P(\delta_n),
\end{align*}
where $\delta_n=O(\sqrt{r_n\cdot p\cdot \log n/n})=o(n^{-1/4})$ defined in \eqref{eq:delta}. Let $\epsilon_n$ be a positive sequence satisfying $\epsilon_n=o(n^{-1/2})$.
For any $g_d\in \{g_d\in\mathcal{G}_n:\|g_d-g_d^*\|_{L^2(dF_X)}\leq \delta_n \}$, we define a local alternative value around $g_d$ as follows:
\begin{align*}
	\overline{g}(g_d,\epsilon_n):=(1-\epsilon_n)\cdot g_d+\epsilon_n\cdot \{u^*+g_d^*\},
\end{align*}
where $u^*(\boldsymbol{X}_{i})=\mathcal{E}_d(\boldsymbol{X}_{i},\beta_d^*)/\pi_d^*(\boldsymbol{X}_i)\in \mathcal{F}_p^m$. Let $\text{Proj}_{\mathcal{G}_n}g$ denote the $L^2(dF_X)$-projection of $g$ on the ANN space $\mathcal{G}_n$. We recall the following notations
\begin{align*}
	&\ell_d(D_{di},\boldsymbol{X}_{i};g_d)=D_{di}g_d(\boldsymbol{X}_i)-\log\left[1+\exp(g_d(\boldsymbol{X}_i))\right],\\
	&L_{d,n}(g_d)=\frac{1}{n}\sum_{i=1}^n\ell_d(D_{di},\boldsymbol{X}_{i};g_d).
\end{align*}
We compute the following quantities that will be used later:
\begin{align*}
	\frac{\partial}{\partial g_d}\ell_d(D_{di},\boldsymbol{X}_{i};g_d)[u]:=&\lim_{t\to 0}\frac{\ell_d(D_{di},\boldsymbol{X}_{i};g_d+t\cdot u)-\ell_d(D_{di},\boldsymbol{X}_{i};g_d)}{t}\\
	=&\left\{D_{di}-L(g_d(\boldsymbol{X}_i))\right\} u(\boldsymbol{X}_{i}),\end{align*}
and
\begin{align}\label{def:r}
	&r(D_{di},\boldsymbol{X}_{i};g_d-g^*_d)\\
	:=&\ell_d(D_{di},\boldsymbol{X}_{i};g_d)-\ell_d(D_{di},\boldsymbol{X}_{i};g^*_d)-\frac{\partial}{\partial g_d}\ell_d(D_{di},\boldsymbol{X}_{i};g^*_d(\cdot))[g_d-g^*_d]\notag\\
	=&D_{di}g_d(\boldsymbol{X}_i)-\log\left[1+\exp(g_d(\boldsymbol{X}_i))\right]-D_{di}g^*_d(\boldsymbol{X}_i)+\log\left[1+\exp(g^*_d(\boldsymbol{X}_i))\right]\notag\\
	&-\left\{D_{di}-\pi^*_d(\boldsymbol{X}_i)\right\} \{g_d(\boldsymbol{X}_{i})-g^*_d(\boldsymbol{X}_{i}) \}\notag\\
	=&-\frac{1}{2}\cdot\pi_d^*(\boldsymbol{X}_i)(1-\pi_d^*(\boldsymbol{X}_i))\{g_d(\boldsymbol{X}_{i})-g^*_d(\boldsymbol{X}_{i}) \}^2+R(D_{di},\boldsymbol{X}_{i};g_d),\notag
\end{align}
where
\begin{align*}
	R(D_{di},\boldsymbol{X}_{i};g_d):=-\frac{1}{2}\int_{g_d^*(\boldsymbol{X}_i)}^{g_d(\boldsymbol{X}_i)}L(t)\{1-L(t)\}\{1-2L(t)\}(g_d(\boldsymbol{X}_i)-t)^2dt,
\end{align*}
and the last equality holds by the following Taylor's theorem
\begin{align*}
	f(v)=f(a)+f'(a)(v-a)+\frac{f''(a)}{2}(v-a)^2+R(v), \ \forall f\in C^3(\mathbb{R}),
\end{align*}
and $R(v)$ is the integral form of the remainder defined by
\begin{align*}
	R(v)=\int_a^v \frac{f'''(t)}{2}(v-t)^2dt.
\end{align*}

{
	\subsection{Lemma \ref{claim:uc}}
	\begin{lemma} \label{claim:uc}
		Suppose  Assumptions \ref{as:rate_rp} and \ref{as:parameter} hold. Then
		\begin{align*}
			\|\widehat{g}_d-g_d^*\|_{\infty}=o_P(1) \ \text{and} \ \|\widehat{\pi}_d-\pi_d^*\|_{\infty}=o_P(1).
		\end{align*}
	\end{lemma}
	\begin{proof}
		We prove this result by applying  \citet[Theorem 3.1]{chen2007large}.
		Recall $\widehat{g}_d$ is the sieve MLE  of $g_d^*\in \mathcal{G}:=\{g_d\in\mathcal{F}^m_p:L^{-1}(\underline{c})-1\leq g_d(x)\leq L^{-1}(\overline{c})+1 \}$ based on the entire sample. That is,
		\begin{align*}
			\frac{1}{n}\sum_{i=1}^n\ell_d(D_{di},\boldsymbol{X}_{i};\widehat{g}_d)>\sup_{g_d\in\mathcal{G}_n}\frac{1}{n}\sum_{i=1}^n\ell_d(D_{di},\boldsymbol{X}_{i};{g}_d)-O(\epsilon_n^2),
		\end{align*}
		where $\ell_d(D_{di},\boldsymbol{X}_{i};g_d):=D_{di}g_d(\boldsymbol{X}_i)-\log\left[1+\exp(g_d(\boldsymbol{X}_i))\right]$. We have the following results:
		\begin{itemize}
			\item [(i)] (Identification) $\mathbb{E}[\ell_d(D_{di},\boldsymbol{X}_{i};g_d)]$ is uniquely maximized at $g^*_{d}(\cdot)=L^{-1}\left(\pi_d^*(\cdot)\right)\in \mathcal{G}=\{g_d\in\mathcal{F}^m_p:L^{-1}(\underline{c})-1\leq g_d(x)\leq L^{-1}(\overline{c})+1 \}$;
			\item [(ii)] (Sieve spaces) The ANNs $\mathcal{G}_n$ increases with sample size $n$ and is dense in $\mathcal{F}_p^m$;
			\item [(iii)] (Continuity) $\mathbb{E}[\ell_d(D_{di},\boldsymbol{X}_{i};g_d)]$ is continuous in $g_d(\cdot)$ under the metric $\|\cdot\|_{\infty}$;
			\item [(iv)] (Compact sieve space) The ANNs $\mathcal{G}_n$ are compact under $\|\cdot\|_{\infty}$;
			\item [(v)] (Uniform convergence over sieves) $\sup_{g_d\in\mathcal{G}_n}|n^{-1}\sum_{i=1}^n\{\ell_d(D_{di},\boldsymbol{X}_{i};g_d)-\mathbb{E}[\ell_d(D_{di},\boldsymbol{X}_{i};g_d)]\}|=o_P(1)$,
		\end{itemize}
		where both results $(iii)$ and $(v)$ are due to the fact that for any $g_d(\cdot)$ and $\widetilde{g}_d(\cdot)$:
		\begin{align*}
			&|\ell_d(D_{di},\boldsymbol{X}_{i};g_d)-\ell_d(D_{di},\boldsymbol{X}_{i};\widetilde{g}_d)|\\
			=&\left|D_{di}g_d(\boldsymbol{X}_i)-\log\left[1+\exp(g_d(\boldsymbol{X}_i))\right]-D_{di}\widetilde{g}_d(\boldsymbol{X}_i)+\log\left[1+\exp(\widetilde{g}_d(\boldsymbol{X}_i))\right]\right|\\
			\leq & 2\cdot |g_d(\boldsymbol{X}_{i})-\widetilde{g}_d(\boldsymbol{X}_{i})|.
		\end{align*}
		Then by  \citet[Theorem 3.1]{chen2007large}, we have $\|\widehat{g}_d-g_d^*\|_{\infty}=o_P(1)$. Furthermore, $\|\widehat{\pi}_d-\pi_d^*\|_{\infty}=\|L(\widehat{g}_d)-L(g_d^*)\|_{\infty}\leq 1/4\cdot \|\widehat{g}_d-g_d^*\|_{\infty}=o_P(1)$ since $L(\cdot)$ is a Lipschitz function with Lipschitz coefficient $1/4$.
\end{proof}}

\clearpage

\subsection{Lemma \ref{lemma:several_results}}
Lemma \ref{lemma:several_results} is a stochastic equicontinuity result will be used in the proof of Lemma \ref{lemma:projection}.
\begin{lemma}\label{lemma:several_results}
	Under Assumption \ref{as:rate_rp},  we have
	\begin{align*}
		\sup_{\{g_d\in \widetilde{\mathcal{G}}_n:\|g_d-g_d^*\|_{L^2(dF_X)}\leq \delta_n\}} \mathbb{G}_n\left(\frac{\partial}{\partial g_d}\ell_d(D_{di},\boldsymbol{X}_{i};g^*_d)[ g_d(\boldsymbol{X}_{i})-g_d^*(\boldsymbol{X}_{i})]\right)=o_P(1),
	\end{align*}
	where $\widetilde{\mathcal{G}}_n=\mathcal{G}_n\cap \{g: L^{-1}(\underline{c})-1\leq g(x)\leq  L^{-1}(\overline{c})+1,\  \forall x\in\mathcal{X}\}$.
\end{lemma}
\begin{proof}
	Let
	\begin{align*}
		\widetilde{\mathcal{L}}_{\delta_n}:=&\left\{\frac{\partial}{\partial g_d}\ell_t(D_{di},\boldsymbol{X}_i;g_d^*)[ g_d(\boldsymbol{X}_{i})-g_d^*(\boldsymbol{X}_{i})]:g_d\in\widetilde{\mathcal{G}}_n,\|g_d-g_d^*\|_{L^2(dF_X)}\leq \delta_n \right\}\\
		=&\left\{\{D_{di}-\pi_d^*(\bs{X}_i)\}[g_d(\boldsymbol{X}_i)-g_d^*(\boldsymbol{X}_i)]:g\in\widetilde{\mathcal{G}}_n,\|g_d-g_d^*\|_{L^2(dF_X)}\leq \delta_n \right\},
	\end{align*}	
	and it is obvious to see that
	\begin{align*}
		\log N_{[\ ]}(\epsilon,\widetilde{\mathcal{L}}_{\delta_n},\|\cdot\|_{L^2(dF_{X})}) \lesssim 	\log N_{[\ ]}(\epsilon,\widetilde{\mathcal{G}}_{n},\|\cdot\|_{L^2(dF_{X})}).
	\end{align*}
	We show that $\|\mathbb{G}_n\|_{\widetilde{\mathcal{L}}_{\delta_n}}=o_P(1)$.	By \eqref{eq:Hbound1} and Assumption \ref{as:rate_rp}, we have
	\begin{align*}
		\log N_{[\ ]}&(\epsilon,\widetilde{\mathcal{G}}_n,\|\cdot\|_{L^2(dF_{X})})\leq  \log N_{[\  ]}(\epsilon,\mathcal{G}_n,\|\cdot\|_{L^2(dF_{X})})\\
		\leq & \text{const}\times r_n(p+1)\log \left(\frac{B_n\cdot(p+1)}{\epsilon}\right)\\
		\leq &\text{const} \times o(1)\times n^{\frac{1}{2}}\{\log n\}^{-1}\left[\sqrt{\log n} +\log \left(\frac{1}{\epsilon}\right)\right]\\
		=&\text{const}\times o(1)\times \left[ n^{\frac{1}{2}}\cdot \{\log n \}^{-1/2}+n^{\frac{1}{2}}\cdot\{\log n\}^{-1}\cdot\log \left(\frac{1}{\epsilon}\right)\right].
	\end{align*}	
	Then
	\begin{align*}
		\widetilde{J}_{[\ ]}(\delta_n,\widetilde{\mathcal{L}}_{\delta_n},\|\cdot\|_{L^2(dF_{X})})	=&\int_0^{\delta_n}\sqrt{1+\log N_{[\ ]}(\epsilon,\widetilde{\mathcal{L}}_{\delta_n},\|\cdot\|_{L^2(dF_{X})})}d\epsilon\\
		\lesssim&\int_0^{\delta_n}\sqrt{1+\log N_{[\ ]}(\epsilon,\widetilde{\mathcal{G}}_{n},\|\cdot\|_{L^2(F_{X})})}d\epsilon\\
		\lesssim&\delta_n\cdot o(1)\times   n^{\frac{1}{4}}\cdot \{\log n \}^{-1/4}
		\lesssim   o\left(\{\log n \}^{-1/4}\right)\cdot \delta_n\cdot n^{\frac{1}{4}}=o\left(1\right),
	\end{align*}
	where the last equality holds since $\delta_n=o(n^{-1/4})$.
	Note that the envelope function of $\widetilde{\mathcal{L}}_{\delta_n}$ is bounded by the universal constant $2\cdot \max\{|L^{-1}(\underline{c})|+1,|L^{-1}(\overline{c})|+1\}$. Then by \citet[Lemma 3.4.2]{van1996weak}, we have
	\begin{align*}
		&\mathbb{E}^*\left[\|\mathbb{G}_n\|_{\widetilde{\mathcal{L}}_{\delta_n}}\right]\\
		\lesssim &\widetilde{J}_{[\ ]}(\delta_n,\widetilde{\mathcal{L}}_{\delta_n},\|\cdot\|_{L^2(dF_{X})})\left(1+\frac{\widetilde{J}_{[\ ]}(\delta_n,\widetilde{\mathcal{L}}_{\delta_n},\|\cdot\|_{L^2(dF_{X})})}{\sqrt{n}\delta_n^2}\cdot 2\cdot \max\{|L^{-1}(\underline{c})|+1,|L^{-1}(\overline{c})|+1\}\right)\\
		\lesssim & o\left(1\right),
	\end{align*}
	which implies that $\|\mathbb{G}_n\|_{\widetilde{\mathcal{L}}_{\delta_n}}=o_P(1)$ by Markov inequality. Finally, we can conclude the desired result.
\end{proof}

%\newpage
\subsection{Lemma \ref{lemma:projection}}
\begin{lemma}\label{lemma:projection}
	Under Assumptions \ref{as:rate_rp}-\ref{as:app_error},  we have
	\begin{align*}
		&\sqrt{n}\cdot\mathbb{E}\left[\{1-\pi_d^*(\boldsymbol{X}_i)\}\left\{\widehat{g}_d(\boldsymbol{X}_{i})-g^*_d(\boldsymbol{X}_{i})\right\}\mathcal{E}_d(\boldsymbol{X}_{i};\beta_d^*)\right]=\frac{1}{\sqrt{n}}\sum_{i=1}^n\left\{\frac{D_{di}-\pi_d^*(\boldsymbol{X}_{i})}{\pi_d^*(\boldsymbol{X}_{i})}\right\}\mathcal{E}_d(\boldsymbol{X}_{i};\beta_d^*)+o_P(1).
	\end{align*}
\end{lemma}
\begin{proof}

	By \eqref{def:r},	for any $g_d\in\mathcal{G}_{n}$, we have
	\begin{align}
		L_{d,n}&(g_d)=L_{d,n}(g^*_d)+\left\{L_{d,n}(g_d)-L_{d,n}(g^*_d)\right\} \notag\\
		=&L_{d,n}(g^*_d)+\frac{1}{n}\sum_{i=1}^n\left\{\ell_d(D_{di},\boldsymbol{X}_{i};g_d)-\ell_d(D_{di},\boldsymbol{X}_{i};g^*_d)\right\} \notag\\
		=&L_{d,n}(g^*_d)+\frac{1}{n}\sum_{i=1}^n\left\{\frac{\partial}{\partial g_d}\ell_d(D_{di},\boldsymbol{X}_{i};g^*_d)[g_d-g^*_d]+r(D_{di},\boldsymbol{X}_{i};g_d-g^*_d)\right\}  \notag\\
		=&L_{d,n}(g^*_d)+\frac{1}{\sqrt{n}}\cdot  \mathbb{G}_n\left(\frac{\partial}{\partial g_d}\ell_d(D_{di},\boldsymbol{X}_{i};g^*_d)[g_d-g^*_d]\right)+\frac{1}{n}\sum_{i=1}^nr(D_{di},\boldsymbol{X}_{i};g_d-g^*_d), \label{eq:L_proj}
	\end{align}
	where the last equality holds because of the following first order condition
	\begin{align*}
		\mathbb{E}\bigg[\frac{\partial}{\partial g_d}\ell_d(D_{di},\boldsymbol{X}_{i};g^*_d)[g_d -g^*_d ]\bigg]=0, \quad \forall g_d\in\mathcal{G}_n.
	\end{align*}
	Using \eqref{eq:L_proj}, by substituting $g_d$ with $\widehat{g}_d$ and $\text{Proj}_{\mathcal{G}_{n}}\overline{g}(\widehat{g}_d,\epsilon_n)$ respectively, we get
	\begin{align}
		L_{d,n}(\widehat{g}_d)=L_{d,n}(g^*_d)+\frac{1}{\sqrt{n}}\cdot  \mathbb{G}_n\left(\frac{\partial}{\partial g_d}\ell_d(D_{di},\boldsymbol{X}_{i};g^*_d)[ \widehat{g}_d-g^*_d]\right)+\frac{1}{n}\sum_{i=1}^nr(D_{di},\boldsymbol{X}_{i};\widehat{g}_d-g^*_d), \label{eq:L_projhat}
	\end{align}
	and
	\begin{align}
		L_{d,n}(\text{Proj}_{\mathcal{G}_{n}}\overline{g}(\widehat{g}_d,\epsilon_n))=&L_{d,n}(g^*_d)+\frac{1}{\sqrt{n}}\cdot  \mathbb{G}_n\left(\frac{\partial}{\partial g_d}\ell_d(D_{di},\boldsymbol{X}_{i};g^*_d)[\text{Proj}_{\mathcal{G}_{n}}\overline{g}(\widehat{g}_d,\epsilon_n)-g^*_d]\right) \notag\\
		&+\frac{1}{n} \sum_{i=1}^nr(D_{di},\boldsymbol{X}_{i};\text{Proj}_{\mathcal{G}_{n}}\overline{g}(\widehat{g}_d,\epsilon_n)-g^*_d). \label{eq:L_projbar}
	\end{align}
	Substracting \eqref{eq:L_projbar} from \eqref{eq:L_projhat} yields:
	\begin{align*}
		&L_{d,n}(\widehat{g}_d)=L_{d,n}(\text{Proj}_{\mathcal{G}_{n}}\overline{g}(\widehat{g}_d,\epsilon_n))+\frac{1}{\sqrt{n}}\cdot  \mathbb{G}_n\left(\frac{\partial}{\partial g_d}\ell_d(D_{di},\boldsymbol{X}_{i};g^*_d)[ \widehat{g}_d-\text{Proj}_{\mathcal{G}_{n}}\overline{g}(\widehat{g}_d,\epsilon_n)]\right) \notag\\
		&+\frac{1}{n}\sum_{i=1}^n\bigg(r(D_{di},\boldsymbol{X}_{i}; \widehat{g}_d-g^*_d)-r(D_{di},\boldsymbol{X}_{i}; \text{Proj}_{\mathcal{G}_{n}}\overline{g}(\widehat{g}_d,\epsilon_n)-g^*_d)\bigg).
	\end{align*}
	We first claim the following result whose proof will be given in Subsection \ref{proof:eq:mu_r}:
	\begin{align}\label{eq:mu_r}
		&\frac{1}{n}\sum_{i=1}^n\left\{r(D_{di},\boldsymbol{X}_{i};\widehat{g}_d-g^*_d)-r(D_{di},\boldsymbol{X}_{i};\text{Proj}_{\mathcal{G}_n}\overline{g}(\widehat{g}_d,\epsilon_n)-g^*_d)\right\}\\
		\leq&\epsilon_n \cdot (1-\epsilon_n)\cdot   \mathbb{E}\left[\pi_d^*(\boldsymbol{X}_i)\{1-\pi_d^*(\boldsymbol{X}_i)\}(\widehat{g}_d(\boldsymbol{X}_{i})-g^*_d(\boldsymbol{X}_{i})) \cdot u^*(\boldsymbol{X}_{i})\right]+o_P\left(\frac{\epsilon_n}{\sqrt{n}}\right). \notag
	\end{align}
	By Assumption \ref{as:pihat_1st} and $L_{d,n}(\widehat{g}_d)\geq L_{d,n}(\text{Proj}_{\mathcal{G}_{n}}\overline{g}(\widehat{g}_d,\epsilon_n))-O(\epsilon^2_n)$,  we get
	\begin{align} \label{eq:Lpihat-Lpibar}
		-O(\epsilon_n^2)\leq & \frac{1}{\sqrt{n}}\cdot  \mathbb{G}_n\left(\frac{\partial}{\partial g_d}\ell_d(D_{di},\boldsymbol{X}_{i};g^*_d)[ \widehat{g}_d(\boldsymbol{X}_{i})-\text{Proj}_{\mathcal{G}_n}\overline{g}(\boldsymbol{X}_{i};\widehat{g}_d,\epsilon_n)]\right) \\
		&+\epsilon_n \cdot (1-\epsilon_n)\cdot   \mathbb{E}\left[\pi_d^*(\boldsymbol{X}_i)\{1-\pi_d^*(\boldsymbol{X}_i)\}(\widehat{g}_d(\boldsymbol{X}_{i})-g^*_d(\boldsymbol{X}_{i})) \cdot u^*(\boldsymbol{X}_{i})\right]+o_P\left(\frac{\epsilon_n}{\sqrt{n}}\right)  \notag\\
		= &\frac{1}{\sqrt{n}}\cdot  \mathbb{G}_n\left(\frac{\partial}{\partial g_d}\ell_d(D_{di},\boldsymbol{X}_{i};g^*_d)[ \widehat{g}_d(\boldsymbol{X}_{i})-\overline{g}(\boldsymbol{X}_{i};\widehat{g}_d,\epsilon_n)]\right) \notag\\
		&+\frac{1}{\sqrt{n}}\cdot  \mathbb{G}_n\left(\frac{\partial}{\partial g_d}\ell_d(D_{di},\boldsymbol{X}_{i};g^*_d)[\overline{g}(\boldsymbol{X}_{i};\widehat{g}_d,\epsilon_n)-\text{Proj}_{\mathcal{G}_n}\overline{g}(\boldsymbol{X}_{i};\widehat{g}_d,\epsilon_n)]\right)\notag\\
		&+\epsilon_n \cdot (1-\epsilon_n)\cdot   \mathbb{E}\left[\pi_d^*(\boldsymbol{X}_i)\{1-\pi_d^*(\boldsymbol{X}_i)\}(\widehat{g}_d(\boldsymbol{X}_{i})-g^*_d(\boldsymbol{X}_{i})) \cdot u^*(\boldsymbol{X}_{i})\right]+o_P\left(\frac{\epsilon_n}{\sqrt{n}}\right)   \notag\\
		=& \frac{1}{\sqrt{n}}\cdot  \mathbb{G}_n\left(\frac{\partial}{\partial g_d}\ell_d(D_{di},\boldsymbol{X}_{i};g^*_d)[ \widehat{g}_d(\boldsymbol{X}_{i})-\overline{g}(\boldsymbol{X}_{i};\widehat{g}_d,\epsilon_n)]\right) \notag\\
		&+\epsilon_n \cdot (1-\epsilon_n)\cdot   \mathbb{E}\left[\pi_d^*(\boldsymbol{X}_i)\{1-\pi_d^*(\boldsymbol{X}_i)\}(\widehat{g}_d(\boldsymbol{X}_{i})-g^*_d(\boldsymbol{X}_{i})) \cdot u^*(\boldsymbol{X}_{i})\right] \notag\\
		&+O_P(\epsilon_n^2)+o_P\left(\frac{\epsilon_n}{\sqrt{n}}\right) \notag \\
		=&\epsilon_n\cdot\frac{1}{\sqrt{n}}\cdot  \mathbb{G}_n\left(\frac{\partial}{\partial g_d}\ell_d(D_{di},\boldsymbol{X}_{i};g^*_d)[ \widehat{g}_d(\boldsymbol{X}_{i})-g_d^*(\boldsymbol{X}_i)]\right) \notag\\
		&-\epsilon_n\cdot\frac{1}{\sqrt{n}}\cdot  \mathbb{G}_n\left(\frac{\partial}{\partial g_d}\ell_d(D_{di},\boldsymbol{X}_{i};g^*_d)[u^*(\boldsymbol{X}_i)]\right) \notag\\
		&+\epsilon_n \cdot (1-\epsilon_n)\cdot   \mathbb{E}\left[\pi_d^*(\boldsymbol{X}_i)\{1-\pi_d^*(\boldsymbol{X}_i)\}(\widehat{g}_d(\boldsymbol{X}_{i})-g^*_d(\boldsymbol{X}_{i})) \cdot u^*(\boldsymbol{X}_{i})\right] \notag\\
		&+O_P(\epsilon_n^2)+o_P\left(\frac{\epsilon_n}{\sqrt{n}}\right)  \notag\\
		=&-\epsilon_n\cdot\frac{1}{\sqrt{n}}\cdot  \mathbb{G}_n\left(\frac{\partial}{\partial g_d}\ell_d(D_{di},\boldsymbol{X}_{i};g^*_d)[u^*(\boldsymbol{X}_i)]\right) \notag\\
		&+\epsilon_n \cdot (1-\epsilon_n)\cdot   \mathbb{E}\left[\pi_d^*(\boldsymbol{X}_i)\{1-\pi_d^*(\boldsymbol{X}_i)\}(\widehat{g}_d(\boldsymbol{X}_{i})-g^*_d(\boldsymbol{X}_{i})) \cdot u^*(\boldsymbol{X}_{i})\right] \notag\\
		&+O_P(\epsilon_n^2)+o_P\left(\frac{\epsilon_n}{\sqrt{n}}\right),  \notag
	\end{align}
	where the second equality holds by Assumption \ref{as:app_error}, and the last equality holds by Lemma \ref{lemma:several_results}, the fact $\|\widehat{g}_d-g_d^*\|_{\infty}=o_P(1)$, and Assumption \ref{as:parameter_beta} (ii). Hence,
	\begin{align*}
		&- (1-\epsilon_n) \cdot \mathbb{E}\left[  \pi^*_d(\boldsymbol{X}_{i})(1-\pi^*_d(\boldsymbol{X}_{i}))\cdot u^*(\boldsymbol{X}_{i}) \cdot (\widehat{g}_d(\boldsymbol{X}_{i})-g^*_d(\boldsymbol{X}_{i}))\right]\\
		&+\frac{1}{\sqrt{n}}  \cdot  \mathbb{G}_n\left(\frac{\partial}{\partial g_d}\ell_d(D_{di},\boldsymbol{X}_{i};g^*_d)[u^*(\boldsymbol{X}_{i})]\right)\leq O_P(\epsilon_n)+o_P\left(\frac{1}{\sqrt{n}}\right)=o_P\left(\frac{1}{\sqrt{n}}\right).
	\end{align*}
	With $u^*(\boldsymbol{X})$ being replaced by $-u^*(\boldsymbol{X})$, we can further obtain our desired result:
	\begin{align*}
		&\sqrt{n}\cdot \mathbb{E}\left[ \pi^*_d(\boldsymbol{X}_{i})(1-\pi^*_d(\boldsymbol{X}_{i}))\cdot u^*(\boldsymbol{X}_{i}) \cdot (\widehat{g}_d(\boldsymbol{X}_{i})-g^*_d(\boldsymbol{X}_{i}))\right]\notag\\
		=&\mathbb{G}_n\left(\frac{\partial}{\partial g_d}\ell_d(D_{di},\boldsymbol{X}_{i};g^*_d)[u^*(\boldsymbol{X}_{i})]\right)+o_P\left(1\right) \notag\\
		=&\frac{1}{\sqrt{n}}\sum_{i=1}^n\left\{D_{di}-\pi^*_d(\boldsymbol{X}_i)\right\} u^*(\boldsymbol{X}_{i}) +o_P\left(1\right),
	\end{align*}
	where $u^*(\boldsymbol{X}_{i})=\mathcal{E}_d(\boldsymbol{X}_{i},\beta_d^*)/\pi_d^*(\boldsymbol{X}_i)$.
	
\end{proof}

%\newpage

\subsection{Proof of \eqref{eq:mu_r}}\label{proof:eq:mu_r}
Note that
\begin{align*}
	&\frac{1}{n}\sum_{i=1}^n\left\{r(D_{di},\boldsymbol{X}_{i};\widehat{g}_d-g^*_d)-r(D_{di},\boldsymbol{X}_{i};\text{Proj}_{\mathcal{G}_n}\overline{g}(\widehat{g}_d,\epsilon_n)-g^*_d)\right\}\\
	=&\frac{1}{2n}\sum_{i=1}^n \pi_d^*(\boldsymbol{X}_i)\{1-\pi_d^*(\boldsymbol{X}_i)\}\left(\{\text{Proj}_{\mathcal{G}_n}\overline{g}(\widehat{g}_d,\epsilon_n)-g^*_d(\boldsymbol{X}_{i}) \}^2-\{\widehat{g}_d(\boldsymbol{X}_{i})-g^*_d(\boldsymbol{X}_{i}) \}^2\right)\\
	&+\frac{1}{n}\sum_{i=1}^n\left(R(D_{di},\boldsymbol{X}_{i};\widehat{g}_d)-R(D_{di},\boldsymbol{X}_{i};\text{Proj}_{\mathcal{G}_n}\overline{g}(\widehat{g}_d,\epsilon_n)\right)\\
	=&(I)+(II),
\end{align*}
where
\begin{align*}
	(I):=&\frac{1}{2n}\sum_{i=1}^n \pi_d^*(\boldsymbol{X}_i)\{1-\pi_d^*(\boldsymbol{X}_i)\}\left(\{\text{Proj}_{\mathcal{G}_n}\overline{g}(\widehat{g}_d,\epsilon_n)-g^*_d(\boldsymbol{X}_{i}) \}^2-\{\widehat{g}_d(\boldsymbol{X}_{i})-g^*_d(\boldsymbol{X}_{i}) \}^2\right),\\
	(II):=&\frac{1}{n}\sum_{i=1}^n\left(R(D_{di},\boldsymbol{X}_{i};\widehat{g}_d)-R(D_{di},\boldsymbol{X}_{i};\text{Proj}_{\mathcal{G}_n}\overline{g}(\widehat{g}_d,\epsilon_n)\right).
\end{align*}
Consider the  first item $(I)$. For the term $\{\text{Proj}_{\mathcal{G}_n}\overline{g}(\boldsymbol{X}_{i};\widehat{g}_d,\epsilon_n)-g^*_d(\boldsymbol{X}_{i})\}^2$, we have
\begin{align*}
	&\{\text{Proj}_{\mathcal{G}_n}\overline{g}(\boldsymbol{X}_{i};\widehat{g}_d,\epsilon_n)-g^*_d(\boldsymbol{X}_{i})\}^2\\
	=&\{\text{Proj}_{\mathcal{G}_n}\overline{g}(\boldsymbol{X}_{i};\widehat{g}_d,\epsilon_n)-\overline{g}(\boldsymbol{X}_{i};\widehat{g}_d,\epsilon_n)+\overline{g}(\boldsymbol{X}_{i};\widehat{g}_d,\epsilon_n)-g^*_d(\boldsymbol{X}_{i})\}^2\\
	=&\{\text{Proj}_{\mathcal{G}_n}\overline{g}(\boldsymbol{X}_{i};\widehat{g}_d,\epsilon_n)-\overline{g}(\boldsymbol{X}_{i};\widehat{g}_d,\epsilon_n)+(1-\epsilon_n)(\widehat{g}_d(\boldsymbol{X}_{i})-g^*_d(\boldsymbol{X}_{i}))+\epsilon_n\cdot u^*(\boldsymbol{X}_{i})\}^2\\
	=&\{\text{Proj}_{\mathcal{G}_n}\overline{g}(\boldsymbol{X}_{i};\widehat{g}_d,\epsilon_n)-\overline{g}(\boldsymbol{X}_{i};\widehat{g}_d,\epsilon_n)\}^2+(1-\epsilon_n)^2\cdot \{\widehat{g}_d(\boldsymbol{X}_{i})-g^*_d(\boldsymbol{X}_{i})\}^2+\epsilon^2_n\cdot |u^*(\boldsymbol{X}_{i})|^2\\
	&+2\cdot(1-\epsilon_n)\cdot (\text{Proj}_{\mathcal{G}_n}\overline{g}(\boldsymbol{X}_{i};\widehat{g}_d,\epsilon_n)-\overline{g}(\boldsymbol{X}_{i};\widehat{g}_d,\epsilon_n)) \cdot (\widehat{g}_d(\boldsymbol{X}_{i})-g^*_d(\boldsymbol{X}_{i})) \\
	&+2\cdot \epsilon_n \cdot (\text{Proj}_{\mathcal{G}_n}\overline{g}(\boldsymbol{X}_{i};\widehat{g}_d,\epsilon_n)-\overline{g}(\boldsymbol{X}_{i};\widehat{g}_d,\epsilon_n)) \cdot u^*(\boldsymbol{X}_{i}) \\
	&+2\cdot \epsilon_n \cdot (1-\epsilon_n)\cdot (\widehat{g}_d(\boldsymbol{X}_{i})-g^*_d(\boldsymbol{X}_{i})) \cdot u^*(\boldsymbol{X}_{i})\\
	\leq & 	\{\text{Proj}_{\mathcal{G}_n}\overline{g}(\boldsymbol{X}_{i};\widehat{g}_d,\epsilon_n)-\overline{g}(\boldsymbol{X}_{i};\widehat{g}_d,\epsilon_n)\}^2+(1-\epsilon_n)^2\cdot \{\widehat{g}_d(\boldsymbol{X}_{i})-g^*_d(\boldsymbol{X}_{i})\}^2+\epsilon^2_n\cdot |u^*(\boldsymbol{X}_{i})|^2\\
	&+2\cdot(1-\epsilon_n)\cdot |\text{Proj}_{\mathcal{G}_n}\overline{g}(\boldsymbol{X}_{i};\widehat{g}_d,\epsilon_n)-\overline{g}(\boldsymbol{X}_{i};\widehat{g}_d,\epsilon_n)| \cdot |\widehat{g}_d(\boldsymbol{X}_{i})-g^*_d(\boldsymbol{X}_{i})| \\
	&+2\cdot \epsilon_n \cdot |\text{Proj}_{\mathcal{G}_n}\overline{g}(\boldsymbol{X}_{i};\widehat{g}_d,\epsilon_n)-\overline{g}(\boldsymbol{X}_{i};\widehat{g}_d,\epsilon_n)|\cdot |u^*(\boldsymbol{X}_{i})| \\
	&+2\cdot \epsilon_n \cdot (1-\epsilon_n)\cdot (\widehat{g}_d(\boldsymbol{X}_{i})-g^*_d(\boldsymbol{X}_{i})) \cdot u^*(\boldsymbol{X}_{i})
\end{align*}
Then by Assumptions \ref{as:parameter_beta} (ii) and \ref{as:app_error}, and Theorem \ref{thm:rate_ANN}, we have
\begin{align*}
	&(I)\leq\frac{1}{2n}\sum_{i=1}^n \pi_d^*(\boldsymbol{X}_i)\{1-\pi_d^*(\boldsymbol{X}_i)\}\cdot \{\text{Proj}_{\mathcal{G}_n}\overline{g}(\boldsymbol{X}_{i};\widehat{g}_d,\epsilon_n)-\overline{g}(\boldsymbol{X}_{i};\widehat{g}_d,\epsilon_n)\}^2\\
	&+\{(1-\epsilon_n)^2-1\}\cdot \frac{1}{2n}\sum_{i=1}^n \pi_d^*(\boldsymbol{X}_i)\{1-\pi_d^*(\boldsymbol{X}_i)\}\{\widehat{g}_d(\boldsymbol{X}_{i})-g^*_d(\boldsymbol{X}_{i})\}^2\\
	&+\epsilon^2_n\cdot \frac{1}{2n}\sum_{i=1}^n \pi_d^*(\boldsymbol{X}_i)\{1-\pi_d^*(\boldsymbol{X}_i)\}\cdot|u^*(\boldsymbol{X}_{i})|^2\\
	&+(1-\epsilon_n)\cdot \frac{1}{n}\sum_{i=1}^n \pi_d^*(\boldsymbol{X}_i)\{1-\pi_d^*(\boldsymbol{X}_i)\}|\text{Proj}_{\mathcal{G}_n}\overline{g}(\boldsymbol{X}_{i};\widehat{g}_d,\epsilon_n)-\overline{g}(\boldsymbol{X}_{i};\widehat{g}_d,\epsilon_n)| \cdot |\widehat{g}_d(\boldsymbol{X}_{i})-g^*_d(\boldsymbol{X}_{i})| \\
	&+\epsilon_n \cdot \frac{1}{n}\sum_{i=1}^n \pi_d^*(\boldsymbol{X}_i)\{1-\pi_d^*(\boldsymbol{X}_i)\}\cdot|\text{Proj}_{\mathcal{G}_n}\overline{g}(\boldsymbol{X}_{i};\widehat{g}_d,\epsilon_n)-\overline{g}(\boldsymbol{X}_{i};\widehat{g}_d,\epsilon_n)|\cdot |u^*(\boldsymbol{X}_{i})| \\
	&+ \epsilon_n \cdot (1-\epsilon_n)\cdot  \frac{1}{n}\sum_{i=1}^n \pi_d^*(\boldsymbol{X}_i)\{1-\pi_d^*(\boldsymbol{X}_i)\}(\widehat{g}_d(\boldsymbol{X}_{i})-g^*_d(\boldsymbol{X}_{i})) \cdot u^*(\boldsymbol{X}_{i})\\
	= & \frac{1}{2}\cdot \mathbb{E} \left[\pi_d^*(\boldsymbol{X}_i)\{1-\pi_d^*(\boldsymbol{X}_i)\}\{\text{Proj}_{\mathcal{G}_n}\overline{g}(\boldsymbol{X}_{i};\widehat{g}_d,\epsilon_n)-\overline{g}(\boldsymbol{X}_{i};\widehat{g}_d,\epsilon_n)\}^2\right]\{1+o_P(1)\}\\
	&+\{-2\epsilon_n+\epsilon_n^2\}\cdot \frac{1}{2}\cdot \mathbb{E} \left[\pi_d^*(\boldsymbol{X}_i)\{1-\pi_d^*(\boldsymbol{X}_i)\}\{\widehat{g}_d(\boldsymbol{X}_{i})-g^*_d(\boldsymbol{X}_{i})\}^2\right]\{1+o_P(1)\}\\
	&+\epsilon^2_n\cdot \frac{1}{2}\cdot \mathbb{E} \left[\pi_d^*(\boldsymbol{X}_i)\{1-\pi_d^*(\boldsymbol{X}_i)\}\cdot|u^*(\boldsymbol{X}_{i})|^2\right]\{1+o_P(1)\}\\
	&+(1-\epsilon_n)\cdot \mathbb{E} \left[ \pi_d^*(\boldsymbol{X}_i)\{1-\pi_d^*(\boldsymbol{X}_i)\}\cdot|\text{Proj}_{\mathcal{G}_n}\overline{g}(\boldsymbol{X}_{i};\widehat{g}_d,\epsilon_n)-\overline{g}(\boldsymbol{X}_{i};\widehat{g}_d,\epsilon_n)| \cdot |\widehat{g}_d(\boldsymbol{X}_{i})-g^*_d(\boldsymbol{X}_{i})| \right]\\
	&\quad\times \{1+o_P(1)\} \\
	&+\epsilon_n \cdot \mathbb{E} \left[\pi_d^*(\boldsymbol{X}_i)\{1-\pi_d^*(\boldsymbol{X}_i)\}\cdot|\text{Proj}_{\mathcal{G}_n}\overline{g}(\boldsymbol{X}_{i};\widehat{g}_d,\epsilon_n)-\overline{g}(\boldsymbol{X}_{i};\widehat{g}_d,\epsilon_n)|\cdot |u^*(\boldsymbol{X}_{i})|\right] \{1+o_P(1)\}\\
	&+ \epsilon_n \cdot (1-\epsilon_n)\cdot  \frac{1}{\sqrt{n}}\mathbb{G}_n\left( \pi_d^*(\boldsymbol{X}_i)\{1-\pi_d^*(\boldsymbol{X}_i)\}(\widehat{g}_d(\boldsymbol{X}_{i})-g^*_d(\boldsymbol{X}_{i})) \cdot u^*(\boldsymbol{X}_{i})\right)\\
	&+ \epsilon_n \cdot (1-\epsilon_n)\cdot \sqrt{n}\cdot  \mathbb{E}\left[\pi_d^*(\boldsymbol{X}_i)\{1-\pi_d^*(\boldsymbol{X}_i)\}(\widehat{g}_d(\boldsymbol{X}_{i})-g^*_d(\boldsymbol{X}_{i})) \cdot u^*(\boldsymbol{X}_{i})\right]\\
	\leq& O_P(\epsilon_n^2\cdot \delta_n^2)+O_P(\epsilon_n^2\cdot \delta_n^2)+O_P(\epsilon_n^2)+O_P(\epsilon_n\cdot \delta_n^2)+O_P(\epsilon_n^2\cdot\delta_n)+o_P\left(\frac{\epsilon_n}{\sqrt{n}}\right)\\
	&+\epsilon_n \cdot (1-\epsilon_n)\cdot  \mathbb{E}\left[\pi_d^*(\boldsymbol{X}_i)\{1-\pi_d^*(\boldsymbol{X}_i)\}(\widehat{g}_d(\boldsymbol{X}_{i})-g^*_d(\boldsymbol{X}_{i})) \cdot u^*(\boldsymbol{X}_{i})\right],
\end{align*}
where the last inequality holds in light of the following result:
\begin{align*}
	\mathbb{G}_n\left( \pi_d^*(\boldsymbol{X}_i)\{1-\pi_d^*(\boldsymbol{X}_i)\}(\widehat{g}_d(\boldsymbol{X}_{i})-g^*_d(\boldsymbol{X}_{i})) \cdot u^*(\boldsymbol{X}_{i})\right)=o_P(1),
\end{align*}
whose proof is similar to that for Lemma \ref{lemma:several_results}. Therefore, we have
\begin{align}
	(I)\leq\epsilon_n \cdot (1-\epsilon_n)\cdot   \mathbb{E}\left[\pi_d^*(\boldsymbol{X}_i)\{1-\pi_d^*(\boldsymbol{X}_i)\}(\widehat{g}_d(\boldsymbol{X}_{i})-g^*_d(\boldsymbol{X}_{i})) \cdot u^*(\boldsymbol{X}_{i})\right]+o_P\left(\frac{\epsilon_n}{\sqrt{n}}\right).\label{eq:I_estimate}
\end{align}
For the term $(II)$, we have
\begin{align}
	(II)=&\frac{1}{n}\sum_{i=1}^n\left(R(D_{di},\boldsymbol{X}_{i};\widehat{g}_d)-R(D_{di},\boldsymbol{X}_{i};\text{Proj}_{\mathcal{G}_n}\overline{g}(\widehat{g}_d,\epsilon_n)\right) \notag	\\
	=&-\frac{1}{n}\sum_{i=1}^n\int_{g_d^*(\boldsymbol{X}_i)}^{\widehat{g}_d(\boldsymbol{X}_i)}L(t)\{1-L(t)\}\{1-2L(t)\}(\widehat{g}_d(\boldsymbol{X}_i)-t)^2dt  \notag\\
	&+\frac{1}{n}\sum_{i=1}^n\int_{g_d^*(\boldsymbol{X}_i)}^{\text{Proj}_{\mathcal{G}_n}\overline{g}(\widehat{g}_d,\epsilon_n)}L(t)\{1-L(t)\}\{1-2L(t)\}(\text{Proj}_{\mathcal{G}_n}\overline{g}(\widehat{g}_d,\epsilon_n)-t)^2dt \notag\\
	=&-\frac{1}{n}\sum_{i=1}^n\int_{g_d^*(\boldsymbol{X}_i)}^{\widehat{g}_d(\boldsymbol{X}_i)}L(t)\{1-L(t)\}\{1-2L(t)\}\cdot\bigg[(\widehat{g}_d(\boldsymbol{X}_i)-t)^2-(\text{Proj}_{\mathcal{G}_n}\overline{g}(\widehat{g}_d,\epsilon_n)-t)^2\bigg]dt \label{eq:II-1}\\
	&+\frac{1}{n}\sum_{i=1}^n\int_{\widehat{g}_d(\boldsymbol{X}_i)}^{\text{Proj}_{\mathcal{G}_n}\overline{g}(\widehat{g}_d,\epsilon_n)}L(t)\{1-L(t)\}\{1-2L(t)\}(\text{Proj}_{\mathcal{G}_n}\overline{g}(\widehat{g}_d,\epsilon_n)-t)^2dt.\label{eq:II-2}
\end{align}
For the term \eqref{eq:II-1}, we have
\begin{align*}
	&|\eqref{eq:II-1}|=\left|\frac{1}{n}\sum_{i=1}^n\int_{g_d^*(\boldsymbol{X}_i)}^{\widehat{g}_d(\boldsymbol{X}_i)}L(t)\{1-L(t)\}\{1-2L(t)\}\cdot\bigg[(\widehat{g}_d(\boldsymbol{X}_i)-t)^2-(\text{Proj}_{\mathcal{G}_n}\overline{g}(\widehat{g}_d,\epsilon_n)-t)^2\bigg]dt\right|\\
	=&\bigg|\frac{1}{n}\sum_{i=1}^n\int_{g_d^*(\boldsymbol{X}_i)}^{\widehat{g}_d(\boldsymbol{X}_i)}L(t)\{1-L(t)\}\{1-2L(t)\}\cdot\bigg[\widehat{g}_d(\boldsymbol{X}_i)-\text{Proj}_{\mathcal{G}_n}\overline{g}(\widehat{g}_d,\epsilon_n)\bigg]\\
	&\times \bigg[\widehat{g}_d(\boldsymbol{X}_i)+\text{Proj}_{\mathcal{G}_n}\overline{g}(\widehat{g}_d,\epsilon_n)-2t\bigg]dt\bigg|\\
	\leq & \frac{1}{n}\sum_{i=1}^n\int_{g_d^*(\boldsymbol{X}_i)}^{\widehat{g}_d(\boldsymbol{X}_i)}\bigg|\widehat{g}_d(\boldsymbol{X}_i)-\text{Proj}_{\mathcal{G}_n}\overline{g}(\widehat{g}_d,\epsilon_n)\bigg|\cdot \bigg|\widehat{g}_d(\boldsymbol{X}_i)+\text{Proj}_{\mathcal{G}_n}\overline{g}(\widehat{g}_d,\epsilon_n)-2t\bigg|dt\\
	\leq & \frac{1}{n}\sum_{i=1}^n|\widehat{g}_d(\boldsymbol{X}_i)-\text{Proj}_{\mathcal{G}_n}\overline{g}(\widehat{g}_d,\epsilon_n)|\cdot|\widehat{g}_d(\boldsymbol{X}_i)-g_d^*(\boldsymbol{X}_i)| \\
	&\qquad \cdot\left|\widehat{g}_d(\boldsymbol{X}_i)+\text{Proj}_{\mathcal{G}_n}\overline{g}(\widehat{g}_d,\epsilon_n)-2(s_i\cdot\widehat{g}_d(\boldsymbol{X}_i)+(1-s_i)g_d^*(\boldsymbol{X}_i))\right|\quad (\text{for some} \ s_i\in (0,1)) \\
	=&	  \frac{1}{n}\sum_{i=1}^n|\widehat{g}_d(\boldsymbol{X}_i)-\text{Proj}_{\mathcal{G}_n}\overline{g}(\widehat{g}_d,\epsilon_n)|\cdot|\widehat{g}_d(\boldsymbol{X}_i)-g_d^*(\boldsymbol{X}_i)| \\
	&\qquad\qquad \cdot\left|\text{Proj}_{\mathcal{G}_n}\overline{g}(\widehat{g}_d,\epsilon_n)-\widehat{g}_d(\boldsymbol{X}_i)+2(1-s_i)(\widehat{g}_d(\boldsymbol{X}_i)-g_d^*(\boldsymbol{X}_i))\right|\\
	\leq &		  \frac{1}{n}\sum_{i=1}^n\left\{|\text{Proj}_{\mathcal{G}_n}\overline{g}(\widehat{g}_d,\epsilon_n)-\overline{g}(\widehat{g}_d,\epsilon_n)|+|\overline{g}(\widehat{g}_d,\epsilon_n)-\widehat{g}_d(\boldsymbol{X}_i)|\right\}\cdot|\widehat{g}_d(\boldsymbol{X}_i)-g_d^*(\boldsymbol{X}_i)| \\
	&\qquad\qquad \cdot\left\{|\text{Proj}_{\mathcal{G}_n}\overline{g}(\widehat{g}_d,\epsilon_n)-\overline{g}(\widehat{g}_d,\epsilon_n)|+|\overline{g}(\widehat{g}_d,\epsilon_n)-\widehat{g}_d(\boldsymbol{X}_i)|+2\cdot|\widehat{g}_d(\boldsymbol{X}_i)-g_d^*(\boldsymbol{X}_i)|\right\}\\
	=&  \frac{1}{n}\sum_{i=1}^n|\text{Proj}_{\mathcal{G}_n}\overline{g}(\widehat{g}_d,\epsilon_n)-\overline{g}(\widehat{g}_d,\epsilon_n)|^2 \cdot|\widehat{g}_d(\boldsymbol{X}_i)-g_d^*(\boldsymbol{X}_i)| \\
	&+ \frac{1}{n}\sum_{i=1}^n|\text{Proj}_{\mathcal{G}_n}\overline{g}(\widehat{g}_d,\epsilon_n)-\overline{g}(\widehat{g}_d,\epsilon_n)| \cdot|\widehat{g}_d(\boldsymbol{X}_i)-g_d^*(\boldsymbol{X}_i)|\cdot |\overline{g}(\widehat{g}_d,\epsilon_n)-\widehat{g}_d(\boldsymbol{X}_i)|\\
	&+  \frac{2}{n}\sum_{i=1}^n|\text{Proj}_{\mathcal{G}_n}\overline{g}(\widehat{g}_d,\epsilon_n)-\overline{g}(\widehat{g}_d,\epsilon_n)| \cdot|\widehat{g}_d(\boldsymbol{X}_i)-g_d^*(\boldsymbol{X}_i)|^2\\
	&+  \frac{1}{n}\sum_{i=1}^n|\overline{g}(\widehat{g}_d,\epsilon_n)-\widehat{g}_d(\boldsymbol{X}_i)|\cdot|\widehat{g}_d(\boldsymbol{X}_i)-g_d^*(\boldsymbol{X}_i)|\cdot |\text{Proj}_{\mathcal{G}_n}\overline{g}(\widehat{g}_d,\epsilon_n)-\overline{g}(\widehat{g}_d,\epsilon_n)|\\
	&+ \frac{1}{n}\sum_{i=1}^n|\overline{g}(\widehat{g}_d,\epsilon_n)-\widehat{g}_d(\boldsymbol{X}_i)|^2\cdot|\widehat{g}_d(\boldsymbol{X}_i)-g_d^*(\boldsymbol{X}_i)|\\
	&+ \frac{2}{n}\sum_{i=1}^n|\overline{g}(\widehat{g}_d,\epsilon_n)-\widehat{g}_d(\boldsymbol{X}_i)|\cdot|\widehat{g}_d(\boldsymbol{X}_i)-g_d^*(\boldsymbol{X}_i)|^2\\
	\leq & \sup_{\{{g}_d\in\mathcal{G}_n:\|{g}_d-g_d^*\|_{L^2(dF_X)}\leq \delta_n\}}\|\text{Proj}_{\mathcal{G}_n}\overline{g}({g}_d,\epsilon_n)-\overline{g}({g}_d,\epsilon_n)\|^2_{n}\cdot  \|\widehat{g}_d-g_d^*\|_{\infty} \\
	&+  \sup_{\{{g}_d\in\mathcal{G}_n:\|{g}_d-g_d^*\|_{L^2(dF_X)}\leq \delta_n\}}\|\text{Proj}_{\mathcal{G}_n}\overline{g}({g}_d,\epsilon_n)-\overline{g}({g}_d,\epsilon_n)\|_{n}\cdot \|\widehat{g}_d-g_d^*\|_{n} \cdot O_P(\epsilon_n)\\
	&+ \sup_{\{{g}_d\in\mathcal{G}_n:\|{g}_d-g_d^*\|_{L^2(dF_X)}\leq \delta_n\}}\|\text{Proj}_{\mathcal{G}_n}\overline{g}({g}_d,\epsilon_n)-\overline{g}({g}_d,\epsilon_n)\|_{n}\cdot  \|\widehat{g}_d-g_d^*\|_{n}\cdot \|\widehat{g}_d-g_d^*\|_{\infty}  \\
	&+ O_P(\epsilon_n)\cdot \sup_{\{{g}_d\in\mathcal{G}_n:\|{g}_d-g_d^*\|_{L^2(dF_X)}\leq \delta_n\}}\|\text{Proj}_{\mathcal{G}_n}\overline{g}({g}_d,\epsilon_n)-\overline{g}({g}_d,\epsilon_n)\|_{n}\cdot\|\widehat{g}_d-g_d^*\|_{n} \\
	&+O_P(\epsilon_n^2) \cdot \|\widehat{g}_d-g_d^*\|_{n}\\
	&+ O_P(\epsilon_n) \cdot \|\widehat{g}_d-g_d^*\|^2_{n} \\
	\leq & o_P\left(\frac{\epsilon_n^4}{\delta_n^2}\right)+O_P\left(\epsilon_n^3\right)+o_P\left(\epsilon_n^2\right)+O_P\left(\epsilon_n^3\right)+O_P(\epsilon_n^2\cdot \delta_n)+O_P(\epsilon_n\cdot \delta_n^2)=o_P\left(\frac{\epsilon_n}{\sqrt{n}}\right),
\end{align*}
where the first inequality holds by Assumption \ref{as:parameter_beta} and the result $\|\widehat{g}_d-g_d^*\|_{\infty}=o_P(1)$, the second inequality holds by the mean value theorem, and the last inequality holds by Cauchy-Schwarz inequality, Assumption \ref{as:app_error} and Theorem \ref{thm:rate_ANN}.

For the term $\eqref{eq:II-2}$, we have
\begin{align*}
	|\eqref{eq:II-2}|=&\left|\frac{1}{2n}\sum_{i=1}^n\int_{\widehat{g}_d(\boldsymbol{X}_i)}^{\text{Proj}_{\mathcal{G}_n}\overline{g}(\widehat{g}_d,\epsilon_n)}L(t)\{1-L(t)\}\{1-2L(t)\}(\text{Proj}_{\mathcal{G}_n}\overline{g}(\widehat{g}_d,\epsilon_n)-t)^2dt\right|\\
	\leq&   \left|\frac{1}{n}\sum_{i=1}^n\int_{\widehat{g}_d(\boldsymbol{X}_i)}^{\text{Proj}_{\mathcal{G}_n}\overline{g}(\widehat{g}_d,\epsilon_n)}(\text{Proj}_{\mathcal{G}_n}\overline{g}(\widehat{g}_d,\epsilon_n)-t)^2dt\right|\\
	\leq &   \frac{1}{3n}\sum_{i=1}^n|\text{Proj}_{\mathcal{G}_n}\overline{g}(\widehat{g}_d,\epsilon_n)-\widehat{g}_d(\boldsymbol{X}_i)|^3\\
	\leq &  \frac{1}{3n}\sum_{i=1}^n\left\{|\text{Proj}_{\mathcal{G}_n}\overline{g}(\widehat{g}_d,\epsilon_n)-\overline{g}(\widehat{g}_d,\epsilon_n)|^3+|\overline{g}(\widehat{g}_d,\epsilon_n)-\widehat{g}_d(\boldsymbol{X}_i)|^3\right\}\\
	\leq &o_P(1)\cdot O_P\left(\frac{\epsilon_n^4}{\delta_n^2}\right) +O_P(\epsilon_n^3)=o_P\left(\frac{\epsilon_n^4}{\delta_n^2}\right)+O_P(\epsilon_n^3)=o_P\left(\frac{\epsilon_n}{\sqrt{n}}\right),
\end{align*}
where the first inequality holds by  Assumption \ref{as:parameter_beta} and the result $\|\widehat{g}_d-g_d^*\|_{\infty}=o_P(1)$ and $\|\text{Proj}_{\mathcal{G}_n}\overline{g}(\widehat{g}_d,\epsilon_n)-g_d^*\|_{\infty}=o_P(1)$, and the second equality holds by Assumption \ref{as:app_error}. Combining the results for \eqref{eq:II-1} and \eqref{eq:II-2}, we have
$$(II)=o_P\left(\frac{\epsilon_n}{\sqrt{n}}\right).$$
Then together with \eqref{eq:I_estimate}, we can conclude the result \eqref{eq:mu_r}.

%\newpage
\section{Proof of Theorem \ref{thm:eff_IPW}}\label{app:thm:eff_IPW}	
We first prove $\widehat{\beta}_d\xrightarrow{p}\beta_d^*$ for every $d\in\{0,1,...,J\}$. Because $\widehat{\beta}_d$  (resp. ${\beta}_d^*$) is a minimizer of $n^{-1}\sum_{i=1}^nD_{di}\mathcal{L}\left(Y_i-\beta_d\right)/\widehat{\pi}_d(\boldsymbol{X}_{i})$  (resp. $\mathbb{E}\left[D_{di}\mathcal{L}\left(Y_i-\beta\right)/\pi^*_d(\boldsymbol{X}_{i})\right]$), from the theory of $M$-estimation \cite[Theorem 5.7]{van1998asymptotic}, if the following uniform convergence holds:
\begin{align}
	\sup_{{\beta}\in\Theta}\left|\frac{1}{n}\sum_{i=1}^n\frac{D_{di}}{\widehat{\pi}_d(\boldsymbol{X}_{i})}\mathcal{L}\left(Y_i-\beta\right)-\mathbb{E}\left[\frac{D_{di}}{{\pi}^*_d(\boldsymbol{X}_{i})}\mathcal{L}\left(Y_i-\beta\right)\right] \right|\xrightarrow{p}0.\notag
\end{align}
then $\widehat{{\beta}}_d\xrightarrow{p}{\beta}^*_d$. We start to verify above condition. Using the triangular inequality, we have
\begin{align}
	&\sup_{{\beta}\in\Theta}\left|\frac{1}{n}\sum_{i=1}^n\frac{D_{di}}{\widehat{\pi}_d(\boldsymbol{X}_{i})}\mathcal{L}\left(Y_i-\beta\right)-\mathbb{E}\left[\frac{D_{di}}{{\pi}^*_d(\boldsymbol{X}_{i})}\mathcal{L}\left(Y_i-\beta\right)\right]  \right|\notag\\
	\leq &\sup_{{\beta}\in\Theta}\left|\frac{1}{n}\sum_{i=1}^n\frac{D_{di}}{\widehat{\pi}_d(\boldsymbol{X}_{i})\pi_d^*(\boldsymbol{X}_{i})}\left\{\widehat{\pi}_d(\boldsymbol{X}_{i})-\pi_d^*(\boldsymbol{X}_{i})\right\}\mathcal{L}\left(Y_i-\beta\right)\right| \label{eq:pihat-pi_L} \\
	&+\sup_{{\beta}\in\Theta}\left|\frac{1}{N}\sum_{i=1}^N\frac{D_{di}}{{\pi}^*_d(\boldsymbol{X}_{i})}\mathcal{L}\left(Y_i-\beta\right)-\mathbb{E}\left[\frac{D_{di}}{{\pi}^*_d(\boldsymbol{X}_{i})}\mathcal{L}\left(Y-\beta\right)\right] \right|.  \label{eq:pi0-Epi0_L}
\end{align}
We first show \eqref{eq:pihat-pi_L} is of $o_P(1)$. By Theorem \ref{thm:rate_ANN},  Cauchy-Scharwz' inequality, and  Assumptions \ref{as:parameter_beta} and \ref{as:L2_continuous}, we have
\begin{align*}
	|\eqref{eq:pihat-pi_L}|\leq&O_p(1)\cdot \left\{\frac{1}{n}\sum_{i=1}^n\left\{\widehat{\pi}_d(\boldsymbol{X}_{i})-\pi_d^*(\boldsymbol{X}_{i})\right\}^2\right\}^{1/2} \cdot \sup_{{\beta}\in\Theta} \left\{  \frac{1}{n}\sum_{i=1}^n\mathcal{L}\left(Y_i-\beta\right)^2\right\}^{1/2}\\
	\leq &o_P(1)\cdot \left\{\sup_{\beta\in\Theta} \mathbb{E}\left[\mathcal{L}\left(Y-\beta\right)^2\right] +o_P(1)\right\}^{1/2}=o_P(1),
\end{align*}
where the first inequality holds because $\widehat{\pi}_d(x)$ is uniformly bounded away from zero with probability approaching to one since $\widehat{\pi}_d(x)\xrightarrow{p} \pi_d^*(x)$ uniformly in $x\in\mathcal{X}$ and $0<\underline{c}\leq \pi_d^*(x)$ by Assumption \ref{as:parameter_beta}. To show \eqref{eq:pi0-Epi0_L} is of $o_P(1)$, it is sufficient to verify the following conditions holds true:
\begin{enumerate}
	\item $\Theta$ is compact;
	\item $\mathcal{L}'(Y_i-\beta)$ is continuous in $\beta$ with probability one;
	\item $\mathbb{E}\left[\sup_{\beta\in \Theta}|\mathcal{L}(Y-\beta)|\right]<\infty$;
\end{enumerate}
which are imposed in  Assumptions \ref{as:parameter_beta} and \ref{as:L2_continuous}. Hence, $\widehat{\beta}_d\xrightarrow{p}\beta_d^*$ holds.

Next, we establish the asymptotic normality for $\sqrt{n}\{\widehat{\beta}_d-\beta_d^*\}$. Since the loss function $\mathcal{L}(\cdot)$ may not be smooth (e.g. $\mathcal{L}(v)=v\{\tau-\mathds{1}(v\leq 0)\}$ in quantile regression),  the Delta method for deriving the large sample property is not applicable in our case. To circumvent this problem, we apply the \emph{nearness of arg mins} argument.  Define
\begin{align}\label{def:Gd}
	{G}_{d,n}(\beta,\widehat{\pi}_d):=\frac{1}{n}\sum_{i=1}^n\frac{D_{di}}{\widehat{\pi}_d(\boldsymbol{X}_{i})}\mathcal{L}(Y_i-\beta).
\end{align}
By definition
\begin{align}\label{def:beta_d}
	\widehat{\beta}_d=\arg\min_{\beta\in\Theta} G_{d,n}(\beta,\widehat{\pi}_d)=\arg\min_{\beta\in\Theta}\frac{1}{n}\sum_{i=1}^n\frac{D_{di}}{\widehat{\pi}_d(\boldsymbol{X}_{i})}\mathcal{L}(Y_i-\beta),
\end{align}
then
\begin{align*}
	\widehat{\beta}_d=&\arg\min_{\beta\in\Theta} n\left\{G_{d,n}(\beta,\widehat{\pi}_d)-G_{d,n}(\beta_d^*,\widehat{\pi}_d) \right\}=\arg\min_{\beta\in\Theta}\sum_{i=1}^n\frac{D_{di}}{\widehat{\pi}_d(\boldsymbol{X}_{i})}\left\{\mathcal{L}(Y_i-\beta)-\mathcal{L}(Y_i-\beta_d^*)\right\}\\
	=&\arg\min_{\beta\in\Theta}\sum_{i=1}^n\frac{D_{di}}{\widehat{\pi}_d(\boldsymbol{X}_{i})}\left[-\mathcal{L}'(Y_i-\beta_d^*)(\beta-\beta_d^*)+\left\{\mathcal{L}(Y_i-\beta)-\mathcal{L}(Y_i-\beta_d^*)+\mathcal{L}'(Y_i-\beta_d^*)(\beta-\beta_d^*)\right\}\right].
\end{align*}
By using change of variables and defining the following functions:
\begin{align*}
	&\widehat{u}_d:=\sqrt{n}(\widehat{\beta}_d-{\beta}_d^*), \  u:=\sqrt{n}(\beta-{\beta}_d^*),\\
	&R_d(Y_i,u):=\mathcal{L}\left(Y_i-\left\{\beta_d^*+\frac{u}{\sqrt{n}}\right\}\right)-\mathcal{L}(Y_i-\beta_d^*)+\mathcal{L}'(Y_i-\beta_d^*)\cdot \frac{u}{\sqrt{n}},\\
	&Q_{d,n}(u,\widehat{\pi}_d):=\sum_{i=1}^n\frac{D_{di}}{\widehat{\pi}_d(\boldsymbol{X}_{i})}\left[-\mathcal{L}'(Y_i-\beta_d^*)\cdot \frac{u}{\sqrt{n}}+R_d(Y_i,u)\right]=n\cdot \left[{G}_{d,n}(\beta,\widehat{\pi}_d)-{G}_{d,n}(\beta_d^*,\widehat{\pi}_d)\right].
\end{align*}
Then we get
\begin{align*}
	\widehat{u}_d=\arg\min_{u} Q_{d,n}(u,\widehat{\pi}_d).
\end{align*}
Next, we define the following quadratic function
\begin{align*}
	\widetilde{Q}_{d,n}(u):=\frac{u}{\sqrt{n}}\sum_{i=1}^n\left[-\frac{D_{di}}{{\pi}^*_d(\boldsymbol{X}_{i})}\mathcal{L}'(Y_i-\beta_d^*)+\left(\frac{D_{di}}{\pi_d^*(\boldsymbol{X}_{i})}-1\right)\mathcal{E}_{d}(\boldsymbol{X}_{i},\beta_d^*) \right]-\partial_{\beta_d}\mathbb{E}[\mathcal{L}'(Y^*_i(d)-\beta_d^*)]\cdot \frac{u^2}{2},
\end{align*}
which does not depend on $\widehat{\pi}_d$, and its minimizer is defined by
$$\widetilde{u}_{d}:=\arg\min_{u} \widetilde{Q}_{d,n}(u).$$
Since $\widetilde{Q}_{d,n}(u)$ is strictly convex and $\partial_{\beta}\mathbb{E}[\mathcal{L}'(Y^*_i(d)-\beta_d^*)]<0$, then the minimizer $\widetilde{u}_d$ is equal to
\begin{align*}
	\widetilde{u}_{d}=\frac{1}{\sqrt{n}}\sum_{i=1}^nH_d^{-1}\cdot S_d(Y_i,D_{di},\boldsymbol{X}_{i};\beta_d^*),
\end{align*}
where
\begin{align*}
	S_d(Y_i,D_{di},\boldsymbol{X}_{i};\beta_d^*):= \frac{D_{di}}{{\pi}^*_d(\boldsymbol{X}_{i})}\mathcal{L}'(Y_i-\beta_d^*)-\left(\frac{D_{di}}{\pi_d^*(\boldsymbol{X}_{i})}-1\right)\mathcal{E}_{d}(\boldsymbol{X}_{i},\beta_d^*)
\end{align*}
is the influence function of $\beta_d^*$ and $H_d:=-\partial_{\beta_d}\mathbb{E}[\mathcal{L}'(Y^*_i(d)-\beta_d^*)]$.

We complete the proof via the following steps:
\begin{itemize}
	\item Step I: showing $\xi_{d,n}(u,\widehat{\pi}_d):=\widetilde{Q}_{d,n}(u)-Q_{d,n}(u,\widehat{\pi}_d)=o_p(1)$ for every fixed $u$;
	\item Step II: showing $|\widehat{u}_{d}-\widetilde{u}_{d}|=o_P(1)$;
	\item Step III: obtaining the desired result $\sqrt{n}\{\widehat{\beta}_d-\beta_d^*\}=\widetilde{u}_{d}+\{\widehat{u}_{d}-\widetilde{u}_{d}\}=n^{-1/2}\sum_{i=1}^nH_d^{-1}\cdot S_d(Y_i,D_{di},\boldsymbol{X}_{i};\beta_d^*)+o_P(1)$.
\end{itemize}
We begin to prove Step I by showing that $\widetilde{Q}_{d,n}(u)$ is a quadratic approximation to the objective function $Q_{d,n}(u,\widehat{\pi}_d)$. We write the absolute value of the difference as follows:
\begin{align*}
	&\left|\widetilde{Q}_{d,n}(u)-Q_{d,n}(u,\widehat{\pi}_d)\right|\\
	=&\bigg|\frac{u}{\sqrt{n}}\sum_{i=1}^n\left[\frac{D_{di}}{\widehat{\pi}_d(\boldsymbol{X}_{i})}\mathcal{L}'(Y_i-\beta_d^*)-\frac{D_{di}}{{\pi}^*_d(\boldsymbol{X}_{i})}\mathcal{L}'(Y_i-\beta_d^*)+\left(\frac{D_{di}}{\pi_d^*(\boldsymbol{X}_{i})}-1\right)\mathcal{E}_{d}(\boldsymbol{X}_{i},\beta_d^*) \right]\\
	&-\sum_{i=1}^n\frac{D_{di}}{\widehat{\pi}_d(\boldsymbol{X}_{i})}R_d(Y_i,u)+\partial_{\beta_d}\mathbb{E}[\mathcal{L}'(Y^*_i(d)-\beta_d^*)]\cdot \frac{u^2}{2}\bigg|\\
	\leq &\left|\xi_{1,d,n}(u,\widehat{\pi}_d)\right|+\left|\xi_{2,d,n}(u,\widehat{\pi}_d)\right|,
\end{align*}
where
\begin{align*}
	&	\xi_{1,d,n}(u,\widehat{\pi}_d):=\frac{u}{\sqrt{n}}\sum_{i=1}^n\left[\frac{D_{di}}{\widehat{\pi}_d(\boldsymbol{X}_{i})}\mathcal{L}'(Y_i-\beta_d^*)-\frac{D_{di}}{{\pi}^*_d(\boldsymbol{X}_{i})}\mathcal{L}'(Y_i-\beta_d^*)+\left(\frac{D_{di}}{\pi_d^*(\boldsymbol{X}_{i})}-1\right)\mathcal{E}_{d}(\boldsymbol{X}_{i},\beta_d^*)\right],\\
	&\xi_{2,d,n}(u,\widehat{\pi}_d):=	\sum_{i=1}^n\frac{D_{di}}{\widehat{\pi}_d(\boldsymbol{X}_{i})}R_d(Y_i,u)-\partial_{\beta_d}\mathbb{E}[\mathcal{L}'(Y^*_i(d)-\beta_d^*)]\cdot \frac{u^2}{2}.
\end{align*}
Then Step I holds if we prove that for every $u$,
\begin{align}
	&\xi_{1,d,n}(u,\widehat{\pi}_d)=o_P(1),\label{eq:xi_1} \\
	&\xi_{2,d,n}(u,\widehat{\pi}_d)=o_P(1).\label{eq:xi_2}
\end{align}
We begin to establish \eqref{eq:xi_1}. By Taylor's expansion, we have
\begin{align}
	\xi_{1,d,n}(u,\widehat{\pi}_d)
	=&-\frac{u}{\sqrt{n}}\sum_{i=1}^n\frac{D_{di}}{\{{\pi}^*_d(\boldsymbol{X}_{i})\}^2}\mathcal{L}'\left(Y_i-{\beta}_d^*\right)\{\widehat{\pi}_d(\boldsymbol{X}_{i})-\pi_d^*(\boldsymbol{X}_{i}) \} \notag\\
	&+\frac{u}{\sqrt{n}}\sum_{i=1}^n\frac{D_{di}}{\{\widetilde{\pi}_d(\boldsymbol{X}_{i})\}^3}\mathcal{L}'\left(Y_i-{\beta}_d^*\right)\{\widehat{\pi}_d(\boldsymbol{X}_{i})-\pi_d^*(\boldsymbol{X}_{i}) \}^2 \notag\\
	&+\frac{u}{\sqrt{n}}\sum_{i=1}^n\left(\frac{D_{di}}{\pi_d^*(\boldsymbol{X}_{i})}-1\right)\mathcal{E}_{d}(\boldsymbol{X}_{i},\beta_d^*)\notag\\
	=&-u\cdot \mathbb{G}_n\left(\frac{D_{di}}{\{{\pi}^*_d(\boldsymbol{X}_{i})\}^2}\mathcal{L}'\left(Y_i-{\beta}_d^*\right)\{\widehat{\pi}_d(\boldsymbol{X}_{i})-\pi_d^*(\boldsymbol{X}_{i}) \}\right)\label{eq:mu_n:pihat-pi}\\
	&+\frac{u}{\sqrt{n}}\sum_{i=1}^n\frac{D_{di}}{\{\widetilde{\pi}_d(\boldsymbol{X}_{i})\}^3}\mathcal{L}'\left(Y_i-{\beta}_d^*\right)\{\widehat{\pi}_d(\boldsymbol{X}_{i})-\pi_d^*(\boldsymbol{X}_{i}) \}^2 \label{eq:pihat-pi^2}\\
	&+\frac{u}{\sqrt{n}}\sum_{i=1}^n\left(\frac{D_{di}}{\pi_d^*(\boldsymbol{X}_{i})}-1\right)\mathcal{E}_{d}(\boldsymbol{X}_{i},\beta_d^*)-u\cdot\sqrt{n}\cdot\mathbb{E}\left[\frac{\mathcal{E}_{d}(\boldsymbol{X}_{i},\beta_d^*)}{{\pi}^*_d(\boldsymbol{X}_{i})}\{\widehat{\pi}_d(\boldsymbol{X}_{i})-\pi_d^*(\boldsymbol{X}_{i}) \}\right] \label{eq:Epihat-pi}
\end{align}	
where $\widetilde{\pi}_d(\boldsymbol{X}_{i})$ is between $\widehat{\pi}_d(\boldsymbol{X}_{i})$ is ${\pi}^*_d(\boldsymbol{X}_{i})$. For \eqref{eq:mu_n:pihat-pi}, similar to the proof of Lemma \ref{lemma:several_results} and using Assumption \ref{as:parameter_beta} (ii), we have that \eqref{eq:mu_n:pihat-pi} is of $o_P(1)$.

For \eqref{eq:pihat-pi^2}, by Theorem \ref{thm:rate_ANN} and Assumptions \ref{as:parameter_beta} (ii) and \ref{as:L2_continuous}, we have
\begin{align}
	&|\eqref{eq:pihat-pi^2}|=\left|\frac{u}{\sqrt{n}}\sum_{i=1}^n\frac{D_{di}}{\{\widetilde{\pi}_d(\boldsymbol{X}_{i})\}^3}\mathcal{L}'\left(Y_i-{\beta}\right)\{\widehat{\pi}_d(\boldsymbol{X}_{i})-\pi_d^*(\boldsymbol{X}_{i}) \}^2\right| \notag\\
	\leq& O_P(\sqrt{n})\cdot\frac{u}{n}\sum_{i=1}^n|\mathcal{L}'(Y_i-{\beta}^*_d)|\cdot \{\widehat{\pi}_d(\boldsymbol{X}_{i})-{\pi}^*_d(\boldsymbol{X}_{i})\}^2 \notag\\
	\leq&O_P(\sqrt{n})\cdot u\cdot O(1)\cdot o_P(n^{-1/2})\cdot O_P(1)=o_P(1). \notag
\end{align}
For the term \eqref{eq:Epihat-pi}, we have
\begin{align}
	\eqref{eq:Epihat-pi}=&\frac{u}{\sqrt{n}}\sum_{i=1}^n\left(\frac{D_{di}}{\pi_d^*(\boldsymbol{X}_{i})}-1\right)\mathcal{E}_{d}(\boldsymbol{X}_{i},\beta_d^*)-u\cdot\sqrt{n}\cdot\mathbb{E}\left[\frac{\mathcal{E}_{d}(\boldsymbol{X}_{i},\beta_d^*)}{{\pi}^*_d(\boldsymbol{X}_{i})}\{L\left(\widehat{g}_d(\boldsymbol{X}_{i})\right)-L\left(g_d^*(\boldsymbol{X}_{i})\right) \}\right]\notag\\
	=& \frac{u}{\sqrt{n}}\sum_{i=1}^n\left(\frac{D_{di}}{\pi_d^*(\boldsymbol{X}_{i})}-1\right)\mathcal{E}_{d}(\boldsymbol{X}_{i},\beta_d^*)-u\cdot\sqrt{n}\cdot\mathbb{E}\left[\frac{\mathcal{E}_{d}(\boldsymbol{X}_{i},\beta_d^*)}{{\pi}^*_d(\boldsymbol{X}_{i})}L'\left(g_d^*(\boldsymbol{X}_{i})\right)\{\widehat{g}_d(\boldsymbol{X}_{i})-g_d^*(\boldsymbol{X}_{i})\}\right]\label{eq:Epihat-pi_1}\\
	&-\frac{u}{2}\cdot\sqrt{n}\cdot\mathbb{E}\left[\frac{\mathcal{E}_{d}(\boldsymbol{X}_{i},\beta_d^*)}{{\pi}^*_d(\boldsymbol{X}_{i})}L''\left(\widetilde{g}_d(\boldsymbol{X}_{i})\right)\{\widehat{g}_d(\boldsymbol{X}_{i})-g_d^*(\boldsymbol{X}_{i})\}^2\right],\label{eq:Epihat-pi_2}
\end{align}
where $\widetilde{g}_d$ is between $\widehat{g}_d$ and ${g}^*_d$.  We have that \eqref{eq:Epihat-pi_1} is of $o_P(1)$ by Lemma \ref{lemma:projection} and the fact $L'\left(g_d^*(\boldsymbol{X}_{i})\right)=L\left(g_d^*(\boldsymbol{X}_{i})\right)(1-L\left(g_d^*(\boldsymbol{X}_{i})\right))=\pi_d^*(\boldsymbol{X}_i)(1-\pi_d^*(\boldsymbol{X}_i))$.  For the term \eqref{eq:Epihat-pi_2}, we have
\begin{align*}
	|\eqref{eq:Epihat-pi_2}|\leq& \frac{|u|}{2}\cdot\sqrt{n}\cdot\mathbb{E}\left[\frac{|\mathcal{E}_{d}(\boldsymbol{X}_{i},\beta_d^*)|}{{\pi}^*_d(\boldsymbol{X}_{i})}L\left(\widetilde{g}_d(\boldsymbol{X}_{i})\right)(1-L\left(\widetilde{g}_d(\boldsymbol{X}_{i})\right))\cdot|1-2L\left(\widetilde{g}_d(\boldsymbol{X}_{i})\right)|\cdot\{\widehat{g}_d(\boldsymbol{X}_{i})-g_d^*(\boldsymbol{X}_{i})\}^2\right]\\
	\leq&|u|\cdot O(1)\cdot \sqrt{n}\cdot \mathbb{E}[\{\widehat{g}_d(\boldsymbol{X}_{i})-g_d^*(\boldsymbol{X}_{i})\}^2]=o_P(1),
\end{align*}
where the second inequality holds by Assumptions \ref{as:parameter_beta} (ii) and \ref{as:L2_continuous}, and the last equality holds by Lemma \ref{app:rate_ANN}. Combining the results for \eqref{eq:Epihat-pi_1} and \eqref{eq:Epihat-pi_2},  we obtain that \eqref{eq:Epihat-pi} is of $o_P(1)$. Combining the results for \eqref{eq:mu_n:pihat-pi}, \eqref{eq:pihat-pi^2}, and \eqref{eq:Epihat-pi}, we obtain  \eqref{eq:xi_1}.

Next, we prove \eqref{eq:xi_2}. Note that
\begin{align}
	|\xi_{2,d,n}(u,\widehat{\pi}_d)|=&\left|	\sum_{i=1}^n\frac{D_{di}}{\widehat{\pi}_d(\boldsymbol{X}_{i})}R_d(Y_i,u)-\partial_{\beta}\mathbb{E}[\mathcal{L}'(Y^*_i(d)-\beta_d^*)]\cdot \frac{u^2}{2}\right| \notag\\
	\leq&\left|	\sum_{i=1}^nD_{di}R_d(Y_i,u)\frac{\widehat{\pi}_d(\boldsymbol{X}_{i})-\pi^*_d(\boldsymbol{X}_{i})}{\widehat{\pi}_d(\boldsymbol{X}_{i})\pi^*_d(\boldsymbol{X}_{i})}\right| \label{eq:xi_2_1} \\
	& +\left|	\sum_{i=1}^n\frac{D_{di}}{{\pi}^*_d(\boldsymbol{X}_{i})}R_d(Y_i,u)-n\cdot\mathbb{E}\left[\frac{D_{di}}{{\pi}^*_d(\boldsymbol{X}_{i})}R_d(Y_i,u)\right] \right| \label{eq:xi_2_2}\\
	&+\left|n\cdot\mathbb{E}\left[\frac{D_{di}}{{\pi}^*_d(\boldsymbol{X}_{i})}R_d(Y_i,u)\right]-\partial_{\beta}\mathbb{E}[\mathcal{L}'(Y^*_i(d)-\beta_d^*)]\cdot \frac{u^2}{2} \right| \label{eq:xi_2_3}.
\end{align}
For \eqref{eq:xi_2_1}, we have
\begin{align*}
	&\left|	\sum_{i=1}^nD_{di}R_d(Y_i,u)\frac{\widehat{\pi}_d(\boldsymbol{X}_{i})-\pi^*_d(\boldsymbol{X}_{i})}{\widehat{\pi}_d(\boldsymbol{X}_{i})\pi^*_d(\boldsymbol{X}_{i})}\right|\\
	\leq &n\cdot		\sqrt{\frac{1}{n}\sum_{i=1}^n\left|\frac{D_{di}R_d(Y_i,u)}{\widehat{\pi}_d(\boldsymbol{X}_{i})\pi^*_d(\boldsymbol{X}_{i})}\right|^2}\cdot \sqrt{\frac{1}{n}\sum_{i=1}^n\left|\widehat{\pi}_d(\boldsymbol{X}_{i})-\pi^*_d(\boldsymbol{X}_{i})\right|^2}\\
	\leq &n\cdot	O_P(1) \cdot 	\sqrt{\frac{1}{n}\sum_{i=1}^n\left|R_d(Y^*_i(d),u)\right|^2}\cdot \sqrt{\frac{1}{n}\sum_{i=1}^n\left|\widehat{\pi}_d(\boldsymbol{X}_{i})-\pi^*_d(\boldsymbol{X}_{i})\right|^2}\\
	=&n\cdot O(1)\cdot O\left(\frac{|u|^{3/2}}{n^{3/4}}\right)\cdot o_P(n^{-1/4})\cdot\{1+o_P(1)\}=|u|^{3/2}\cdot o_P(1),
\end{align*}
where the second inequality holds by Assumption \ref{as:parameter_beta} (ii) and $\|\widehat{\pi}_d-\pi_d^*\|_{\infty}=o_P(1)$, the first equality holds because of Lemma \ref{lemma:projection}, the first and second moments $R_d(Y^*_i(d),u)$ given by
\begin{align*}
	\mathbb{E}\left[R_d(Y^*_i(d),u)\right]=&\mathbb{E}\left[\mathcal{L}\left(Y^*(d)-\left\{\beta_d^*+\frac{u}{\sqrt{n}}\right\}\right)\right]-\mathbb{E}\left[\mathcal{L}(Y^*(d)-\beta_d^*)\right]+\mathbb{E}[\mathcal{L}'(Y^*(d)-\beta_d^*)]\cdot \frac{u}{\sqrt{n}}\\
	=&\frac{u^2}{2n}\cdot \partial_{\beta_d}\mathbb{E}\left[\mathcal{L}'\left(Y^*(d)-\left\{\beta_d^*+\frac{u^*}{\sqrt{n}}\right\}\right)\right]
\end{align*}
where $u^*$ is between $0$ and $u$, and
\begin{align}
	&\mathbb{E}\left[R_d(Y^*_i(d),u)^2\right]\notag\\
	=&\mathbb{E}\left[\left\{\mathcal{L}\left(Y^*_i(d)-\left\{\beta_d^*+\frac{u}{\sqrt{n}}\right\}\right)-\mathcal{L}(Y^*_i(d)-\beta_d^*)+\mathcal{L}'(Y^*_i(d)-\beta_d^*)\cdot \frac{u}{\sqrt{n}}\right\}^2\right] \notag\\
	=& \mathbb{E}\left[\left\{-\mathcal{L}'\left(Y^*_i(d)-\left\{\beta_d^*+\frac{\overline{u}}{\sqrt{n}}\right\}\right)\cdot \frac{u}{\sqrt{n}} +\mathcal{L}'(Y^*_i(d)-\beta_d^*)\cdot \frac{u}{\sqrt{n}}\right\}^2\right]\notag\\
	\leq & \text{const}\times \left(\frac{|\overline{u}|}{\sqrt{n}}\right)^{2\kappa} \frac{u^2}{n} \leq \text{const}\times  \frac{|u|^3}{n^{3/2}}, \label{eq:apply_L^2}
\end{align}
where $\overline{u}$ is between $0$ and $u$, and \eqref{eq:apply_L^2} holds by Assumption \ref{as:L2_continuous} (i) and $\kappa\geq 1/2$.

For \eqref{eq:xi_2_2}, by computing its second moment and using Chebyshev's inequality, we have
\begin{align*}
	\eqref{eq:xi_2_2}=O_P\left(\sqrt{n}\cdot O\left(\frac{|u|^{3/2}}{n^{\kappa/4+1/2}}\right) \right)=O_P(n^{-\kappa/4}\cdot |u|^{3/2})=o_P(|u|^{3/2}).
\end{align*}
For \eqref{eq:xi_2_3}, we have
\begin{align*}
	\eqref{eq:xi_2_3}=&\left|n\cdot\mathbb{E}\left[\frac{D_{di}}{{\pi}^*_d(\boldsymbol{X}_{i})}R_d(Y_i,u)\right]-\partial_{\beta_d}\mathbb{E}[\mathcal{L}'(Y^*_i(d)-\beta_d^*)]\cdot \frac{u^2}{2} \right|\\
	=&\left|n\cdot\mathbb{E}\left[R_d(Y^*_i(d),u)\right]-\partial_{\beta_d}\mathbb{E}[\mathcal{L}'(Y^*_i(d)-\beta_d^*)]\cdot \frac{u^2}{2} \right|\\
	=&\left|\frac{u^2}{2}\cdot \partial_{\beta_d}\mathbb{E}\left[\mathcal{L}'\left(Y_i^*(d)-\left\{\beta_d^*+\frac{\theta\cdot u}{\sqrt{n}}\right\}\right)\right]-\partial_{\beta_d}\mathbb{E}[\mathcal{L}'(Y^*_i(d)-\beta_d^*)]\cdot \frac{u^2}{2}\right| \\
	=&o(u^2),
\end{align*}
where $\theta\in (0,1)$ arises from the application of the mean value theorem, the last equality holds because Assumption \ref{as:density_smooth} implies that  $\partial_{\beta_d}\mathbb{E}[\mathcal{L}'(Y^*_i(d)-\beta_d)]$ is continuous in $\beta_d$.  Therefore, we obtain \eqref{eq:xi_2}. So for fixed $u$,
\begin{align}\label{eq:Qtilde-Q}
	\xi_{d,n}(u,\widehat{\pi}_d)=\widetilde{Q}_{d,n}(u)-Q_{d,n}(u,\widehat{\pi}_d)=o_P(1).
\end{align}

To establish Step II, it is sufficient to prove that for each $\varepsilon>0$,
\begin{align}\label{eq:Puhat-utilde>e}
	\mathbb{P}\left(|\widehat{u}_d-\widetilde{u}_d|\geq \varepsilon\right)\leq \mathbb{P}\left(\sup_{|u-\widetilde{u}|\leq \varepsilon}|\xi_{d,n}(u,\widehat{\pi}_d)|\geq -\frac{1}{4}\cdot \partial_{\beta_d}\mathbb{E}[\mathcal{L}'(Y^*(d)-\beta_d^*)] \cdot \varepsilon^2 \right)=o(1).
\end{align}
Note that $G_{d,n}(\beta,\widehat{\pi}_d)=n^{-1}\sum_{i=1}^nD_{di}\mathcal{L}(Y_i-\beta)/\widehat{\pi}_d(\boldsymbol{X}_{i})$ is convex in $\beta$ with probability approaching one, as it is a sum of zeros and convex functions in $\beta$. As a result, both $Q_{d,n}(u,\widehat{\pi}_d)$  and the following function $B_{d,n}(u,\widehat{\pi}_d)$ are also convex in $u$:
\begin{align*}
	B_{d,n}(u,\widehat{\pi}_d):=&Q_{d,n}(u,\widehat{\pi}_d)+\frac{u}{\sqrt{n}}\sum_{i=1}^n\left[\frac{D_{di}}{\pi_d^*(\boldsymbol{X}_i)}\mathcal{L}'(Y_i-\beta_d^*)-\left\{\frac{D_{di}}{\pi_d^*(\boldsymbol{X}_i)}-1\right\}\mathcal{E}_d^*(\boldsymbol{X}_i,\beta_d^*)\right]\\
	=&-\frac{u^2}{2}\cdot \partial_{\beta_d}\mathbb{E}[\mathcal{L}'(Y_i^*(d)-\beta_d^*)]-\xi_{d,n}(u,\widehat{\pi}_d).
\end{align*}
Let $B_{d}(u):=-\frac{u^2}{2}\cdot \partial_{\beta_d}\mathbb{E}[\mathcal{L}'(Y_i^*(d)-\beta_d^*)]$ be the quadratic function. In light of the convexity of $B_{d,n}(u,\widehat{\pi}_d)$, we have that for any $u$ such that $|u-\widetilde{u}|=a>\varepsilon$,
\begin{align*}
	\left(1-\frac{\varepsilon}{a}\right)\cdot B_{d,n}(\widetilde{u}_d,\widehat{\pi}_d)+\frac{\varepsilon}{a}\cdot B_{d,n}(u,\widehat{\pi}_d)\geq B_{d,n}(\widetilde{u}_d+\varepsilon,\widehat{\pi}_d).
\end{align*}
Then we have that
\begin{align*}
	&\frac{\varepsilon}{a}\left( B_{d,n}(u,\widehat{\pi}_d)-B_{d,n}(\widetilde{u}_d,\widehat{\pi}_d)\right)\\
	\geq& B_d(\widetilde{u}_d+\varepsilon)-\xi_{d,n}(\widetilde{u}_d+\varepsilon,\widehat{\pi}_d)-\left(B_d(\widetilde{u}_d)-\xi_{d,n}(\widetilde{u}_d,\widehat{\pi}_d)\right)\\
	\geq &-2\sup_{|u-\widetilde{u}_d|\leq \varepsilon}|\xi_{d,n}(u,\widehat{\pi}_d)|+\inf_{|u-\widetilde{u}_d|=\varepsilon}|B_d(u)-B_d(\widetilde{u}_d)|.
\end{align*}
Since $\inf_{|u-\widetilde{u}_d|=\varepsilon}|B_d(u)-B_d(\widetilde{u}_d)|=-\frac{1}{2}\cdot \partial_{\beta_d}\mathbb{E}[\mathcal{L}'(Y_i^*(d)-\beta_d^*)]\cdot \varepsilon^2$. Therefore, for all $u$ out of the $\varepsilon$ interval around $\widetilde{u}_d$, if
\begin{align}\
	-2\sup_{|u-\widetilde{u}_d|\leq \varepsilon}|\xi_{d,n}(u,\widehat{\pi}_d)|-\frac{1}{2}\cdot \partial_{\beta_d}\mathbb{E}[\mathcal{L}'(Y_i^*(d)-\beta_d^*)]\cdot \varepsilon^2>0,
\end{align}
then $\widehat{u}_d$, the minimizer of $Q_{d,n}(u,\widehat{\pi}_d)$, will be located in the $\varepsilon$ interval around $\widetilde{u}_d$, i.e. the inequality part of \eqref{eq:Puhat-utilde>e} holds.  For the equality part of \eqref{eq:Puhat-utilde>e}, since $\{u:|u-\widetilde{u}_d|\leq \varepsilon\}$ is compact, by the \cite{hjort2011asymptotics} version of the convexity lemma, we have $\sup_{|u-\widetilde{u}_d|\leq \varepsilon}|\xi_{d,n}(u,\widehat{\pi}_d)|=o_P(1)$, which gives that for each $\varepsilon>0$,
$$\mathbb{P}\left(\sup_{|u-\widetilde{u}_d|\leq \varepsilon}|\xi_{d,n}(u,\widehat{\pi}_d)|\geq -\frac{1}{4}\cdot \partial_{\beta_d}\mathbb{E}[\mathcal{L}'(Y^*(d)-\beta_d^*)] \cdot \varepsilon^2 \right)=o(1),$$
as $-\partial_{\beta_d}\mathbb{E}[\mathcal{L}'(Y^*(d)-\beta_d^*)]>0$ by Assumption \ref{as:density_smooth} (iii). Hence, \eqref{eq:Puhat-utilde>e} holds and the proof of Step II is done.  Finally, we  conclude our desired result in Step III:
\begin{align*}
	\sqrt{n}\{\widehat{\beta}_d-\beta_d^*\}=\widetilde{u}_d+\{\widehat{u}_d-\widetilde{u}_d\}=\frac{1}{\sqrt{n}}\sum_{i=1}^nH_d^{-1}\cdot S_d(Y_i,D_{di},\boldsymbol{X}_{i};\beta_d^*)+o_P(1).
\end{align*}

%\newpage

\section{Proof of Theorem \ref{thm:regression}}\label{proof:thm:regression}
Let	$\widehat{F}_X(x):=n^{-1}\sum_{i=1}^n\mathds{1}(\boldsymbol{X}_{i}\leq x)$ be the empirical cumulative distribution function of $\boldsymbol{X}$. Note that
\begin{align}
	\sqrt{n}\left\{\widehat{\beta}_d^{OR} -\beta^*_d\right\}=&\sqrt{n}\left\{\int
	\widehat{z}_d(x)d\widehat{F}_X(x) -\int z_d^*(x)dF_X(x) \right\}\notag \\
	=&\sqrt{n}\int [\widehat{z}_d(x)-z^*_d(x)]d[\widehat{F}_X(x)-F_X(x)]\label{ghat-g*_Fhat-F}\\
	&+\sqrt{n}\int z^*_d(x)d[\widehat{F}_X(x)-F_X(x)] \label{g*_Fhat-F}\\
	&+\sqrt{n}\int [\widehat{z}_d(x)-z^*_d(x)]dF_X(x). \label{ghat-g*_F}
\end{align}
For the term \eqref{ghat-g*_Fhat-F}, it can be written as $\eqref{ghat-g*_Fhat-F}=\mathbb{G}_n(\widehat{z}_d-z^*_d)$. Since $z^*_d(x)\in\mathcal{F}_p^m$, by Theorem \ref{thm:rate_ANN},  $$\|\widehat{z}_d-z^*_d\|_{L^2(dF_X)}=O_P(\delta_n)=o_P(n^{-1/4}).$$ Using a similar argument of proving Lemma \ref{lemma:several_results}, we have
\begin{align*}
	\sup_{\{z_d\in \mathcal{G}_n:\|z_d-z^*_d\|_{L^2(dF_X)}\leq  \delta_n\}}\mathbb{G}_n(z_d-z^*_d)=o_P(1),
\end{align*}
which gives that \eqref{ghat-g*_Fhat-F} is of $o_P(1)$. By definition, \eqref{g*_Fhat-F} can be written as follows:
\begin{align*}
	\eqref{g*_Fhat-F}=\frac{1}{\sqrt{n}}\sum_{i=1}^n\left\{z_d^*(\boldsymbol{X}_{i})-\mathbb{E}[z_d^*(\boldsymbol{X}_{i})]\right\}.
\end{align*}	
For the term \eqref{ghat-g*_F},	it can be written as follows:
\begin{align*}
	\eqref{ghat-g*_F}=\sqrt{n}\int [\widehat{z}_d(x)-z^*_d(x)]w_d^*(x)dF_{X|D}(x|d)
\end{align*}
where
\begin{align*}
	w^*_d(x):=\frac{f_X(x)}{f_{X|D}(x|d)}.
\end{align*}
To complete the proof, it is sufficient to establish
\begin{align}\label{eq:ghat-g*_Fd}
	\sqrt{n} \int [\widehat{z}_d(x)-z^*_d(x)]w^*_d(x)dF_{X|D}(x|d)=- \frac{1}{\sqrt{n}}\sum_{i=1}^n\frac{D_{di}}{\pi_d^*(\boldsymbol{X}_{i})}\{Y_i-z^*_d(\boldsymbol{X}_{i})\}+o_P(1).
\end{align}
The proof of \eqref{eq:ghat-g*_Fd} is similar to that of Lemma \ref{lemma:projection}.	Note that
\begin{align*}
	L^{OR}_{d,n}(z_d):= \frac{1}{n}\sum_{i=1}^{n}\ell^{OR}_d(Y_i,D_{di},\boldsymbol{X}_{i};z_d)=- \frac{1}{n}\sum_{i=1}^{n}D_{di}\{Y_i-z_d(\boldsymbol{X}_{i})\}^2,
\end{align*}
where $\ell_d(Y_i,D_{di},\boldsymbol{X}_{i};z_d):=-D_{di}\{Y_i-z_d(\boldsymbol{X}_{i})\}^2$.  Let $\epsilon_n$ be a positive sequence satisfying $\epsilon_n=o(n^{-1/2})$.
For any $z_d\in \{z_d\in\mathcal{G}_n:\|z_d-z_d^*\|_{L^2(dF_X)}\leq \delta_n \}$, we consider a local alternative value
\begin{align*}
	\overline{z}(z_d,\epsilon_n):=(1-\epsilon_n)\cdot z_d+\epsilon_n\cdot \{w_d^*+z_d^*\}.
\end{align*}
Then by applying a similar argument  of establishing Lemma \ref{lemma:projection},  we can get
\begin{align}
	&-\sqrt{n}\cdot 2\cdot \mathbb{E}\left[ D_{di}\cdot w_d^*(\boldsymbol{X}_{i}) \cdot (\widehat{z}_d(\boldsymbol{X}_{i})-z^*_d(\boldsymbol{X}_{i}))\right]\notag\\
	=&\mathbb{G}_n\left(\frac{d}{dz_d}\ell_d(D_{di},\boldsymbol{X}_{i},Y_i;z^*_d)[w_d^*(\boldsymbol{X}_{i})]\right)+o_P\left(1\right) \notag\\
	=&\frac{2}{\sqrt{n}}\sum_{i=1}^nD_{di}\left\{Y_i-z^*_d(\boldsymbol{X}_{i})\right\} w_d^*(\boldsymbol{X}_{i}) +o_P\left(1\right),\label{eq:ghat-g*_u*_0}
\end{align}
which implies \eqref{eq:ghat-g*_Fd} after simple calculation.

%\newpage
\section{Statistical inference}\label{appendix:eff_IPW_bootsrap}
\subsection{Proof of Theorem \ref{thm:eff_IPW_bootsrap}}
We prove the part (i), and a similar argument can be applied to obtain the part (ii). We repeat the proofs of the consistency
and the convergence rates of Theorem \ref{thm:rate_ANN}, except using $\omega_{di}\ell_d(D_{di},\boldsymbol{X}_{i};g_d(\cdot))$
instead of $\ell_d(D_{di},\boldsymbol{X}_{i};g_d(\cdot))$. We can show that
the weighted bootstrap estimators $\|\widehat{g}^{B}_d-{g}_d^*\|_{L^2(dF_X)}=o_P(n^{-1/4})$ and $\|\widehat{\pi}^{B}_d-{\pi}_d^*\|_{L^2(dF_X)}=o_P(n^{-1/4})$   with probability approaching one. We shall establish the limiting
distribution in two steps.
\begin{itemize}
	\item STEP 1: We first derive the asymptotic representation of $\sqrt{n}\cdot\mathbb{E}[\{1-\pi_d^*(\boldsymbol{X})\}\{\widehat{g}^{B}_d(\boldsymbol{X})-g_d^*(\boldsymbol{X})\}\mathcal{E}_d(\boldsymbol{X},\beta_d^*)]$ by mimicking the proof of Lemma \ref{lemma:projection}. Then we further repeat the proof of Theorem \ref{thm:eff_IPW}, and obtain:
	\begin{align}\label{eq:betahat*-beta*}
		\sqrt{n}\{\widehat{\beta}^{B}_{d}-\beta^*_{d}\}=H_d^{-1}\cdot \frac{1}{\sqrt{n}}\sum_{i=1}^n\omega_{di}S_{d}(Y_i,D_{di},\boldsymbol{X}_{i};\beta_d^*)+o_P(1).
	\end{align}
	This implies that $\sqrt{n}\{\widehat{\beta}^{B}_{d}-\beta^*_{d}\}$ is asymptotically
	normal with zero mean and variance $V_d=Var(S_{d}(Y_i,D_{di},\boldsymbol{X}_{i};\beta_d^*))$.
	\item STEP 2: Subtracting \eqref{eq:betahat-beta*} from \eqref{eq:betahat*-beta*}, we obtain
	\begin{align*}
		\sqrt{n}\{\widehat{\beta}^{B}_{d}-\widehat{\beta}_{d}\}=H_d^{-1}\cdot \frac{1}{\sqrt{n}}\sum_{i=1}^n\{\omega_{di}-1\}S_{d}(Y_i,D_{di},\boldsymbol{X}_{i};\beta_d^*)+o_P(1).
	\end{align*}
	Given that $\mathbb{E}(\omega_{di})=1$, $Var(\omega_{di}-1)=Var(\omega_{di})=1$ and that $\{\omega_{di}\}_{i=1}^n$ is independent of $\{D_{di},\boldsymbol{X}_{i},Y_i\}_{i=1}^n$, it follows that, conditional on the data $\{D_{di},\boldsymbol{X}_{i},Y_i\}_{i=1}^n$, $\sqrt{n/V_{d}}\{\widehat{\beta}^{B}_{d}-\widehat{\beta}_{d}\}$ is asymptotically a standard normal random variable, the same limiting distribution as that of $\sqrt{n/V_{d}}\{\widehat{\beta}_{d}-{\beta}^*_{d}\}$.
\end{itemize}

\subsection{Variance estimation}\label{app:variance}
This section studies the estimation of $\boldsymbol{V}$ in Theorem \ref{thm:eff_IPW}.
\begin{assumption}\label{as:density_smooth}    For every $d\in\mathcal{D}=\{0,1,...,J\}$,
	the conditional probability density  $f_{Y|X,D}(y|x,d)$  is continuously differentiable in $y\in \mathbb{R}$ and the derivative is uniformly bounded.
\end{assumption}
Since the nonsmooth loss function may invalidate the
exchangeability between the expectation and derivative operator, some care
in the estimation of $H_{d}$ is warranted. By Assumption \ref{as:density_smooth}, the tower property of
conditional expectation and Leibniz integration rule, we rewrite $H_{d}$ as:
\begin{align}\label{appendix:integration_by_parts}
	H_{d}= -\mathbb{E}\left[\frac{D_{di}}{\pi_d^*(\boldsymbol{X}_{i})} \mathcal{L}^{\prime }(Y_i-\beta_d^*) \frac{\partial }{\partial y}\log f_{Y,X|D}(Y_i,\boldsymbol{X}_{i}|d) \right] .
\end{align}
The proof of \eqref{appendix:integration_by_parts} will be given later. The log density $\log f_{Y,X|D}(y,x|d)$ can be estimated via the widely used ANN extremum estimator:
\begin{equation*}
	\widehat{f}_{Y,X|D}(y,x|d):=\frac{\exp \left( \widehat{a}_d(y,x)\right) }{\int_{\mathcal{Y}%
			\times  \mathcal{X}}\exp \left(\widehat{a}_d (y,x)\right) dydx},
\end{equation*}%
where $\widehat{a}_d \in \widetilde{\mathcal{G}}_n$
approximately maximizes the following criterion function:
\begin{align*}
	\widetilde{L}_{d,n}\left(\widehat{a}_d\right)\geq \sup_{a\in\widetilde{G}_n}\widetilde{L}_{d,n}\left(a\right)-O(\epsilon_n^2),
\end{align*}
where
$\widetilde{L}_{d,n}(a):=\frac{1}{n}\sum_{i=1}^{n}D_{di}\left[ a(Y_{i},\boldsymbol{X}_{i})-\log \int_{
	\mathcal{Y}\times \mathcal{X}}\exp \left( a(y,x)\right) dydx\right],$
and $\widetilde{\mathcal{G}}_n$ is a neural network similar to $\mathcal{G}_n$ with input variable $(x,y)$. Then $H_{d}$ can be estimated by
\begin{align}\label{def:Hdhat}
	\widehat{H}_d:=-\frac{1}{n}\sum_{i=1}^{n}\frac{D_{di}}{\widehat{\pi}_d(\boldsymbol{X}_{i})}\mathcal{L}^{\prime }(Y_{i}-\widehat{\beta}_d)\cdot \left\{\frac{\partial }{\partial y} \widehat{a}
	_{d}(Y_{i},
	\boldsymbol{X}_{i})\right\}.
\end{align}

Under Assumption \ref{as:CIA}, $\mathcal{E}_d(x,{\beta}^*_d)=\mathbb{E}[ \mathcal{L}'(Y_i-{\beta}^*_d)|\boldsymbol{X}_{i}=x,D_i=d]$, hence  $\mathcal{E}_d(x,{\beta}^*_d)$ can be estimated by using ANN extremum estimates:
\begin{align}\label{eq:Ed_hat}
	\widehat{\mathcal{E}}_d(\cdot):=\arg\min_{g(\cdot)\in\mathcal{G}_n }\frac{1}{n}\sum_{i=1}^n D_{di}\left\{ \mathcal{L}'\left(Y_i-\widehat{\beta}_d\right)-g(\boldsymbol{X}_{i}) \right\}^2.
\end{align}
Therefore, the plug-in estimates of $S_d(Y_i,D_{di},\boldsymbol{X}_{i};{\beta}^*_d)$ is \begin{align}\label{def:psidhat}
	\widehat{S}_{di}=	\frac{D_{di}}{\widehat{\pi}_d(\boldsymbol{X}_{i})}\mathcal{L}'\{Y_i-\widehat{\beta}_d\}-\left\{\frac{D_{di}-\widehat{\pi}_d(\boldsymbol{X}_{i})}{\widehat{\pi}_d(\boldsymbol{X}_{i})}\right\}\widehat{\mathcal{E}}_d(\boldsymbol{X}_{i}).
\end{align}	
Finally, by \eqref{def:Hdhat} and \eqref{def:psidhat}, the asymptotic
covariance matrix of the estimator is estimated by
$$\widehat{\boldsymbol{V}}:=\widehat{\boldsymbol{H}}^{-1}\left\{ \frac{1}{n}\sum_{i=1}^{n}\widehat{\boldsymbol{S}}_i\widehat{\boldsymbol{S}}_i^{\top }\right\} (%
\widehat{\boldsymbol{H}}^{\top })^{-1},$$
where $\widehat{\boldsymbol{H}}=\text{Diag}\{\widehat{H}_0,...,\widehat{H}_J\}$ and $\widehat{\boldsymbol{S}}_i=(\widehat{S}_{0i},...,\widehat{S}_{Ji})^\top$.
Since $| \widehat{{\beta }}_d-{\beta }_d
^{\ast }| \xrightarrow{P} 0$, $\widehat{\pi}_d(\cdot)\xrightarrow{n}\pi_d^*(\cdot)$ and $\widehat{\mathcal{E}}_d(\cdot)\xrightarrow{n}\mathcal{E}_d(\cdot,\beta_d^*)$ for all $d\in \{0,1,...,J\}$.  Based on these results, the consistency of $\widehat{\boldsymbol{V}}$, i.e. $\widehat{\boldsymbol{V}}\xrightarrow{p}\boldsymbol{V}$ follows
from standard arguments.

The asymptotic normality of $\widehat{\boldsymbol{\beta }}=(%
\widehat{\beta }_{0},\widehat{\beta }_{1},...,\widehat{\beta }_{J})^{\top }$
established in Theorem \ref{thm:eff_IPW} together with the consistency of $%
\widehat{\boldsymbol{V}}$ provides a theoretical support for conducting
statistical inference of the TE parameter vector $\boldsymbol{\beta }^{\ast
}=(\beta _{0}^{\ast },\beta _{1}^{\ast },....,\beta _{J}^{\ast })^{\top }$.
For instance, based on these results, we can construct a $100(1-\alpha )\%$
confidence interval for each ${\beta }_{d}^{\ast}$, $d\in\{0,1,...,J\}$,
given by
\begin{equation*}
	\left[ \widehat{\beta }_{d}-n^{-1/2}z_{\alpha /2}\widehat{V}%
	_{dd}^{1/2},\quad \widehat{\beta}_{d}+n^{-1/2}z_{\alpha /2}\widehat{V}_{dd}^{1/2}%
	\right] ,
\end{equation*}%
where $\widehat{V}_{dd}$ is the $(d,d)$-element of the estimated covariance
matrix $\widehat{\boldsymbol{V}}$, and $z_{\alpha /2}$ is the $100(1-\alpha
/2)$ percentile of the standard normal. We can also construct confidence
intervals for a contrast of {$\boldsymbol{\beta }^{\ast }$ for a comparison
	of different TE parameters. That is, }for any given {$\boldsymbol{a}$}$\in
\mathbb{R}^{J+1}$, a $100(1-\alpha )\%$ confidence interval for $%
\boldsymbol{a^{\top }\beta }^{\ast }$ is given by
\begin{equation*}
	\left[ {\boldsymbol{a^{\top }}}\widehat{\boldsymbol{\beta }}%
	-n^{-1/2}z_{\alpha /2}({\boldsymbol{a^{\top }}\widehat{\boldsymbol{V}}%
		\boldsymbol{a}})^{1/2},\quad {\boldsymbol{a^{\top }}}\widehat{\boldsymbol{%
			\beta }}+n^{-1/2}z_{\alpha /2}({\boldsymbol{a^{\top }}\widehat{\boldsymbol{V}%
		}\boldsymbol{a}})^{1/2}\right] .
\end{equation*}

It is worth noting that estimation of the asymptotic variance is straightforward for average TE, but it can be difficult for other types of TEs, such as quantile TEs. Thus, the weighted bootstrap method is recommended for conducting inference of $\boldsymbol{\beta }^{\ast }$, and it can yield a more stable inferential result. Since our TE estimator is
obtained by optimizing a generalized objective function, it is very convenient to apply the weighted bootstrap in our estimation procedure.

Finally, we complete the proof by establishing \eqref{appendix:integration_by_parts}. Using the tower property of
conditional expectation, we rewrite $H_{d}$ as:
\begin{align*}
	H_{d}=& -\partial _{\beta}\mathbb{E}\left[ \frac{D_{di}}{\pi_d^*(\boldsymbol{X}_{i})}\cdot \mathbb{E%
	}\left[ \mathcal{L}^{\prime }(Y_i-\beta)|D_{di},\boldsymbol{X}_{i}\right] \right] \Bigg|_{\beta=\beta_d^*} \\
	=& -\mathbb{E}\left[\frac{D_{di}}{\pi_d^*(\boldsymbol{X}_{i})}\cdot \partial _{\beta }\mathbb{E}\left[ \mathcal{L}^{\prime }(Y_i-\beta)|D_{di},\boldsymbol{X}_{i}\right] \Big|
	_{{\beta }={\beta }_d^{\ast }}\right]  .
\end{align*}
By Assumption \ref{as:density_smooth} and Leibniz integration rule, we obtain
\begin{align*}
	&\partial _{\beta}\mathbb{E}\left[ \mathcal{L}^{\prime }(Y-\beta)|D=d,\boldsymbol{X}=x\right] \Big|_{{\beta }=\beta_d^*} \notag \\
	=&\partial _{\beta }\left[\int_{\mathbb{R}}\mathcal{L}'(y-\beta)f_{Y|D,X}(y|d,x)dy\right]\Bigg|_{{\beta}={\beta }_{d}^*} \\
	=&\partial _{\beta }\left[\int_{\mathbb{R}}\mathcal{L}'(z)f_{Y|D,X}(z+\beta|d,x)dz\right]\Big|_{{\beta }=\beta_{d}^*}\  (\text{use}\ z=y-\beta)) \\
	=& \int_{\mathbb{R}}\mathcal{L}'(z)\cdot \frac{\partial}{\partial y}f_{Y|D,X}(z+\beta_d^*|d,x)dz\\
	=& \int_{\mathbb{R}}\mathcal{L}'(y-\beta_d^*)\cdot \frac{\partial}{\partial y}f_{Y|D,X}(y|d,x)dy \\
	=& \int_{\mathbb{R}}\mathcal{L}'(y-\beta_d^*)\cdot \frac{\frac{\partial}{\partial y}f_{Y|D,X}(y|d,x)}{f_{Y|D,X}(y|d,x)}f_{Y|D,X}(y|d,x)dy\\
	=& \int_{\mathbb{R}}\mathcal{L}'(y-\beta_d^*)\cdot \frac{\frac{\partial}{\partial y}f_{Y,X|D}(y,x|d)}{f_{Y,X|D}(y,x|d)}f_{Y|D,X}(y|d,x)dy \\
	=& \mathbb{E}\left[ \mathcal{L}^{\prime }(Y-\beta_d^*)\frac{\frac{
			\partial }{\partial y}f_{Y,X|D}(Y,\boldsymbol{X}|d)}{f_{Y,X|D}(Y,\boldsymbol{X}|d)}\bigg|D=d,\boldsymbol{X}=x
	\right],
\end{align*}
and consequently
\begin{equation*}
	H_{d}=-\mathbb{E}\left[\frac{D_{di}}{\pi_d^*(\boldsymbol{X}_{i})}\mathcal{L}^{\prime }(Y_i-\beta_d^*) \frac{\partial }{\partial y}\log f_{Y,X|D}(Y_i,\boldsymbol{X}_{i}|d) \right] .
\end{equation*}

%\newpage
\section{Proof of Theorem \ref{thm:att}}
%-----------------------------
The proof is very similar to that of Theorem \ref{thm:eff_IPW} so we will be brief here.
Similar to \eqref{def:Gd} and \eqref{def:beta_d}, the estimator of $\beta^*_{d,d'}$ for $d\in\{0,1,...,J\}$ can be written as
\begin{align*}
	\widehat{\beta}_{d,d'}=\arg\min_{\beta\in\Theta}G_{d,d',n}(\beta,\widehat{\pi}_d,\widehat{\pi}_{d'})=\arg\min_{\beta\in\Theta}\frac{1}{n}\sum_{i=1}^n\frac{D_{di}}{\widehat{\pi}_{d}(\boldsymbol{X}_{i})}\widehat{\pi}_{d'}(\boldsymbol{X}_{i}) \mathcal{L}\left(Y_i-\beta\right),
\end{align*}
where the definition of $G_{d,d',n}(\beta,\widehat{\pi}_d,\widehat{\pi}_{d'})$ is obvious. We can rewrite $ \widehat{\beta}_{d,d'}$ as follows:
\begin{align*}
	\widehat{\beta}_{d,d'}=&\arg\min_{\beta\in\Theta} n\left\{G_{d,d',n}(\beta,\widehat{\pi}_d,\widehat{\pi}_{d'})-G_{d,n}(\beta_{d,d'}^*,\widehat{\pi}_d,\widehat{\pi}_{d'}) \right\}\\
	=&\arg\min_{\beta\in\Theta}\sum_{i=1}^n\frac{D_{di}}{\widehat{\pi}_d(\boldsymbol{X}_{i})}\widehat{\pi}_{d'}(\boldsymbol{X}_{i})\left\{\mathcal{L}(Y_i-\beta)-\mathcal{L}(Y_i-\beta_{d,d'}^*)\right\}\\
	=&\arg\min_{\beta\in\Theta}\sum_{i=1}^n\frac{D_{di}}{\widehat{\pi}_d(\boldsymbol{X}_{i})}\widehat{\pi}_{d'}(\boldsymbol{X}_{i})\bigg[-\mathcal{L}'(Y_i-\beta_{d,d'}^*)(\beta-\beta_{d,d'}^*)\\
	&\qquad\qquad+\left\{\mathcal{L}(Y_i-\beta)-\mathcal{L}(Y_i-\beta_{d,d'}^*)+\mathcal{L}'(Y_i-\beta_{d,d'}^*)(\beta-\beta_{d,d'}^*)\right\}\bigg].
\end{align*}
By using change of variables and defining the following functions:
\begin{align*}
	&\widehat{u}_{d,d'}:=\sqrt{n}(\widehat{\beta}_{d,d'}-{\beta}_{d,d'}^*), \  u:=\sqrt{n}(\beta-{\beta}_{d,d'}^*),\\
	&R_{d,d'}(Y_i,u):=\mathcal{L}\left(Y_i-\left\{\beta_{d,d'}^*+\frac{u}{\sqrt{n}}\right\}\right)-\mathcal{L}(Y_i-\beta_{d,d'}^*)+\mathcal{L}'(Y_i-\beta_{d,d'}^*)\cdot \frac{u}{\sqrt{n}},\\
	&Q_{d,d',n}(u,\widehat{\pi}_d,\widehat{\pi}_{d'}):=\sum_{i=1}^n\frac{D_{di}}{\widehat{\pi}_d(\boldsymbol{X}_{i})}\widehat{\pi}_{d'}(\boldsymbol{X}_{i})\left[-\mathcal{L}'(Y_i-\beta_{d,d'}^*)\cdot \frac{u}{\sqrt{n}}+R_{d,d'}(Y_i,u)\right].
\end{align*}
Then we get
\begin{align*}
	\widehat{u}_{d,d'}=\arg\min_{u} Q_{d,d',n}(u,\widehat{\pi}_d,\widehat{\pi}_{d'}).
\end{align*}
Similar to the proof of Theorem \ref{thm:eff_IPW}, we can construct the quadratic approximation function for $Q_{d,d',n}(u,\widehat{\pi}_d,\widehat{\pi}_{d'})$. Note that
\begin{align}
	&Q_{d,d',n}(u,\widehat{\pi}_d,\widehat{\pi}_{d'})\notag\\
	&=-\frac{u}{\sqrt{n}}\sum_{i=1}^n\frac{D_{di}}{\widehat{\pi}_{d}(\boldsymbol{X}_{i})}{\pi}^*_{d'}(\boldsymbol{X}_{i}) \mathcal{L}'\left(Y_i-\beta_{d,d'}^*\right)+\sum_{i=1}^n\frac{D_{di}}{\widehat{\pi}_{d}(\boldsymbol{X}_{i})}\widehat{\pi}_{d'}(\boldsymbol{X}_{i})R_{d,d'}(Y_i,u)\label{eq:Q1} \\
	&+\sqrt{n}\cdot\mathbb{E}\left[\frac{D_{di}}{{\pi}^*_{d}(\boldsymbol{X}_{i})}\{\widehat{\pi}_{d'}(\boldsymbol{X}_{i})-{\pi}^*_{d'}(\boldsymbol{X}_{i})\} \mathcal{L}'\left(Y_i-\beta_{d,d'}^*\right)\right] \label{eq:Q2}\\
	&+\mathbb{G}_n\left(\frac{D_{di}}{{\pi}^*_{d}(\boldsymbol{X}_{i})}\{\widehat{\pi}_{d'}(\boldsymbol{X}_{i})-{\pi}^*_{d'}(\boldsymbol{X}_{i})\} \mathcal{L}'\left(Y_i-\beta_{d,d'}^*\right)\right) \label{eq:Q3}\\
	&-\frac{1}{\sqrt{n}}\sum_{i=1}^n\frac{D_{di}}{{\pi}^*_{d}(\boldsymbol{X}_{i})\widehat{\pi}_{d}(\boldsymbol{X}_{i})}\{\widehat{\pi}_{d'}(\boldsymbol{X}_{i})-{\pi}^*_{d'}(\boldsymbol{X}_{i})\}\{\widehat{\pi}_{d}(\boldsymbol{X}_{i})-{\pi}^*_{d}(\boldsymbol{X}_{i})\} \mathcal{L}'\left(Y_i-\beta_{d,d'}^*\right). \label{eq:Q4}
\end{align}
For the term  \eqref{eq:Q1}, similar to the proof of establishing $|\widetilde{Q}_{d,n}(u)-Q_{d,n}(u,\widehat{\pi}_d)|=o_P(u)$ in \eqref{eq:Qtilde-Q}, we have
\begin{align*}
	\eqref{eq:Q1}=&\frac{u}{\sqrt{n}}\sum_{i=1}^n\bigg[-\frac{D_{di}}{{\pi}^*_d(\boldsymbol{X}_{i})}\pi_{d'}^*(\boldsymbol{X}_{i})\mathcal{L}'(Y_i-\beta_{d,d'}^*)+\left(\frac{D_{di}}{\pi_d^*(\boldsymbol{X}_{i})}-1\right)\pi_{d'}^*(\boldsymbol{X}_{i})\mathcal{E}_{d}(\boldsymbol{X}_{i},\beta_{d,d'}^*) \bigg]\\
	&-\partial_{\beta}\mathbb{E}[\pi_{d'}^*(\boldsymbol{X}_{i})\mathcal{L}'(Y^*_i(d)-\beta_{d,d'}^*)]\cdot \frac{u^2}{2}.
\end{align*}
For the term \eqref{eq:Q2}, similar to Lemma \ref{lemma:projection} with $u^*(\boldsymbol{X})$ replaced by $\mathcal{E}_{d}(\boldsymbol{X},\beta_{d,d'}^*)$,  we can have
\begin{align*}
	\eqref{eq:Q2}=&\sqrt{n}\cdot\mathbb{E}\left[\{\widehat{\pi}_{d'}(\boldsymbol{X}_{i})-{\pi}^*_{d'}(\boldsymbol{X}_{i})\} \mathcal{E}_{d}(\boldsymbol{X}_{i},\beta_{d,d'}^*)\right]\\
	=&\frac{1}{\sqrt{n}}\sum_{i=1}^n\{D_{d'i}-{\pi}^*_{d'}(\boldsymbol{X}_{i})\}\mathcal{E}_{d}(\boldsymbol{X}_{i},\beta_{d,d'}^*)+o_P(1).
\end{align*}
Similar to \eqref{eq:mu_n:pihat-pi} and \eqref{eq:pihat-pi^2}, it can be shown that \eqref{eq:Q3} and \eqref{eq:Q4} are of $o_P(1)$.

We define the following quadratic function
\begin{align*}
	\widetilde{Q}_{d,d',n}(u):=&\frac{u}{\sqrt{n}}\sum_{i=1}^n\bigg[-\frac{D_{di}}{{\pi}^*_d(\boldsymbol{X}_{i})}\pi_{d'}^*(\boldsymbol{X}_{i})\mathcal{L}'(Y_i-\beta_{d,d'}^*)+\left(\frac{D_{di}}{\pi_d^*(\boldsymbol{X}_{i})}-1\right)\pi_{d'}^*(\boldsymbol{X}_{i})\mathcal{E}_{d}(\boldsymbol{X}_{i},\beta_{d,d'}^*) \\
	&\qquad\qquad-\left(D_{d'i}-\pi_{d'}^*(\boldsymbol{X}_{i})\right)\mathcal{E}_d(\boldsymbol{X}_{i},\beta_{d,d'}^*) \bigg]-\partial_{\beta}\mathbb{E}[\pi_{d'}^*(\boldsymbol{X}_{i})\mathcal{L}'(Y^*_i(d)-\beta_{d,d'}^*)]\cdot \frac{u^2}{2},
\end{align*}
which is an approximation for ${Q}_{d,d',n}(u)$, i.e. $\widetilde{Q}_{d,d',n}(u)-Q_{d,d',n}(u,\widehat{\pi}_d,\widehat{\pi}_{d'})=o_p(1)$ for every fixed $u$, and it does not depend on $\widehat{\pi}_d$ and $\widehat{\pi}_{d'}$. The minimizer of $\widetilde{Q}_{d,d',n}(u)$ is defined by
$$\widetilde{u}_{d,d'}:=\arg\min_{u} \widetilde{Q}_{d,d',n}(u).$$
Since $\widetilde{Q}_{d,d',n}(u)$ strictly convex and $-\partial_{\beta}\mathbb{E}[{\pi}^*_{d'}(\boldsymbol{X}_{i})\mathcal{L}'(Y^*_i(d)-\beta_{d,d'}^*)]>0$, then the minimizer $\widetilde{u}_{d,d'}$ is equal to
\begin{align*}
	\widetilde{u}_{d,d'}=\frac{1}{\sqrt{n}}\sum_{i=1}^nS_{d,d'}(Y_i,D_{di},\boldsymbol{X}_{i};\beta_{d,d'}^*),
\end{align*}
where  $H_{d,d'}:=\partial_{\beta}\mathbb{E}[{\pi}^*_{d'}(\boldsymbol{X}_{i})\mathcal{L}'(Y^*_i(d)-\beta_{d,d'}^*)]$ and
\begin{align*}
	&S_{d,d'}(Y_i,D_{di},\boldsymbol{X}_{i};\beta_{d,d'}^*)\\
	&:=H_{d,d'}^{-1} \left[\frac{D_{di}}{{\pi}^*_d(\boldsymbol{X}_{i})}{\pi}^*_{d'}(\boldsymbol{X}_{i})\mathcal{L}'(Y_i-\beta_{d,d'}^*)-\left\{\frac{D_{di}}{\pi_d^*(\boldsymbol{X}_{i})}\pi^*_{d'}(\boldsymbol{X}_{i})-D_{d'i}\right\}\mathcal{E}_{d}(\boldsymbol{X}_{i},\beta_{d,d'}^*)\right]
\end{align*}
is the influence function of $\beta_{d,d'}^*$.

Similar to the proof of Theorem \ref{thm:eff_IPW},  we can complete the following steps:
\begin{itemize}
	\item Step I: showing $\widetilde{Q}_{d,d',n}(u)-Q_{d,d',n}(u,\widehat{\pi}_d,\widehat{\pi}_{d'})=o_p(1)$ for every fixed $u$;
	\item Step II: showing $|\widehat{u}_{d,d'}-\widetilde{u}_{d,d'}|=o_P(1)$;
	\item Step III: obtaining the desired result $\sqrt{n}\{\widehat{\beta}_{d,d'}-\beta_{d,d'}^*\}=\widetilde{u}_{d,d'}+\{\widehat{u}_{d,d'}-\widetilde{u}_{d,d'}\}=n^{-1/2}\sum_{i=1}^n$ $S_{d,d'}(Y_i,D_{di},\boldsymbol{X}_{i};\beta_{d,d'}^*)+o_P(1)$.
\end{itemize}

%------------------------------

%\newpage
\section{Extension to Deep Neural Networks}\label{sec:deep}
%\newpage
%\subsection{Deep neural networks}
The proposed method can be extended to networks with multiple hidden layers. To motivate the idea, we consider the following ReLU feed-forward network function space indexed by the number of parameter $W$,
\begin{align*}
	\mathcal{G}_{W}=\left\{h_{L+1,1}(x)\right\},
\end{align*}
where $h_{u,j}(x)$ is the output of the $j^{th}$ node of the layer $u$ in the network with input $x$, $u=0$
or $u=L+1$ correspond to the input and output layers, respectively, and $1 \leq u\leq L$ correspond
to the $u^{th}$ hidden layer. We also have $j\in\{1,2,...,H_u\}$, where $H_u$ is the number of nodes (width) in the $u^{th}$ layer, $H_0=p$, and $H_{L+1}=1$. For $1\leq u \leq L$, the formula for $h_{u,j}(x)$ is
\begin{align*}
	h_{u,j}(x)=\text{ReLU}\left(\sum_{k=1}^{H_{u-1}}\gamma_{u,j,k}\cdot h_{u-1,k}(x)+\gamma_{u,j,0}\right).
\end{align*}
where $h_{0,j}(x)=x_k$, the $k^{th}$ element of $x$.

We use the upper bound $$\max_{1\leq j\leq H_u}\sum_{k=0}^{H_{u-1}}|\gamma_{u,j,k}|\leq M_u,\ \text{for all}\ 1\leq u\leq L+1,$$ where $M_u>1$, $M_u$ can depend on $n$, and $M_0=1$. Let $W$ be the number of parameters $\gamma_{u,j,k}$ in the network, with $W=\sum_{u=0}^{L}(H_u+1)H_{u+1}$. Replacing $\mathcal{G}_{n}$ in Assumption \ref{as:pihat_1st} by $\mathcal{G}_{W}$,  the estimator of the propensity score, denoted by $\widehat{\pi}^{DNN}_d(x)$, can be constructed. Then the estimator of $\beta_d^*$, denoted by $\widehat{\beta}^{DNN}_d$, can be obtained through \eqref{def:beta_d}.

We consider the Sobolev space $\mathcal{W}^{s,\infty }(\mathcal{X})$ for $s\geq 1$ defined by \citet[Section 3.2]{yarotsky2017error}
\begin{equation*}
	\mathcal{W}^{s,\infty }(\mathcal{X}):=\left\{ f:\mathcal{X}\rightarrow
	\mathbb{R}:\max_{|\boldsymbol{\alpha }|_{1}\leq s}\sup_{x\in \mathcal{X}%
	}\left\vert D^{\boldsymbol{\alpha }}f(x)\right\vert \leq 1\right\} ,
\end{equation*}%
where $\boldsymbol{\alpha }:=(\alpha _{1},...,\alpha _{p})$,  $|\boldsymbol{\alpha }|_{1}:=\sum_{j=1}^{p}\alpha _{j}$, and $D^{\boldsymbol{\alpha }}f(x):=\frac{\partial^{|\boldsymbol{\alpha }|_{1}}}{\partial ^{\alpha_{1}}x_{1}\cdots \partial ^{\alpha _{p}}x_{p}}f(x)$.  Suppose $\pi_{d}^{\ast }(\cdot )\in \mathcal{W}^{s,\infty }(\mathcal{X})$, by \citet[Proposition 1]{yarotsky2018optimal}, there exists $\text{Proj}_{\mathcal{G}_{W}}\pi _{d}^{\ast }\in \mathcal{G}_{W}$ s.t.
\begin{align}\label{eq:app_multi}
	\|\text{Proj}_{\mathcal{G}_W}\pi_d^*-\pi_d^*\|_{\infty} =O\left(W^{-\frac{s}{p}}\right).
\end{align}
Unlike Theorem \ref{thm:rate_ANN} where a ``dimension free" approximation rate can be obtained by using ANNs with one hidden layers, \eqref{eq:app_multi} indicates that the approximation rate decreases as the dimension of $X$ grows, when the propensity score function is estimated by ANNs with multiple hidden layers in which neurons between two adjacent layers are fully-connected. Hence, the curse of dimensionality problem still exists.  We treat $p$ as a fixed integer in this subsection.

Let
\begin{align*}
	M_{L+1}^*=\prod_{i=1}^{L+1}M_u,\ M_{L+1}^{all}=\max_{1\leq u\leq L+1} M_u, \ \text{and} \ C_{L+1,W}^*=W\cdot \log \left(p\cdot M_{L+1}^*\cdot W \cdot (M_{L+1}^{all})^{L} \right).
\end{align*}
Similar to the proof of Theorem \ref{thm:rate_ANN},  the convergence rate of estimated $\pi_d^*$, denoted by $\widehat{\pi}^{DNN}_d$, is
\begin{align*}
	\|\widehat{\pi}^{DNN}_d-\pi_d^*\|_{L^2(dF_X)}=O_P\left(\max\{\delta_n,\|\pi_d^*-\text{Proj}_{\mathcal{G}_W}\pi_d^*\|_{L^2(dF_X)}\}\right),
\end{align*}
where
\begin{align}\label{eq:delta_n}
	\delta_n=\inf \left\{ \delta>0:\delta^{-2}\int_{\delta^2}^{\delta}[\log N_{[\ ]}(w,n,\mathcal{G}_W)]^{1/2}dw\leq \text{const}\times n^{1/2} \right\}.
\end{align}
By \cite{van1998asymptotic}, the bracketing number $\log N_{[\ ]}\left(w,n,\mathcal{G}_W\right)$ has the following upper bound:
\begin{align*}
	\log N_{[\ ]}\left(w,n,\mathcal{G}_W\right)\leq \log N\left(\frac{1}{2}w,L^{\infty},\mathcal{G}_W\right),
\end{align*}
where $N_{[\ ]}(w,n,\mathcal{G}_W)$ denotes the bracketing number,  $N\left(\frac{1}{2}w,L^{\infty},\mathcal{G}_W\right)$ denotes the  covering number of $\mathcal{G}_W$ by balls with radius $2^{-1}w$ under $L^{\infty}$-metric. By \citet[Theorem 14.5]{anthony2009neural}, the covering number has the following upper bound:
\begin{align*}
	N\left(\frac{1}{2}w,L^{\infty},\mathcal{G}_W\right)\leq \left(\frac{8\cdot e\cdot p\cdot M^*_{L+1}\cdot W\cdot (M_{L+1}^{all})^{L+1}}{w\cdot (M_{L+1}^{all}-1)}\right)^{W}.
\end{align*}
Then
\begin{align*}
	\log N\left(\frac{1}{2}w,L^{\infty},\mathcal{G}_W\right)\leq W \cdot \log \left(\frac{8\cdot e\cdot p\cdot M^*_{L+1}\cdot W\cdot (M_{L+1}^{all})^{L+1}}{w\cdot (M_{L+1}^{all}-1)}\right).
\end{align*}
We choose
\begin{align*}
	\delta_n=\text{const}\times \{W\cdot \log W\}^{1/2}\cdot n^{-1/2},
\end{align*}
such that \eqref{eq:delta_n} is satisfied.  Setting
$\delta_n=\|\pi_d^*-\text{Proj}_{\mathcal{G}_W}\pi_d^*\|_{L^2(dF_X)}$ yields:
\begin{align*}
	W^{1+\frac{2s}{p}}\log W=O(n).
\end{align*}
Assume that the width of the $u^{\text{th}}$ layer satisfies $H_{u}=H$ for
all $1\leq u\leq L$. Then $W=\{p+1\}\cdot H+\sum_{u=1}^{L-1}\{H+1\}H+\{H+1\}=(L-1)H^2+(L+p+1)H+1$. The above
condition implies that $L$ and $H$ need to satisfy
\begin{equation*}
	L^{1+\frac{2s}{p}}H^{2(1+\frac{2s}{p})}\log (LH^{2})=O(n).
\end{equation*}%
We see that if using a deep ANNs with $L\rightarrow \infty $,  the width of
each hidden layer $H$ can be much smaller than the width of the single
hidden layer ANNs $r_{n}$ that needs to satisfy Assumption \ref{as:rate_rp}.

As a consequence,
\begin{align*}
	\|\widehat{\pi}^{DNN}_d-\pi_d^*\|_{L^2(dF_X)}=O_P\left(\left[\frac{n}{\log n}\right]^{-\frac{s}{p+2s}}\right).
\end{align*}
If $p<2s$, then we can obtain
\begin{align*}
	\|\widehat{\pi}^{DNN}_d-\pi_d^*\|_{L^2(dF_X)}=o_P(n^{-1/4}).
\end{align*}
With these results, the arguments in Lemma \ref{lemma:projection} and Theorem \ref{thm:eff_IPW} are still valid. Therefore, the asymptotic results for the proposded estimator $\widehat{\beta}^{DNN}_d$ can be similarly established.

\newpage
\section{Additional simulation results}\label{S6section}

Tables \ref{tab:simu_average_att_p5} - \ref{tab:simu_quantile_qtt_p10} present the numerical results of different methods for the estimation of the average and quantile treatment effects on the treated (ATT, QTT) for Model 1 given in Section \ref{subsec:simu_models}.  Tables \ref{tab:simu_average_ate_p5_2} - \ref{tab:simu_quantile_qtt_p10_2} show the numerical results for different methods of the estimation of all four treatment effects (ATE, ATT, QTE and QTT) for Model 2 given in Section \ref{subsec:simu_models}. The discussions of the results in these tables are given in Section \ref{simulationresults}.

\begin{table}[]
	\centering
	\caption{The summary statistics of the estimated ATTs for Model 1 with p=5}
	\renewcommand{\arraystretch}{1}
	\resizebox{\textwidth}{!}{%
\begin{tabular}{lcccccccccccccc}
			\hline
			& \multicolumn{6}{c}{PS} & \multicolumn{6}{c}{OR} & \multirow{2}{*}{DR} & \multirow{2}{*}{Oracle} \\ \cmidrule(r){2-7} \cmidrule(r){8-13}
			& \textbf{ANN} & GLM & GAM & RF & GBM & DNN & \textbf{ANN} & GLM & GAM & RF & GBM & DNN &  &  \\ \hline
			\\
			n=1000 &  &  &  &  &  &  &  &  &  &  &  &  &  &  \\
			\hline
			bias & \textbf{0.0002} & 0.0720 & 0.0767 & 0.0982 & 0.1182 & 0.0034 & \textbf{0.0009} & 0.0725 & 0.0700 & 0.0188 & 0.0198 & 0.0012 & 0.0045 & 0.0009 \\
			emp\_sd & \textbf{0.0905} & 0.0912 & 0.1024 & 0.0814 & 0.0877 & 0.0873 & \textbf{0.0813} & 0.0910 & 0.1018 & 0.0818 & 0.0975 & 0.0809 & 0.0826 & 0.0723 \\
			est\_sd & \textbf{0.0804} & 0.0933 & 0.1021 & 0.0632 & 0.0677 & 0.0799 & \textbf{0.0804} & 0.0933 & 0.1021 & 0.0632 & 0.0677 & 0.0799 & 0.0804 & 0.0733 \\
			cover\_rate & \textbf{0.9225} & 0.8850 & 0.8875 & 0.6025 & 0.5400 & 0.9250 & \textbf{0.9500} & 0.8875 & 0.8875 & 0.8700 & 0.8375 & 0.9500 & 0.9400 & 0.9500 \\
\hline
		\end{tabular}
		\label{tab:simu_average_att_p5}}
\end{table}

\begin{table}[]
	\centering
	\caption{The summary statistics of the estimated ATTs for Model 1 with p=10}
	\renewcommand{\arraystretch}{1}
	\resizebox{\textwidth}{!}{%
		\begin{tabular}{lcccccccccccccc}
			\hline
			& \multicolumn{6}{c}{PS} & \multicolumn{6}{c}{OR} & \multirow{2}{*}{DR} & \multirow{2}{*}{Oracle} \\ \cmidrule(r){2-7} \cmidrule(r){8-13}
			&\textbf{ANN} & GLM & GAM & RF & GBM & DNN &\textbf{ANN} & GLM & GAM & RF & GBM & DNN &  &  \\ \hline
			\\
			n=1000 &  &  &  &  &  &  &  &  &  &  &  &  &  &  \\
			\hline
			bias & \textbf{0.0003} & 0.0805 & 0.0920 & 0.0922 & 0.0954 & 0.0006 & \textbf{0.0105} & 0.0812 & 0.0847 & 0.0350 & 0.0380 & 0.0620 & 0. 0111 & 0.0085 \\
			emp\_sd & \textbf{0.1062} & 0.0903 & 0.1437 & 0.0897 & 0.1086 & 0.1033 & \textbf{0.0931} & 0.0905 & 0.1271 & 0.0825 & 0.0904 & 0.0928 & 0.0989 & 0.0723 \\
			est\_sd & \textbf{0.0936} & 0.0961 & 0.1220 & 0.0681 & 0.0759 & 0.0931 & \textbf{0.0936} & 0.0961 & 0.1220 & 0.0681 & 0.0759 & 0.0931 & 0.0936 & 0.0750 \\
			cover\_rate & \textbf{0.9050} & 0.8875 & 0.8350 & 0.6725 & 0.6525 & 0.9100 & \textbf{0.9425} & 0.8825 & 0.8725 & 0.8525 & 0.8750 & 0.9525 & 0.9275 & 0.9600 \\
\hline
		\end{tabular}
		\label{tab:simu_average_att_p10}
	}
	
\end{table}

\begin{table}[]
	\centering
	\caption{The summary statistics of the estimated QTTs by the IPW method for  Model 1 with p=5}
	\renewcommand{\arraystretch}{1}
	\resizebox{\textwidth}{!}{%
		\begin{tabular}{lcccccccccccccccccc}
			\hline
			& \multicolumn{6}{c}{Q1} & \multicolumn{6}{c}{Q2} & \multicolumn{6}{c}{Q3} \\ \cmidrule(r){2-7} \cmidrule(r){8-13} \cmidrule(r){14-19}
			& \textbf{ANN} & GLM & GAM & RF & GBM & DNN & \textbf{ANN} & GLM & GAM & RF & GBM & DNN & \textbf{ANN} & GLM & GAM & RF & GBM & DNN \\ \hline
			\\
			n=1000 &  &  &  &  &  &  &  &  &  &  &  &  &  &  &  &  &  &  \\
			\hline
			bias & \textbf{0.0074} & 0.0751 & 0.0794 & 0.1311 & 0.1623 & 0.0056 & \textbf{0.0019} & 0.0644 & 0.0686 & 0.0973 & 0.1151 & 0.0024 & \textbf{0.0093} & 0.0445 & 0.0513 & 0.0502 & 0.0610 & 0.0076 \\
			emp\_sd & \textbf{0.1503} & 0.1360 & 0.1527 & 0.1289 & 0.1372 & 0.1515 & \textbf{0.1203} & 0.1124 & 0.1295 & 0.1048 & 0.1116 & 0.1241 & \textbf{0.1293} & 0.1162 & 0.1306 & 0.1116 & 0.1199 & 0.1299 \\
			est\_sd & \textbf{0.1660} & 0.1409 & 0.1638 & 0.1334 & 0.1352 & 0.1679 & \textbf{0.1517} & 0.1245 & 0.1481 & 0.1199 & 0.1218 & 0.1567 & \textbf{0.1360} & 0.1238 & 0.1447 & 0.1204 & 0.1227 & 0.1378 \\
			cover\_rate & \textbf{0.9550} & 0.9050 & 0.9125 & 0.8350 & 0.7625 & 0.9525 & \textbf{0.9650} & 0.9425 & 0.9475 & 0.9150 & 0.8575 & 0.9600 & \textbf{0.9375} & 0.9575 & 0.9550 & 0.9425 & 0.9350 & 0.9400 \\
\hline
		\end{tabular}
		\label{tab:simu_quantile_qtt_p5}
	}
	
\end{table}

\begin{table}[]
	\centering
	\caption{The summary statistics of the estimated QTTs by the IPW method for  Model 1 with p=10}
	\renewcommand{\arraystretch}{1}
	\resizebox{\textwidth}{!}{%
		\begin{tabular}{lcccccccccccccccccc}
			\hline
			& \multicolumn{6}{c}{Q1} & \multicolumn{6}{c}{Q2} & \multicolumn{6}{c}{Q3} \\ \cmidrule(r){2-7} \cmidrule(r){8-13} \cmidrule(r){14-19}
			& \textbf{ANN} & GLM & GAM & RF & GBM & DNN &\textbf{ANN} & GLM & GAM & RF & GBM & DNN & \textbf{ANN} & GLM & GAM & RF & GBM & DNN \\ \hline
			\\
			n=1000 &  &  &  &  &  &  &  &  &  &  &  &  &  &  &  &  &  &  \\
			\hline
			bias & \textbf{0.0061} & 0.0871 & 0.1015 & 0.1230 & 0.1359 & 0.0081 & \textbf{0.0098} & 0.0659 & 0.0732 & 0.0815 & 0.0907 & 0.0046 & \textbf{0.0022} & 0.0621 & 0.0701 & 0.0594 & 0.0572 & 0.0027 \\
			emp\_sd & \textbf{0.2019} & 0.1370 & 0.2086 & 0.1362 & 0.1684 & 0.2025 & \textbf{0.1749} & 0.1200 & 0.1860 & 0.1210 & 0.1435 & 0.1799 & \textbf{0.1549} & 0.1181 & 0.1800 & 0.1211 & 0.1369 & 0.1555 \\
			est\_sd & \textbf{0.3678} & 0.1458 & 0.4513 & 0.1355 & 0.1481 & 0.3421 & \textbf{0.3018} & 0.1281 & 0.4455 & 0.1209 & 0.1317 & 0.2975 & \textbf{0.2504} & 0.1264 & 0.4221 & 0.1209 & 0.1309 & 0.2321 \\
			cover\_rate & \textbf{0.9600} & 0.9150 & 0.9750 & 0.8475 & 0.8175 & 0.9575 & \textbf{0.9425} & 0.9225 & 0.9975 & 0.8875 & 0.8475 & 0.9450 & \textbf{0.9675} & 0.9275 & 0.9925 & 0.9075 & 0.9175 & 0.9575 \\
\hline
		\end{tabular}
		\label{tab:simu_quantile_qtt_p10}
	}
	
\end{table}

\begin{table}[]
	\centering
	\caption{The summary statistics of the estimated ATEs for Model 2 with p=5}
	\renewcommand{\arraystretch}{1}
	\resizebox{\textwidth}{!}{%
		\begin{tabular}{lcccccccccccccc}
			\hline
			& \multicolumn{6}{c}{PS} & \multicolumn{6}{c}{OR} & \multirow{2}{*}{DR} & \multirow{2}{*}{Oracle} \\ \cmidrule(r){2-7} \cmidrule(r){8-13}
			& \textbf{ANN} & GLM & GAM & RF & GBM & DNN & \textbf{ANN} & GLM & GAM & RF & GBM & DNN &  &  \\ \hline
			\\
			n=1000 &  &  &  &  &  &  &  &  &  &  &  &  &  &  \\
			\hline
			bias & \textbf{0.0043} & 0.0207 & 0.0214 & 0.2254 & 0.1886 & 0.0037 & \textbf{0.0203} & 0.0194 & 0.0195 & 0.0499 & 0.0385 & 0.0189 & 0.0200 & 0.0190 \\
			emp\_sd & \textbf{0.1676} & 0.1182 & 0.1424 & 0.1933 & 0.1839 & 0.1701 & \textbf{0.1235} & 0.1185 & 0.1204 & 0.1367 & 0.1345 & 0.1227 & 0.1219 & 0.1190 \\
			est\_sd & \textbf{0.1207} & 0.1203 & 0.1205 & 0.1198 & 0.1178 & 0.1213 & \textbf{0.1207} & 0.1203 & 0.1205 & 0.1198 & 0.1178 & 0.1213 & 0.1208 & 0.1203 \\
			cover\_rate & \textbf{0.9225} & 0.9525 & 0.9000 & 0.4950 & 0.5700 & 0.9200 & \textbf{0.9475} & 0.9525 & 0.9525 & 0.8850 & 0.8850 & 0.9475 & 0.9500 & 0.9575 \\
\hline
		\end{tabular}
		\label{tab:simu_average_ate_p5_2}}
\end{table}

\begin{table}[]
	\centering
	\caption{The summary statistics of the estimated ATEs for Model 2 with p=10}
	\renewcommand{\arraystretch}{1}
	\resizebox{\textwidth}{!}{%
		\begin{tabular}{lcccccccccccccc}
			\hline
			& \multicolumn{6}{c}{PS} & \multicolumn{6}{c}{OR} & \multirow{2}{*}{DR} & \multirow{2}{*}{Oracle} \\ \cmidrule(r){2-7} \cmidrule(r){8-13}
			& \textbf{ANN} & GLM & GAM & RF & GBM & DNN & \textbf{ANN} & GLM & GAM & RF & GBM & DNN &  &  \\ \hline
			\\
			n=1000 &  &  &  &  &  &  &  &  &  &  &  &  &  &  \\
			\hline
			bias & \textbf{0.0107} & 0.0004 & 0.0067 & 0.3506 & 0.1334 & 0.0980 & \textbf{0.0013} & 0.0020 & 0.0049 & 0.1367 & 0.0679 & 0.0045 & 0.0073 & 0.0020 \\
			emp\_sd & \textbf{0.1582} & 0.1285 & 0.2410 & 0.2218 & 0.1641 & 0.1632 & \textbf{0.1323} & 0.1265 & 0.1283 & 0.1832 & 0.1635 & 0.1330 & 0.1382 & 0.1262 \\
			est\_sd & \textbf{0.1359} & 0.1293 & 0.1317 & 0.1427 & 0.1369 & 0.1360 & \textbf{0.1359} & 0.1293 & 0.1317 & 0.1427 & 0.1369 & 0.1360 & 0.1357 & 0.1293 \\
			cover\_rate & \textbf{0.9275} & 0.9575 & 0.7225 & 0.3700 & 0.7675 & 0.9250 & \textbf{0.9525} & 0.9550 & 0.9525 & 0.7650 & 0.8850 & 0.9525 & 0.9500 & 0.9500 \\
\hline
		\end{tabular}
		\label{tab:simu_average_ate_p10_2}
	}
	
\end{table}

\begin{table}[]
	\centering
	\caption{The summary statistics of the estimated ATTs for Model 2 with p=5}
	\renewcommand{\arraystretch}{1}
	\resizebox{\textwidth}{!}{%
		\begin{tabular}{lcccccccccccccc}
			\hline
			& \multicolumn{6}{c}{PS} & \multicolumn{6}{c}{OR} & \multirow{2}{*}{DR} & \multirow{2}{*}{Oracle} \\ \cmidrule(r){2-7} \cmidrule(r){8-13}
			& \textbf{ANN} & GLM & GAM & RF & GBM & DNN & \textbf{ANN} & GLM & GAM & RF & GBM & DNN &  &  \\ \hline
			\\
			n=1000 &  &  &  &  &  &  &  &  &  &  &  &  &  &  \\
			\hline
			bias & \textbf{0.0184} & 0.0029 & 0.0018 & 0.1538 & 0.0628 & 0.0134 & \textbf{0.0033} & 0.0017 & 0.0003 & 0.1098 & 0.0642 & 0.0045 & 0.0044 & 0.0023 \\
			emp\_sd & \textbf{0.2037} & 0.1776 & 0.3042 & 0.2421 & 0.2083 & 0.2137 & \textbf{0.1781} & 0.1681 & 0.1724 & 0.2277 & 0.2207 & 0.1809 & 0.1823 & 0.1690 \\
			est\_sd & \textbf{0.1855} & 0.1731 & 0.1809 & 0.1778 & 0.1834 & 0.1905 & \textbf{0.1855} & 0.1731 & 0.1809 & 0.1778 & 0.1834 & 0.1905 & 0.1836 & 0.1725 \\
			cover\_rate & \textbf{0.9125} & 0.9350 & 0.7575 & 0.7800 & 0.9025 & 0.9125 & \textbf{0.9450} & 0.9500 & 0.9575 & 0.8050 & 0.9000 & 0.9450 & 0.9475 & 0.9550 \\
\hline
		\end{tabular}
		\label{tab:simu_average_att_p5_2}}
\end{table}

\begin{table}[]
	\centering
	\caption{The summary statistics of the estimated ATTs for Model 2 with p=10}
	\renewcommand{\arraystretch}{1}
	\resizebox{\textwidth}{!}{%
		\begin{tabular}{lcccccccccccccc}
			\hline
			& \multicolumn{6}{c}{PS} & \multicolumn{6}{c}{OR} & \multirow{2}{*}{DR} & \multirow{2}{*}{Oracle} \\ \cmidrule(r){2-7} \cmidrule(r){8-13}
			&\textbf{ANN} & GLM & GAM & RF & GBM & DNN & \textbf{ANN} & GLM & GAM & RF & GBM & DNN &  &  \\ \hline
			\\
			n=1000 &  &  &  &  &  &  &  &  &  &  &  &  &  &  \\
			\hline
			bias & \textbf{0.0107} & 0.0004 & 0.0067 & 0.3506 & 0.1334 & 0.0980 & \textbf{0.0013} & 0.0020 & 0.0049 & 0.1367 & 0.0679 & 0.0045 & 0.0073 & 0.0020 \\
			emp\_sd & \textbf{0.1582} & 0.1285 & 0.2410 & 0.2218 & 0.1641 & 0.1632 & \textbf{0.1323} & 0.1265 & 0.1283 & 0.1832 & 0.1635 & 0.1330 & 0.1382 & 0.1262 \\
			est\_sd & \textbf{0.1359} & 0.1293 & 0.1317 & 0.1427 & 0.1369 & 0.1360 & \textbf{0.1359} & 0.1293 & 0.1317 & 0.1427 & 0.1369 & 0.1360 & 0.1357 & 0.1293 \\
			cover\_rate & \textbf{0.9275} & 0.9575 & 0.7225 & 0.3700 & 0.7675 & 0.9250 & \textbf{0.9525} & 0.9550 & 0.9525 & 0.7650 & 0.8850 & 0.9525 & 0.9500 & 0.9500 \\
\hline
		\end{tabular}
		\label{tab:simu_average_att_p10_2}
	}
	
\end{table}

\begin{table}[]
	\centering
	\caption{The summary statistics of the estimated QTEs by the IPW method for Model 2 with p=5}
	\renewcommand{\arraystretch}{1}
	\resizebox{\textwidth}{!}{%
		\begin{tabular}{lcccccccccccccccccc}
			\hline
			& \multicolumn{6}{c}{Q1} & \multicolumn{6}{c}{Q2} & \multicolumn{6}{c}{Q3} \\ \cmidrule(r){2-7} \cmidrule(r){8-13} \cmidrule(r){14-19}
			& \textbf{ANN} & GLM & GAM & RF & GBM & DNN & \textbf{ANN} & GLM & GAM & RF & GBM & DNN & \textbf{ANN} & GLM & GAM & RF & GBM & DNN \\ \hline
			\\
			n=1000 &  &  &  &  &  &  &  &  &  &  &  &  &  &  &  &  &  &  \\
			\hline
			bias & \textbf{0.0098} & 0.0046 & 0.0104 & 0.2192 & 0.1814 & 0.0113 & \textbf{0.0075} & 0.0211 & 0.0242 & 0.2209 & 0.1772 & 0.0046 & \textbf{0.0175} & 0.0174 & 0.0262 & 0.2362 & 0.1955 & 0.0099 \\
			emp\_sd & \textbf{0.3349} & 0.3357 & 0.3618 & 0.3462 & 0.3479 & 0.3531 & \textbf{0.3236} & 0.3031 & 0.3299 & 0.3180 & 0.3255 & 0.3481 & \textbf{0.5221} & 0.3401 & 0.3609 & 0.3504 & 0.3496 & 0.4981 \\
			est\_sd & \textbf{0.4023} & 0.3315 & 0.4511 & 0.3507 & 0.3680 & 0.3988 & \textbf{0.6874} & 0.2882 & 0.4017 & 0.3124 & 0.3304 & 0.7012 & \textbf{0.3824} & 0.3366 & 0.4260 & 0.3535 & 0.3729 & 0.4221 \\
			cover\_rate & \textbf{0.9525} & 0.9475 & 0.9775 & 0.9100 & 0.9400 & 0.9500 & \textbf{0.9500} & 0.9300 & 0.9550 & 0.8850 & 0.9150 & 0.9600 & \textbf{0.9550} & 0.9450 & 0.9650 & 0.9075 & 0.9300 & 0.9500 \\
\hline
		\end{tabular}
		\label{tab:simu_quantile_qte_p5_2}
	}
\end{table}

\begin{table}[]
	\centering
	\caption{The summary statistics of the estimated QTEs by the IPW method for Model 2 with  p=10}
	\renewcommand{\arraystretch}{1}
	\resizebox{\textwidth}{!}{%
		\begin{tabular}{lcccccccccccccccccc}
			\hline
			& \multicolumn{6}{c}{Q1} & \multicolumn{6}{c}{Q2} & \multicolumn{6}{c}{Q3} \\ \cmidrule(r){2-7} \cmidrule(r){8-13} \cmidrule(r){14-19}
			& \textbf{ANN} & GLM & GAM & RF & GBM & DNN & \textbf{ANN} & GLM & GAM & RF & GBM & DNN & \textbf{ANN} & GLM & GAM & RF & GBM & DNN \\ \hline
			\\
			n=1000 &  &  &  &  &  &  &  &  &  &  &  &  &  &  &  &  &  &  \\
			\hline
			bias & \textbf{0.0153} & 0.0263 & 0.0011 & 0.3586 & 0.1531 & 0.0143 & \textbf{0.0055} & 0.0052 & 0.0216 & 0.3419 & 0.1266 & 0.0055 & \textbf{0.0141} & 0.0063 & 0.0196 & 0.3714 & 0.1319 & 0.0183 \\
			emp\_sd & \textbf{0.4930} & 0.3731 & 0.4678 & 0.3979 & 0.3882 & 0.5320 & \textbf{0.4603} & 0.2955 & 0.4150 & 0.3514 & 0.3194 & 0.4980 & \textbf{0.5110} & 0.3733 & 0.5267 & 0.4044 & 0.3845 & 0.5980 \\
			est\_sd & \textbf{0.6360} & 0.3657 & 1.9238 & 0.3921 & 0.3707 & 0.7050 & \textbf{0.6659} & 0.3171 & 1.6744 & 0.3495 & 0.3217 & 0.7053 & \textbf{0.6370} & 0.3726 & 1.4469 & 0.3940 & 0.3758 & 0.6510 \\
			cover\_rate & \textbf{0.9700} & 0.9275 & 1.0000 & 0.8600 & 0.9175 & 0.9800 & \textbf{0.9750} & 0.9625 & 0.9975 & 0.8275 & 0.9300 & 0.9775 & \textbf{0.9675} & 0.9475 & 1.0000 & 0.8375 & 0.9300 & 0.9600 \\
\hline
		\end{tabular}
		\label{tab:simu_quantile_qte_p10_2}
	}
\end{table}

\begin{table}[]
	\centering
	\caption{The summary statistics of the estimated QTTs by the IPW method for Model 2 with p=5}
	\renewcommand{\arraystretch}{1}
	\resizebox{\textwidth}{!}{%
		\begin{tabular}{lcccccccccccccccccc}
			\hline
			& \multicolumn{6}{c}{Q1} & \multicolumn{6}{c}{Q2} & \multicolumn{6}{c}{Q3} \\ \cmidrule(r){2-7} \cmidrule(r){8-13} \cmidrule(r){14-19}
			& \textbf{ANN} & GLM & GAM & RF & GBM & DNN & \textbf{ANN} & GLM & GAM & RF & GBM & DNN & \textbf{ANN} & GLM & GAM & RF & GBM & DNN \\ \hline
			\\
			n=1000 &  &  &  &  &  &  &  &  &  &  &  &  &  &  &  &  &  &  \\
			\hline
			bias & \textbf{0.0118} & 0.0085 & 0.0052 & 0.1674 & 0.1446 & 0.0068 & \textbf{0.0119} & 0.0093 & 0.0274 & 0.0916 & 0.0478 & 0.0112 & \textbf{0.0170} & 0.0045 & 0.0297 & 0.0051 & 0.0463 & 0.0167 \\
			emp\_sd & \textbf{0.3751} & 0.3662 & 0.4243 & 0.3725 & 0.3939 & 0.3751 & \textbf{0.3429} & 0.3248 & 0.3628 & 0.3519 & 0.3806 & 0.3522 & \textbf{0.3784} & 0.3618 & 0.3859 & 0.3697 & 0.3927 & 0.4612 \\
			est\_sd & \textbf{0.3867} & 0.3513 & 0.4605 & 0.3985 & 0.4366 & 0.3867 & \textbf{0.3441} & 0.2939 & 0.3992 & 0.3503 & 0.3858 & 0.3551 & \textbf{0.4489} & 0.3260 & 0.4126 & 0.3807 & 0.4136 & 0.4531 \\
			cover\_rate & \textbf{0.9300} & 0.9475 & 0.9625 & 0.9550 & 0.9650 & 0.9300 & \textbf{0.9325} & 0.9200 & 0.9475 & 0.9450 & 0.9600 & 0.9325 & \textbf{0.9400} & 0.9150 & 0.9600 & 0.9675 & 0.9600 & 0.9500 \\
\hline
		\end{tabular}
		\label{tab:simu_quantile_qtt_p5_2}
	}
\end{table}

\begin{table}[]
	\centering
	\caption{The summary statistics of the estimated QTTs by the IPW method for Model 2 with p=10}
	\renewcommand{\arraystretch}{1}
	\resizebox{\textwidth}{!}{%
		\begin{tabular}{lcccccccccccccccccc}
			\hline
			& \multicolumn{6}{c}{Q1} & \multicolumn{6}{c}{Q2} & \multicolumn{6}{c}{Q3} \\ \cmidrule(r){2-7} \cmidrule(r){8-13} \cmidrule(r){14-19}
			& \textbf{ANN} & GLM & GAM & RF & GBM & DNN & \textbf{ANN} & GLM & GAM & RF & GBM & DNN & \textbf{ANN} & GLM & GAM & RF & GBM & DNN \\ \hline
			\\
			n=1000 &  &  &  &  &  &  &  &  &  &  &  &  &  &  &  &  &  &  \\
			\hline
			bias & \textbf{0.0072} & 0.0029 & 0.0042 & 0.2291 & 0.0772 & 0.0109 & \textbf{0.0508} & 0.0146 & 0.0313 & 0.1348 & 0.0501 & 0.0443 & \textbf{0.0125} & 0.0188 & 0.0158 & 0.0932 & 0.0597 & 0.0143 \\
			emp\_sd & \textbf{0.6496} & 0.4152 & 0.5834 & 0.4378 & 0.4260 & 0.7649 & \textbf{0.6306} & 0.3256 & 0.4781 & 0.3725 & 0.3483 & 0.6896 & \textbf{0.5388} & 0.3874 & 0.5469 & 0.4051 & 0.4045 & 0.5788 \\
			est\_sd & \textbf{0.8153} & 0.3965 & 1.8390 & 0.4587 & 0.4094 & 0.9054 & \textbf{0.8247} & 0.3262 & 1.5928 & 0.4008 & 0.3414 & 0.8643 & \textbf{0.6777} & 0.3577 & 1.3671 & 0.4341 & 0.3742 & 0.7009 \\
			cover\_rate & \textbf{0.9575} & 0.9400 & 0.9900 & 0.9275 & 0.9350 & 0.9650 & \textbf{0.9650} & 0.9450 & 0.9950 & 0.9425 & 0.9225 & 0.9675 & \textbf{0.9525} & 0.9425 & 0.9900 & 0.9600 & 0.9500 & 0.9625 \\
\hline
		\end{tabular}
		\label{tab:simu_quantile_qtt_p10_2}
	}
\end{table}

\singlespacing
\bibliographystyle{econometrica}
\bibliography{Semiparametric}

%%%%%%%%%%%%%%%%%%%%%%%%%%%%%%%%%%%%%%%%%%%%%%%%%%%%%%%%%%%%%%%%%%%%%%%

\end{document}